\documentclass[12pt,dvipsnames]{article}

\usepackage[utf8]{inputenc}
\usepackage{amsmath,amsthm,amssymb}
\usepackage{mathtools}
\usepackage{nicefrac}

\usepackage[dvipsnames]{xcolor}
\usepackage{array}
\usepackage{graphicx}
\usepackage{natbib}
\setcitestyle{comma}
\usepackage{booktabs}
\usepackage{adjustbox}
\usepackage{setspace}
\usepackage{ragged2e}
\usepackage{caption}
\usepackage{subcaption}
\usepackage{url}
\usepackage{pdflscape}
\usepackage{placeins}
\usepackage{float}

\usepackage{hyperref}
\definecolor{linkblue}{RGB}{0,64,160}
\hypersetup{
    hidelinks,
    hypertexnames = false,
    breaklinks = true,
    pdftitle = {Huge If True: Learning from the Size of the Surprise}
}
\topmargin -0.25in \textheight 8.5in \hfuzz=50pt  \def\sym#1{\ifmmode^{#1}\else$^{#1}$\fi}
\newcommand{\fittable}[1]{\adjustbox{max width=\linewidth}{#1}}
\newcolumntype{Y}{>{\centering\arraybackslash}p{2.15cm}}
\newcommand{\fitpage}[1]{\adjustbox{max width=\linewidth, max totalheight=0.62\textheight}{#1}}

\providecommand{\Exp}{\mathbb{E}}
\providecommand{\Var}{\operatorname{Var}}

\DeclareMathOperator{\sign}{sign}
\providecommand{\SUE}{\mathrm{SUE}}

\newtheorem{proposition}{Proposition}
\newtheorem{corollary}{Corollary}

\makeatletter
\renewcommand\@biblabel[1]{}
\makeatother
\begin{document}

\author{
\begin{tabular}{@{}c@{\hspace{2.2cm}}c@{}}
Thomas Graeber & Christopher Roth \\[0.35em]
Marco Sammon & Constantin Schesch
\end{tabular}}

\title{\vspace*{0cm}{\Large\textsc{Huge If True: \\ Learning from the Size of the Surprise}}\thanks{\protect\singlespacing\scriptsize
\vspace*{-2.5em} \hspace*{1.1em} This paper was previously circulated under the title \textit{Coarse Categories in a Complex World}. We thank Nick Barberis, Cary Frydman, David Hirshleifer, Eben Lazarus, Jesse Shapiro, Andrei Shleifer, Chloe Xie, Florian Zimmermann and seminar audiences in Berlin, Booth School of Business, CESifo behavioral economics, Harvard Business School, NBER Behavioral Finance, and the Zurich Workshop on Economics and Psychology for helpful comments and suggestions. We thank Dustin Fichmann, Milena Jessen, Julian König, Felix Mao, Max Müller and Nicolas Röver for outstanding research assistance. We thank Andrey Golubov and Sonia Konstantinidi for sharing their valuation uncertainty measure. The experiment received IRB approval at Harvard Business School. Roth acknowledges funding by the Deutsche Forschungsgemeinschaft (DFG, German Research Foundation) under Germany’s Excellence Strategy – EXC 2126/1-390838866. This work was partially supported by the RCN through its CoE Scheme, FAIR project No 262675. This research program and related results were made possible by the support of the NOMIS Foundation. Graeber: University of Zurich, \texttt{thomas.graeber@econ.uzh.edu}. Roth: LMU Munich, \texttt{christopher.roth@econ.lmu.de}. Sammon: Harvard Business School, \texttt{mcsammon@gmail.com}. Schesch: Sciences Po, \texttt{constantin.schesch@sciencespo.fr}.}}

\date{\small \today}
\maketitle

\vspace{-0.35in}

\begin{abstract}

\hyphenpenalty=750
\begin{spacing}{1}
\noindent Responses to earnings surprises depend on an unknown multiplier mapping earnings to value. Investors, however, learn about the news and the multiplier from the same number. Large surprises may indicate small multipliers and tiny surprises large ones. We model how greater uncertainty about the multiplier attenuates responses on average but makes them steeper near zero surprises. In an incentivized experiment, we find that greater processing constraints lower average sensitivity but raise it near zero. We confirm this with real-world earnings announcements: stocks with higher valuation uncertainty, our proxy for multiplier uncertainty, respond less on average but more steeply near zero.
\end{spacing}
\end{abstract}

\thispagestyle{empty}

\vfill

\vspace{-1in}

\pagebreak
\onehalfspacing \addtocounter{page}{-1}

\normalsize

\section{Introduction}\label{sec:intro}

Suppose a firm reports earnings that exceed the analyst consensus by 1\% of its share price. Investors will tend to conclude that, all else equal, the firm is worth a little more than they thought. Now suppose instead that the firm reports earnings that exceed the consensus by 100\% of its share price. Taken at face value, a surprise this large could imply an enormous increase in value. Firms, however, rarely see such significant changes in market capitalization overnight, so investors will wonder how informative this earnings surprise is about firm value in the first place. That is, a surprise that would be huge if true is likely to receive less weight.

This paper studies how investors solve the problem of mapping earnings news into a price response, and how uncertainty about this mapping shapes their reaction. We argue that investors use a multiplier to map earnings surprises into changes in value -- for example, a price-to-earnings ratio -- but are uncertain about its magnitude.  This is because the multiplier depends on hard-to-assess factors, including the persistence of the earnings surprise, the expected growth rate of cash flows, and firm risk. In turn, this \textit{multiplier uncertainty} means that the size of the surprise itself, combined with knowledge about the typical scale of returns, contains information about the magnitude of the multiplier. In particular, large surprises would imply implausibly large returns if multipliers were large, suggesting instead that multipliers are likely to be small. This pushes investors to react more strongly to small than to large surprises, making their overall response to earnings news $S$-shaped: steep around the consensus and flat for large surprises.

Our central result is that greater multiplier uncertainty steepens this S-shape, i.e., it leads investors to respond less on average, but more to surprises near the consensus. This contrast is what distinguishes multiplier uncertainty from standard accounts of decision difficulty, in which harder problems or tighter information processing constraints weaken the response to news everywhere \citep{charles2024insensitive,woodford2020modeling,enke2025behavioral,gabaix2019behavioral}. We formalize this insight in a model and test this prediction in two empirical approaches. First, we run a controlled experiment and find that processing constraints shape responses to news as predicted by our model. Second, we use field data and a measure of how dispersed multipliers are in a valuation model to show that the model's predictions also describe market responses to earnings news.

Multiplier uncertainty has several sources, which can reinforce one another. Valuation problems are inherently difficult, as the mapping from news to value depends on many unobservable quantities. Consider, for example, the Gordon growth formula, $P = CF/(\rho-g)$ where $P$ is the price, $CF$ the next-period cash flow, $g$ is the cash flow growth rate and $\rho$ the discount rate.  Here, $1/(\rho-g)$ can be viewed as the ``multiplier'' translating an earnings surprise $\Delta CF$ into a price change $\Delta P$. Applying this multiplier, however, requires investors to assess the persistence of the cash flow shock, the growth rate of future cash flows ($g$), and firm risk ($\rho$). Further, these judgments require time and effort, yet investors must often act quickly after an announcement. Processing constraints can therefore amplify the uncertainty inherent in any valuation problem, making it harder for investors to determine exactly how strongly they should respond to a given piece of earnings news.

We formalize this mechanism in a model in which an investor observes an earnings surprise $s$ and forecasts the return $r=\beta s$. They know that $\beta$ is positive and know the distribution of returns, but are uncertain about the magnitude of $\beta$ and must infer it from the surprise itself. Under these assumptions, the Bayesian response to news takes a simple closed form and is $S$-shaped. We then show that, as uncertainty about the multiplier increases, average sensitivity decreases, but sensitivity to small surprises goes up, widening the gap between small and large surprises. In the limit, the investor's response becomes categorical: they react to whether earnings beat or missed the consensus, but not to the size of the surprise.

We first test this prediction in a novel, preregistered experiment in which we manipulate processing constraints while holding valuation-relevant information fixed. We have almost 900 participants, all US adults with stock trading accounts, forecast announcement-day returns for five real firms actually about to report their earnings. Respondents are given the firms' stock prices and consensus forecasts, along with randomized earnings realizations. The variation in earnings realizations allows us to incentivize decisions using the firm's earnings realization later on: if a participant sees an earnings realization that happens to be in a window around the subsequently realized earnings per share (EPS), we pay participants based on the accuracy of their forecast relative to the realized stock price reaction. To study the role of information processing constraints, we randomly assign about half of participants to the \textit{High Constraints} treatment, which adds neutral additional text on the firms' history and requires a response within 40 seconds. This experimental manipulation directly constrains the participants' ability to map a given surprise into their forecast of the announcement-day return, and should therefore raise multiplier uncertainty.

The treatment affects responses as the model predicts. Under tighter processing constraints, respondents react \emph{less} strongly to large earnings surprises but \emph{more} strongly to small ones, which makes the response function more $S$-shaped. In a simple piecewise-linear specification, tighter constraints reduce sensitivity over the full range of earnings news by about one fourth, while increasing sensitivity around the consensus by about one fourth. The result also holds for raw percentage forecasts, under various sample restrictions, and with smooth functional forms.

Second, we take our model's predictions to the field.  Specifically, we analyze nearly a quarter of a million earnings announcements by more than 7,500 US firms between 1984 and 2023. We define standardized unexpected earnings (SUE) as reported earnings per share minus the analyst consensus, divided by the pre-announcement stock price \citep{dellavigna2009investor}, and study the five-day market-adjusted return around the announcement. Our main proxy for multiplier uncertainty is the valuation uncertainty (VU) measure of \citet{golubov2026accounting}.

Conceptually, VU asks how tightly a firm's fundamentals pin down its market value: it applies multipliers to each input in an accounting valuation model, where the multipliers are estimated at the lower and upper quartiles of the industry-year valuation distribution.  These multipliers are then applied to each individual firm's fundamentals at that point in time.  Finally, VU is defined as the gap between the resulting implied values, and is therefore designed to capture the difficulty of valuing a firm given uncertainty about the multipliers and a given firm's fundamentals at that time. To summarize, VU is highest when the same fundamentals can support a very wide range of different valuations. This makes it particularly well-suited to test our mechanism: rather than capturing uncertainty about what the earnings number will be, it reflects uncertainty about how the announced number maps into value.

Empirically, we find that valuation uncertainty is associated with the same change in the shape of the response function predicted by the model. Specifically, high-VU firms have a steeper return response around the analyst consensus and a flatter response for larger surprises. In terms of magnitudes, on average across our entire data, a 0.1 percentage-point increase in SUE is associated with a 0.197 percentage-point higher five-day post-announcement return. A one-standard-deviation increase in VU \textit{lowers} this response by 0.029 percentage points, consistent with the familiar prediction of global attenuation in the face of increased uncertainty. 

The key sign flip appears when we allow sensitivity to differ between small and large surprises. For the same 0.1 percentage-point increase in SUE, a one-standard-deviation increase in VU \textit{lowers} the return response by 0.013 percentage points among large surprises, but \textit{raises} it by 0.088 percentage points among small surprises. Thus, VU leads to increased sensitivity near the consensus (i.e., for SUE near zero).  Further, VU's effect on the earnings response is 0.101 percentage points larger for small than for large surprises. To summarize, higher valuation uncertainty is associated with weaker responsiveness to earnings news on average, but stronger responsiveness precisely around the consensus, as predicted by the model.

We provide several additional pieces of evidence to bolster our preferred interpretation of these empirical patterns. First, we show that we obtain similar results using textual uncertainty and textual complexity from firms' regulatory filings \citep{loughran2011liability,loughran2024measuring}. These measures of valuation difficulty do not rely on market data and are only weakly correlated with VU.  Despite this, we find that both textual measures predict greater sensitivity near the consensus. 

Second, we show that the change in the shape of the response function associated with VU looks very different from other familiar factors known to attenuate earnings responses. Analyst disagreement, Friday announcements, dispersed peer news, conflicting management guidance, and prior return variance all reduce earnings sensitivity on average, but also \textit{reduce} responsiveness near the consensus. We argue this is evidence that these measures capture the overall attention of investors or uncertainty about the signal realization itself, rather than the difficulty of mapping signals into values. 

Next, we show that our results are robust to a variety of alternative specifications.  Specifically, they are not sensitive to how we measure abnormal returns, how we define the earnings surprise, and how we quantify the shape of the earnings response function. Further, we show that our results are not driven by differential earnings persistence \citep{freeman1992nonlinear} and are not driven by contemporaneous management guidance \citep{hand2022explaining}.

Lastly, we look at longer-horizon returns, and ask whether the stronger response to small surprises simply reflects overreaction, or the weaker response to large surprises underreaction. If our findings were driven by the fact that high VU predicts over- and underreaction at different points in the SUE distribution, we would expect to observe subsequent reversal or continuation.  Our model, on the other hand, suggests that investors react optimally given the information they have, so it does not imply post-announcement return predictability. 

Empirically, we do not find evidence of continuation or reversal. Specifically, while we find post-earnings-announcement drift in our sample, there is no evidence that VU differentially affects the post-announcement response. We view this as consistent with the model: when investors are uncertain about the earnings-to-value mapping, greater sensitivity to small surprises and greater discounting of large ones need not represent mistakes, but can instead reflect how the news itself changes beliefs about how informative earnings are for value.

\paragraph{Related Literature.}

Our paper connects a number of strands in the literature. 

First, it contributes to research on how cognitive constraints shape responses to new information. Models of inattention and imperfect cognition typically predict attenuated reactions in beliefs and choices \citep{gabaix2019behavioral,woodford2020modeling,ilut2023economic,enke2023cognitive,gabaix2024complexity}. \citet{enke2025behavioral} document this pattern across a broad range of experiments. Other mechanisms instead generate amplified responses, like salience \citep{kahneman1973psychology,bordalo2013salience,bordalo2020overreaction}, while some lead to under- and overreaction in different contexts \citep{barberis1998model,daniel1998investor,koszegi2013model,afrouzi2023overreaction}. We show that both can emerge in the same informational environment across large or small signal realizations.

Our mechanism is most closely related to work on signals of uncertain strength. \citet{ba2024underreaction} show how simplified mental representations can lead to overreaction to surprising news. Closer to us, \citet{augenblick2025overinference} consider decision makers who perfectly observe a signal but only imperfectly infer its significance from a second signal, so that shrinkage toward an intermediate strength leads to overinference from weak signals and underinference from strong ones.  We share the premise that the direction of a response can be easier to determine than its magnitude, but study an inference problem in which the same observed surprise is used to infer both fundamental news and its significance, leading to amplification of small signals and attenuation of large ones.\footnote{Coarse representations can generate similar non-linearities \citep{mullainathan2008coarse}. Relatedly, \citet{dean2026reference} show how comparisons with reference points can amplify sensitivity around them, while our mechanism does not require an additional comparative signal.} A close antecedent is \citet{subramanyam1996uncertain}, where rational investors are uncertain about signal precision and use the realization itself to infer how informative the signal is, so that extreme realizations can be discounted as evidence of lower precision. A model of this type can generate an $S$-shaped response much like ours, and we share its basic inference logic. However, we place uncertainty on the multiplier that maps an observed signal into the appropriate response, which allows for a behavioral interpretation of the mechanism in terms of processing constraints. Since $S$-shaped responses to news can arise from many sources, however, we view the shape itself as only weak evidence for any particular mechanism, and instead derive and test a comparative static that this literature has not formulated or tested: greater multiplier uncertainty lowers average sensitivity while raising sensitivity to small surprises. Taking this comparative static to the data requires empirical counterparts to a change in multiplier uncertainty, which earlier work did not propose, and which we obtain from randomized processing constraints in the experiment and from valuation uncertainty in the field.

Moreover, our empirical analysis contributes to research on reactions to financial information under processing constraints. Friday announcements and competing news releases are associated with weaker immediate earnings responses and stronger subsequent post-earnings announcement drift \citep{dellavigna2009investor,hirshleifer2009driven}. In line with our mechanism of multiplier uncertainty, more recent work documents failures to scale market reactions proportionally \citep{shue2021can} and noisy transmission of subjective expectations into experimental valuations \citep{charles2024insensitive}. \citet{laarits2025retail} show that retail-favored, hard-to-value stocks exhibit \textit{lower} overall earnings sensitivity, while \citet{hong2025uncertain} show that uncertain language in corporate filings is associated with \textit{stronger} immediate responses and less subsequent drift on average. We show that the non-linear shape of the reaction function, predicted by our model and observed in practice, is critical to identify the contrasting effects of processing constraints across small and large surprises. To that end, we use valuation uncertainty from \citet{golubov2026accounting} as a measure of uncertainty about the earnings-to-value mapping. The central finding is that similar average attenuation can accompany opposite changes over the range of surprises: whereas several proxies for overall inattention produce uniform attenuation, valuation uncertainty reduces overall sensitivity while amplifying it around the analyst consensus.

Finally, we build on the literature about non-linear price responses to earnings news. \citet{freeman1992nonlinear} documented declining marginal responses as surprises grow and argued that larger earnings surprises are less persistent \citep[see also][]{das1994nonlinearity}. A second strand emphasizes heterogeneity in the information content of earnings across firms and states, including the weaker response to losses \citep{hayn1995information}, differences across growth and value firms \citep{skinner2002earnings}, and variation with analyst disagreement \citep{kinney2002earnings}. A third strand focuses on the special role of the analyst-consensus threshold: firms have incentives to meet or beat forecasts, and investors may therefore treat small beats and misses differently \citep{burgstahler2006management,bhojraj2009making}. We view these forces as determinants of the unconditional earnings-response function and ask instead how uncertainty about the earnings-to-value mapping changes its shape. Our model predicts that greater multiplier uncertainty steepens the response around the consensus while flattening it farther away: accounting for the non-linear nature of the response is therefore critical to test this comparative static both in the experiment and in the field.

\section{A Model of Multiplier Uncertainty}\label{sec:model}

In this section, we develop a model to describe the optimal decision rule of an investor who knows whether earnings news is good or bad but is uncertain about exactly how much it should change the value of the underlying asset. An observed earnings surprise of 1\% of the share price may imply that the firm is worth more than previously expected. If the investor observes a 100\% surprise, however, they may question whether earnings surprises are informative about firm value at all. In other words, large surprises can reflect either large implied changes in firm value or a weak mapping between earnings surprises and valuations. When large changes in firm value are unconditionally less likely, the investor increasingly attributes extreme surprises to weak mappings. As a result, even under Bayesian updating, the model generates an $S$-shaped response to news.

\subsection{Setup}

Consider a decision maker (DM) who observes a signal $s$ about a firm's fundamental value. This signal contains new information about the value of the firm, and the DM must determine how much the new information should change that value. We represent this relationship with a response function that maps the signal into the return implied by the change in value. We assume that, under what we call the full-information benchmark where the DM knows this mapping perfectly, the return is
\begin{equation}
r=\beta\cdot s,
\label{eq:multiplier}
\end{equation}
where $\beta>0$ is the multiplier that maps the signal into the full-information return. Of course, stocks with positive earnings surprises do not always have positive returns. This may reflect noise in the reference point used to measure the surprise or other contemporaneous drivers of returns, both of which are outside of our model. What matters for our model is that, given these disturbances, a larger earnings surprise still predicts a larger return.

Now suppose that the DM knows that $\beta$ is positive but is uncertain about its magnitude. In the context of an earnings announcement, the DM may be uncertain about how persistent the news contained in the earnings surprise will be, what that news implies for future cash flows and growth, or how those expected cash flows should be discounted given their risk.

We assume that the DM has independent priors over returns $r$ and multipliers $\beta$. Once these are drawn, the DM observes the corresponding surprise $s=r/\beta$.\footnote{A complementary view of the model is that a fundamental shock $z$ determines the earnings surprise via $s=\beta_s z$ and returns via $r=\beta_r z$. Since $z$ is unobserved, while $r=\beta s$ with $\beta=\beta_r/\beta_s$, it is equivalent, more parsimonious and arguably more intuitive to let fundamental shocks occur directly in the space of returns.} We take returns to be normally distributed,
$$
r\sim\mathcal{N}(0,\sigma_r^2).
$$

Ultimately, the unknown multiplier $\beta$ will determine the scale of the distribution of surprises $s$. Thus, $\beta^2$ will be the precision parameter in the DM's updating problem. We therefore assume the DM has a Gamma prior on $\beta^2$,
$$
\beta^2\sim \text{Gamma}
\left(
\frac{1}{\kappa},
\frac{1}{\kappa \cdot \bar{\beta}^2}
\right).
$$
This parametrization implies that the mean squared multiplier is $\Exp(\beta^2) = \bar{\beta}^2$ while its squared coefficient of variation is simply $\kappa$:
$$
\text{CV}(\beta^2)^2
\coloneqq
\frac{\Var(\beta^2)}{\Exp(\beta^2)^2}
=
\kappa.
$$
Increasing $\kappa$, therefore, changes the DM's uncertainty about the mapping from $s$ to $r$ while holding the mean squared multiplier fixed.  Hereafter, we refer to $\kappa$ as multiplier uncertainty.

Under these assumptions, $s = r / \beta$ is a Normal-Gamma scale mixture and thus follows a centered Student distribution with scale $\sigma_r / \bar{\beta}$ and $2 / \kappa$ degrees of freedom. Importantly, this framework allows us to clearly see the force governing the DM's updating rule: unusually large surprises shift posterior weight toward weaker mappings between $s$ and $r$, which generates the non-linear response of prices to earnings news.

\subsection{Bayesian Response}

The posterior mean after observing $s$ is $S$-shaped.\footnote{See Appendix~\ref{app:model} for all proofs.}
\begin{proposition}\label{prop:rootsign_posterior_mean}
After observing $s$, the DM's posterior mean is
\begin{equation}\label{eq:response}
\hat{r}(s)
\coloneqq \Exp(r \mid s)
= A(\kappa) \cdot \bar{\beta} \cdot
\frac{s}{
\sqrt{1+\tfrac{\kappa\bar{\beta}^2}{2\sigma_r^2} s^2}
},
\end{equation}
where $A(\kappa)\coloneqq \sqrt{\kappa}\,\Gamma(\kappa^{-1}+1) / \Gamma(\kappa^{-1}+\tfrac12)$. Moreover, $\hat{r}(s)$ is odd, strictly increasing and $C^\infty$.
\end{proposition}

We call the response function, which in its simplest form is $x\to\nicefrac{x}{\sqrt{1+x^2}}$, \textit{rootsign}, in reference to the \textit{softsign} activation function $x\to\nicefrac{x}{1+|x|}$ used in machine learning \citep{elliott1993better}. Panel A of Figure~\ref{fig:model_response_shape} illustrates this response function. The response is approximately linear and steepest around $s=0$, but becomes progressively flatter as the surprise grows in absolute value. It eventually approaches finite limits: increasingly extreme earnings news produces little additional change in the assessed posterior value of the firm.

This shape reflects the fact that the DM learns about the multiplier from the surprise itself. Indeed, this response is simply $\hat r(s) = \Exp(\beta\mid s) \cdot s$, where the posterior mean estimate of the multiplier is $\Exp(\beta\mid s) = A(\kappa) \cdot \bar\beta \; / \; \sqrt{1+\tfrac{\kappa\bar\beta^2}{2\sigma_r^2}s^2}$. For surprises near zero, the posterior change in the multiplier is only second order in the magnitude of the surprise, so that the return response is approximately linear. Since smaller surprises are relatively more likely under stronger mappings characterized by larger multipliers, the local slope is $A(\kappa)\bar\beta$. Farther away from zero, however, an increase in the magnitude of the surprise shifts posterior weight toward smaller multipliers. The effect of more extreme news on prices, therefore, is partly offset by a lower inferred multiplier. For sufficiently large surprises, the posterior multiplier declines approximately as $1/|s|$, so the response levels off, even though it always remains strictly increasing.

\subsection{Effect of Multiplier Uncertainty}

In this subsection, we ask how greater multiplier uncertainty changes the sensitivity of prices to earnings news. Many models predict that greater uncertainty lowers the average sensitivity of a DM's response to news \citep[see e.g.,][]{gabaix2019behavioral,woodford2020modeling}. We first ask whether this prediction holds in our model. We then investigate whether this also applies near the consensus (i.e., a surprise of $s=0$), and whether sensitivity to small surprises changes relative to sensitivity to large ones.

To measure the overall response to news, we define the population coefficient from a linear regression of returns on surprises,
\begin{equation}\label{eq:model_pooled_slope}
b_\mathrm{OLS} \coloneqq \frac{\Exp[s\hat r(s)]}{\Exp[s^2]}.
\end{equation}
This coefficient is finite if the distribution of surprises has a non-zero finite variance, i.e., if $0<\kappa<1$. Local sensitivity is instead given by $\hat r'(s)$. In our model, both can be computed in closed form, allowing us to characterize how they change with multiplier uncertainty.

\begin{proposition}[Attenuation, amplification, and steepening]
\label{prop:uncertainty_effect}
Increasing multiplier uncertainty $\kappa$ has three effects:

\begin{enumerate}
\item It lowers average sensitivity: for $\kappa<1$,
\[
b_{\mathrm{OLS}}
=
\frac{1-\kappa}{1-\kappa/2}
\frac{\bar\beta}{A(\kappa)},
\qquad \qquad \qquad
\frac{\partial b_{\mathrm{OLS}}}{\partial\kappa}<0.
\]
\item It raises sensitivity to small surprises:
\[
\hat r'(0)
= A(\kappa)\bar\beta,
\qquad \qquad \qquad
\frac{\partial \hat r'(0)}{\partial\kappa}>0.
\]
In particular, $b_{\mathrm{OLS}} < \bar\beta < \hat r'(0)$ for $0<\kappa<1$, so multiplier uncertainty attenuates sensitivity on average while amplifying it locally around small surprises.

\item It increases sensitivity to small surprises relative to large ones: for $\kappa>0$ and $s\neq0$,
\[
\frac{\hat r'(0)}{\hat r'(s)}
=
\left(
1+\frac{\kappa\bar\beta^2}{2\sigma_r^2}s^2
\right)^{3/2}
> 1,
\qquad \qquad \qquad
\frac{\partial}{\partial\kappa}
\left[
\frac{\hat r'(0)}{\hat r'(s)}
\right]
>0.
\]
The level difference in sensitivities, given in Corollary~\ref{cor:slope_difference}, exhibits the same steepening.
\end{enumerate}
As uncertainty vanishes, i.e. as $\kappa \to 0$, $\hat{r}(s) \to \bar{\beta}s$ so $b_\mathrm{OLS}$, $\hat{r}'(0)$ and $\hat{r}'(s)$ all converge to $\bar{\beta}$.
\end{proposition}

The first result shows that the standard result of uncertainty attenuating the overall response to news also holds in our model. The second and third results are the central comparative statics we test in both the experimental and field data: sensitivity increases near the consensus, and it increases \textit{relative} to sensitivity farther away from the consensus.  In other words, the local and relative effects identify the predicted \textit{change} in the shape of the response as a function of multiplier uncertainty, which is what we will quantify empirically.

\begin{figure}[!htbp]
\centering
\caption{Effect of Multiplier Uncertainty on the Reaction Function}
\label{fig:model_response_shape}
\includegraphics[width=\linewidth]{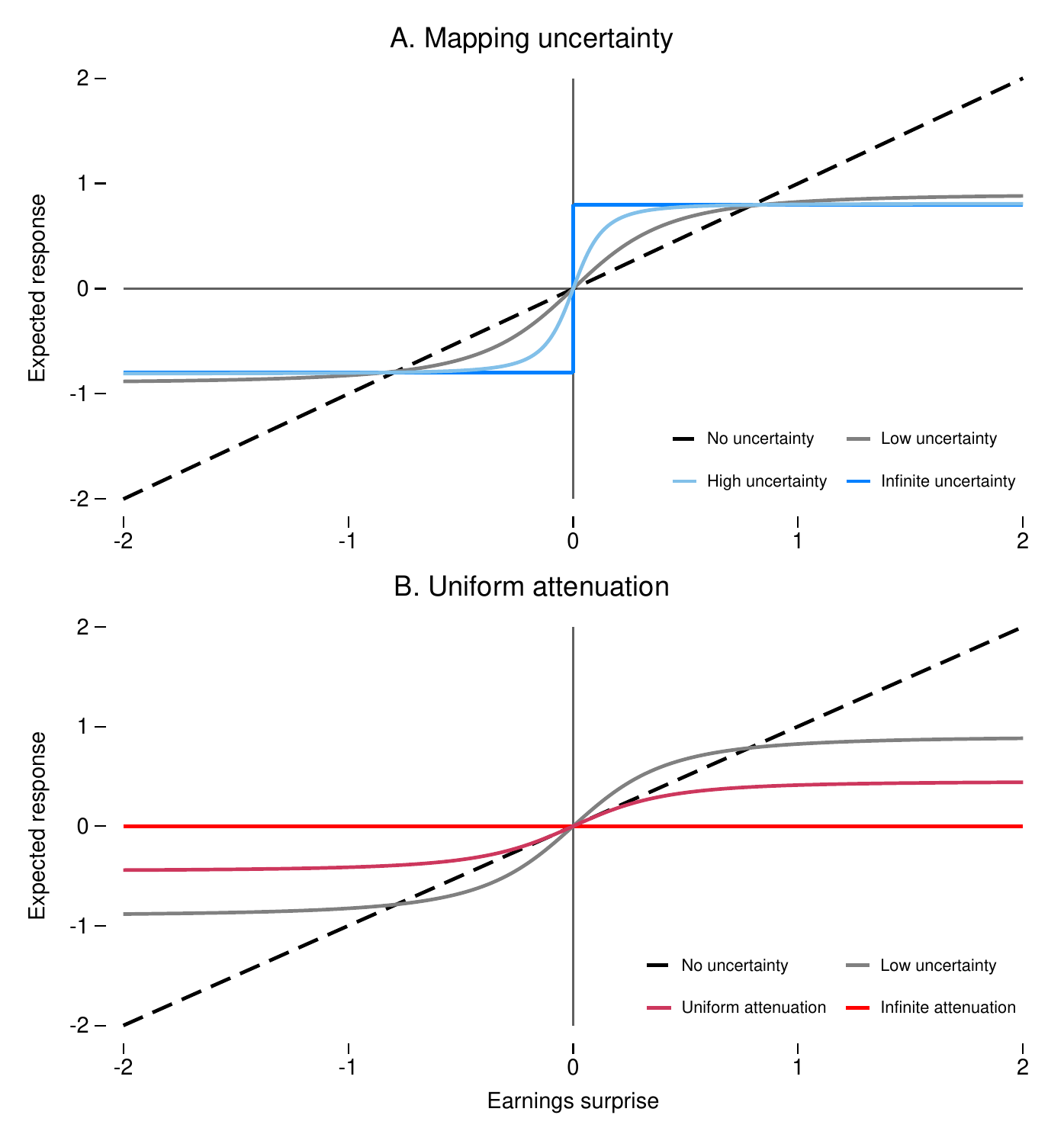}
\caption*{\footnotesize \textit{Notes:} Reaction $\hat r(s)$ to an earnings surprise $s$ as in~\eqref{eq:response}. The black dashed line shows the full-information response $\bar\beta s$. Panel A compares low and high multiplier uncertainty, with $\kappa=10$ in grey and $\kappa=100$ in dark blue, holding $\bar\beta=\sigma_r=1$ fixed: greater multiplier uncertainty steepens the response near zero and flattens it farther away. The light blue step function shows the limiting categorical response as $\kappa \to \infty$. Panel B contrasts this change in shape with uniform attenuation, holding $\kappa=10$ fixed and adding $\bar\beta=\sigma_r=1/2$ in dark red, which multiplies the posterior response by $1/2$ at every signal: uniform attenuation lowers sensitivity everywhere, whereas multiplier uncertainty increasingly concentrates it near zero. In turn, the pink horizontal line shows the limiting non-response under infinite attenuation.}
\end{figure}

Panel A of Figure~\ref{fig:model_response_shape} visually shows the implications of these predictions. The response under increased uncertainty is steeper near zero, so small beats and misses produce larger reactions. Farther from zero, high uncertainty leads the DM to update more strongly toward weak multipliers, so that they update less from large surprises when mapping uncertainty is higher. Eventually, the high-uncertainty response even falls below the low-uncertainty one: hence the curves cross once on each side of the consensus. Corollary~\ref{corollary:slope_single_crossing} establishes this single-crossing result formally: a unique boundary separates surprises whose marginal response rises with uncertainty and those whose marginal response falls.

\begin{figure}[!htb]
\centering
\caption{Effect of Multiplier Uncertainty on Average and Local Sensitivity}
\label{fig:model_sensitivities}
\includegraphics[width=\linewidth]{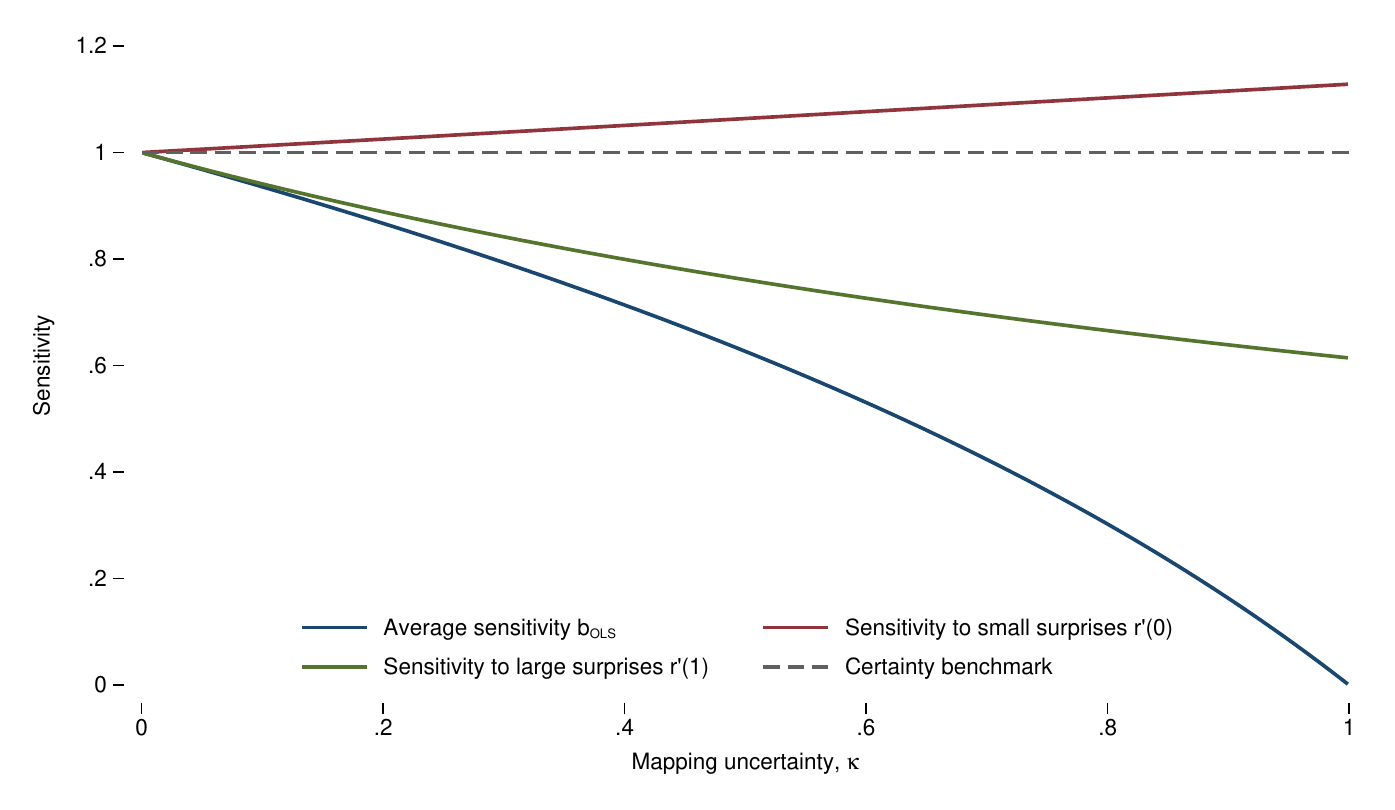}
\caption*{\footnotesize \textit{Notes:} The figure plots the comparative statics in Proposition~\ref{prop:uncertainty_effect} as multiplier uncertainty $\kappa$ varies, holding $\bar\beta=\sigma_r=1$. The solid lines show the population OLS sensitivity $b_{\mathrm{OLS}}$, local sensitivity at the prior mean $\hat r'(0)$, and local sensitivity at $s=1$, $\hat r'(1)$. The dashed horizontal line is the certainty benchmark $\bar\beta=1$. Greater multiplier uncertainty lowers average sensitivity, raises sensitivity at the prior mean, and lowers sensitivity at $s=1$ relative to sensitivity at the prior mean. The plot restricts to $\kappa<1$, where $b_{\mathrm{OLS}}$ exists.}
\end{figure}

Figure~\ref{fig:model_sensitivities} illustrates these changes in sensitivity by varying $\kappa$ over $(0,1)$. As uncertainty increases, the average response coefficient $b_{\mathrm{OLS}}$ falls while sensitivity at zero $\hat r'(0)$ rises. Using $s=1$ as an example of larger surprise, we see that marginal sensitivity $\hat r'(1)$ also falls, so that the gap with sensitivity at zero widens.

The parameter $\kappa$ hence spans a continuum between two interpretable benchmarks. As $\kappa\to0$, the response approaches the full-information straight line $\bar\beta s$. The following corollary establishes the opposite limit.

\begin{corollary}[Categorical limit]\label{cor:categorical_limit}
As multiplier uncertainty becomes infinite, i.e., as $\kappa\to\infty$, for every $s\neq0$,
\[
\hat r(s)\; \to \;\sqrt{\tfrac{2}{\pi}}\,\sigma_r\,\sign(s)
\;=\;\Exp\!\left[\,r\mid \sign(r)=\sign(s)\,\right].
\]
\end{corollary}

Under extreme multiplier uncertainty, the magnitude of the surprise becomes completely uninformative, and the DM only updates based on its direction. The response hence converges to a step function that depends only on the sign of the surprise, whose height is the expected return conditional on knowing only whether earnings beat or missed the consensus. Increasing $\kappa$ thus moves the optimal response continuously from reacting to the number toward reacting to the category: at low multiplier uncertainty the investor prices the size of the news, and at high multiplier uncertainty she mostly prices its sign.

An illuminating contrast is the idea that signal processing difficulty could decrease responses to news everywhere, as in models where processing constraints attenuate sensitivity to decision-relevant inputs. Applied to our setting, this would imply that the reaction function shrinks \textit{proportionally} at every signal. Corollary~\ref{cor:uniform_attenuation} shows that, in our model, such proportional shrinking arises when the perceived magnitude of the multiplier and the scale of returns are jointly scaled down, which rescales the response without changing its shape. Said a different way, a harder valuation problem would have to make investors expect proportionally smaller price responses to the same earnings surprise. For example, suppose an investor believes a given earnings surprise could cause either a 2\% or a 4\% stock return. As the valuation problem becomes harder, this proportional shrinking would have to reduce these possible returns to 1\% and 2\%, respectively, without changing the probability assigned to either outcome. This would mean investors believe the same earnings surprise has an unconditionally smaller effect on valuation, while their uncertainty about the resulting return, measured relative to its expected size, remains unchanged.

Panel B of Figure~\ref{fig:model_response_shape} applies this uniform attenuation to the low-uncertainty response from Panel A. As expected, reactions and marginal sensitivities fall everywhere, including around the consensus. The ratio of small- to large-surprise sensitivity is unchanged, while their absolute difference shrinks. Multiplier uncertainty, on the other hand, predicts both average attenuation and local amplification, instead of imposing a lower response to news everywhere. With infinite attenuation, prices do not respond to earnings surprises \textit{at all}. With unbounded multiplier uncertainty, investors respond only to whether earnings beat or miss expectations, with the magnitude of the surprise having no marginal effect on returns.

\subsection{Extensions}

Our baseline model is intentionally streamlined to isolate the effects of multiplier uncertainty on the shape of the reaction function. In Appendix~\ref{app:model}, we extend it to allow for a non-linear full-information response, asymmetric responses to beats and misses, as well as uncertainty about the scale of the return distribution.

Our baseline model in Equation~\eqref{eq:multiplier} assumes a linear full-information response $r = \beta s$, which under multiplier uncertainty leads to an $S$-shaped reaction function. We show that this mechanism is complementary to other explanations for the $S$-shape put forward in the literature, e.g., that the earnings growth from more extreme surprises is less persistent \citep{freeman1992nonlinear}. Specifically, in Appendix~\ref{app:model_nonlinear} we allow the full-information return itself to be a non-linear function of the underlying news, $r=r^*(\beta s)$, where $\beta s$ is Gaussian and independent of the multiplier. We show that greater multiplier uncertainty still lowers the unconditional average sensitivity, while amplifying sensitivity to surprises near the consensus. Moreover, the reaction function can still be obtained in closed form for any fractional polynomial response function, or for a rootsign response function as it emerges from our own model. Thus, the $S$-shaped reaction function driven by multiplier uncertainty is a general consequence of our framework, rather than a specific outcome when the full-information response function is assumed to be linear.

We also allow the model to be asymmetric across beats and misses. This is designed to capture the fact that  managers' incentives and investors' beliefs may differ between positive and negative earnings news \citep{gul1991theory,huang2025strategic}. To capture this, Appendix~\ref{app:model_asymmetric} allows separate multiplier distributions for beats and misses. We find that under this assumption, the reaction function has the same $S$-shape as in the baseline, but with potentially differing slopes for beats and misses. Even allowing for this additional flexibility, we find that, on either side of zero, greater multiplier uncertainty lowers the overall sensitivity, increases it around $s=0$ and steepens the response to small surprises relative to larger ones.

Finally, we consider an extension where the DM has uncertainty about the scale of returns, since our baseline model assumes that the DM knows the distribution of returns. Appendix~\ref{app:model_return_uncertainty} relaxes that assumption by allowing for uncertainty about $\sigma_r$. Crucially, the likelihood of a surprise of a given size depends on uncertainty about $\sigma_r$ and about $\beta$ only through $\beta^2/\sigma_r^2$. With a Gamma prior on this ratio and an independent prior on the scale of returns, the posterior response still obtains as a rootsign shape. Average attenuation, near-zero amplification, and relative steepening then all go through under uncertainty about $\sigma_r$. We find that changing the expected return scale produces uniform attenuation or amplification, whereas greater uncertainty about the mapping-to-scale ratio steepens the shape of the response. This extension hence further helps distinguish changes that merely rescale the reaction function to news from uncertainty that alters how the news is interpreted.

\subsection{Discussion}

In our model, the magnitude of the signal is informative about the mapping between the signal and returns used to interpret it. This logic -- the signal itself teaching the DM about the nature of the distribution from which it was drawn -- is closely related to \citet{subramanyam1996uncertain}, where extreme realizations lead investors to infer lower signal precision and are therefore discounted more heavily. While he considers uncertainty about additive noise, we instead model uncertainty about a positive multiplier, which we argue is a more natural way to study the difficulty of processing a signal. We moreover obtain transparent closed forms, with a monotonic response that converges to a positive finite limit, whereas in \citet{subramanyam1996uncertain} the response is in general characterized only implicitly and, in his tractable truncated-Gamma example, its slope eventually turns negative, so that the reaction to increasingly positive or negative news declines back toward zero.

The premise that the direction of news is easier to interpret than magnitude is shared with \citet{augenblick2025overinference}. In their model, the DM observes a signal and receives an independent, noisy signal about the signal's significance.  For example, the significance of a point being scored certainly depends on the number of minutes left in a basketball game, but the exact mapping from minutes left to the importance of that basket is not known. Learning from this noisy significance signal pulls perceived signal strength toward the middle: weak signals are interpreted as stronger than they are, while strong signals are interpreted as weaker.\footnote{\citet{bastianello2026biases} extend this to multi-feature environments, where signal strength also depends on dimensions such as shock size, persistence, and horizon, and limited attention to those features can modify or even overturn the usual pattern.} Given an inferred strength, however, the reaction to the signal itself remains linear, so a separate significance cue cannot bend the response within a fixed firm and a fixed distribution of surprises, which is what our experiment produces. In our case, there is only one signal, from which the DM must infer both its magnitude and its significance, and this is what induces the characteristic $S$-shape we observe in the experiment and in the field.

\section{Experimental Evidence}\label{sec:experiment}

Testing our main predictions empirically poses two problems that cannot be addressed with field data alone. The shape of a response function is a statement about counterfactuals, namely how the same investor would have reacted had the same firm announced a bigger or smaller surprise. For each real-world earnings announcement, however, we observe only one earnings surprise and one stock return. We cannot, therefore, observe how investors would have responded to a different surprise in the same announcement. Moreover, firms with different levels of multiplier uncertainty may also differ in other ways. For example, biotech and utility firms likely differ in terms of investor attention, disclosure and institutional ownership, making it difficult to attribute differences in their price responses to multiplier uncertainty alone. 

Our experiment tackles both of these issues directly. First, we recruit investors days before actual earnings announcements and have them forecast the announcement-day return given the firm’s actual stock price and consensus forecast, but present them with a randomly drawn earnings number. Randomizing the surprises lets us trace out each firm's price response over a range of possible news. Second, randomly assigning the severity of processing constraints, which make the same valuation problem harder while leaving all valuation-relevant information untouched, should vary multiplier uncertainty exogenously for a given piece of news.

\subsection{Design}

Participants read a hypothetical earnings article about a real company scheduled to announce quarterly earnings within five days of the study. The article follows the format of standard earnings coverage and uses the company's current stock price and consensus EPS forecast. We randomize reported EPS, and state both in the headline and the article whether the company beat or missed the forecast. It also provides firm background.

We draw SUE from $\pm\{0.01\%,0.05\%,0.1\%,0.5\%,1\%\}$ and calculate the corresponding EPS realization from the firm's actual price and consensus forecast. These values cover more than 85\% of the empirical SUE distribution.\footnote{For implementability, the experiment holds the surprise distribution fixed, whereas the model's population slope averages over a distribution that changes with uncertainty, so that we test this prediction qualitatively. Crucially, the local and relative regressions test the model's predicted steepening directly.} The following example shows the article for Darden Restaurants when SUE equals $-0.05\%$:
\begin{quote}
Darden Restaurants, Inc. is an American multi-brand restaurant operator headquartered in Orlando, Florida. In their earnings announcement for the third quarter of 2024, Darden reported earnings below market expectations. Trading at a stock price of \$164.73 prior to the announcement, Darden reported earnings per share of \$1.94. Darden therefore earned 3.96\% less than analysts expected, given the consensus forecast of \$2.02 earnings per share.
\end{quote}
Participants are asked to treat the announcement as occurring at that moment and to predict the percentage change from the pre-announcement price to the same-day closing price.\footnote{Data collection took place between late morning and early afternoon Eastern time.} Each participant completes five independent rounds involving five different real US firms. We randomize both the order of the firms and the SUE realization in each round. Screenshots of the main instructions are available on the authors' website.\footnote{{\fontsize{9.5pt}{9pt}\selectfont \url{https://raw.githubusercontent.com/cproth/papers/master/CoarseCategories_instructions.pdf}}.}

As incentives, participants receive a \$1.70 base payment. One in ten is selected for a \$50 bonus, for one round chosen at random. The bonus is paid if the firm's actually announced earnings per share are within 10\% of the earnings per share in the experimental scenario and the participant's price forecast is within one percentage point of the realized announcement-day return.\footnote{Participants assigned a mean probability of 56.54\% to satisfying the first condition.}

Our experiment contains two treatment arms. Participants are assigned with equal probability to the \textit{Baseline} or \textit{High Constraints} conditions. Both groups receive the same information about the earnings announcement: hypothetical earnings per share, the analyst consensus forecast and the pre-announcement stock price. The \textit{High Constraints} treatment also presents participants with background material about the firm's history. In other words, \textit{High Constraints} gives participants more information to process without adding additional information relevant to evaluating the earnings news. For example, for Darden Restaurants, the additional passage reads:
\begin{quote}
Darden is an American multi-brand restaurant operator headquartered in Orlando, Florida. Darden has more than 1,800 restaurant locations and more than 175,000 employees, making it the world's largest full-service restaurant company. The company began as an extension of Red Lobster, founded by William Darden and initially backed by General Mills. Red Lobster was later sold in July 2014.
\end{quote}

Further, in the \textit{High Constraints} treatment, we require participants to answer within 40 seconds to remain eligible for the bonus. In \textit{Baseline}, the median response time is 25 seconds, while the 25th percentile is 15 seconds and the 75th percentile is 45 seconds, so this limit creates meaningful time pressure without making completion infeasible (i.e., it is still above the median completion time). The timeout rate in \textit{High Constraints} is relatively low at 6.7\%.

In each round, participants are asked to predict the same-day percentage price change on a scale from $-15\%$ to $15\%$. Before making these predictions, they are asked to report their beliefs about the historical average return associated with an earnings beat and with an earnings miss.\footnote{The question is: ``Historically, what do you think was a company's average stock price change on a day where announced earnings [exceeded/fell below] the consensus forecast?''} We call these estimates the participants' \textit{category defaults}, and they are designed to capture the return a respondent associates with the sign of the news before considering its exact magnitude.

As specified in our preregistration, our primary outcome divides each participant's forecast by the absolute value of their miss default for negative surprises and beat default for positive surprises. A forecast equal to the miss default is therefore coded as $-1$, and a forecast equal to the beat default is coded as $+1$. This lets us compare forecasts across respondents with different unconditional beliefs about average announcement returns. Because these defaults may be measured with error, however, we also report results using raw forecasts and broadly find similar results.

The experiment was conducted in December 2024. We recruited 1,000 US adults on Prolific, all of whom were at least 18 years old and held an account with a trading platform. Participants had two attempts to answer three comprehension questions whose answers were contained in the instructions. Nine percent of those who started the study did not pass. As preregistered, we also exclude participants who reported a nonpositive historical return for earnings beats or a nonnegative return for misses. The sample has 896 respondents. The study was preregistered on AsPredicted (\#205080; \url{https://aspredicted.org/n3zm-md9t.pdf}), covering the design, hypotheses, outcomes, sample size, and exclusions.

\subsection{Results}

\begin{figure}[!htb]
\centering
\caption{Processing Constraints Steepen the Experimental Response}
\label{fig:experiment_median_forecasts}
\includegraphics[width=\linewidth]{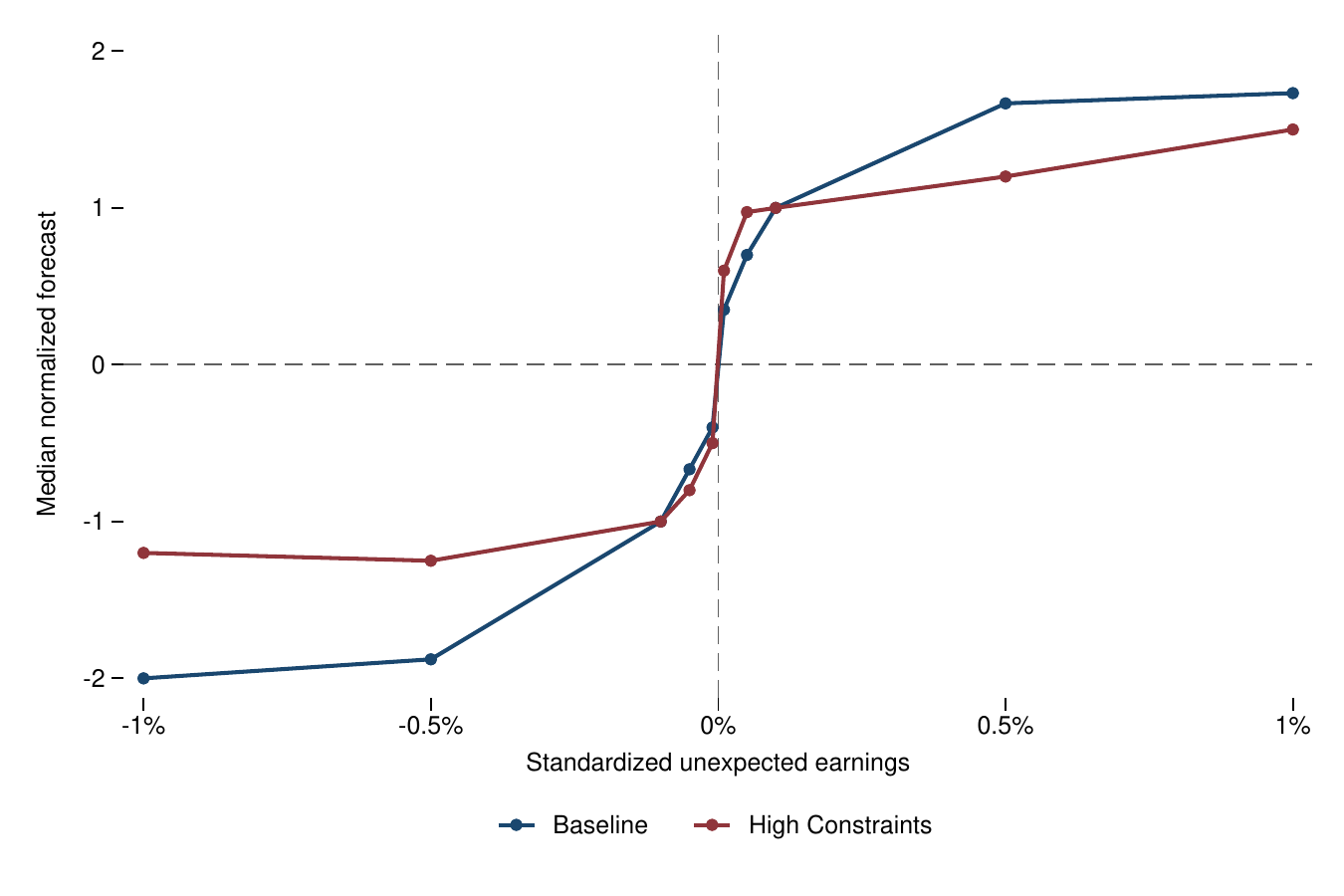}
\caption*{\footnotesize \textit{Notes:} Median normalized forecasts in each of the ten randomized SUE cells, by treatment. A forecast equal to the respondent's category default is coded $-1$ for a miss and $+1$ for a beat. There are 896 respondents.}
\end{figure}

Even in a controlled experiment and before increasing processing constraints, respondents' return forecasts exhibit the characteristic $S$-shape observed for market prices. Figure~\ref{fig:experiment_median_forecasts} plots median normalized forecasts in each SUE cell: already in \textit{Baseline}, forecasts are steeply increasing around zero and much less sensitive to variation in SUE for large surprises. Median forecasts in the experiment therefore reproduce the $S$-shaped response observed in market prices.

We now turn to the effect of increasing processing constraints. In the \textit{High Constraints} condition, the reaction becomes more sensitive around zero, while the sensitivity to large surprises further diminishes, leading the $S$-shape to steepen. Small positive and negative surprises generate more extreme forecasts than in \textit{Baseline}, whereas sufficiently large surprises generate less extreme forecasts. The two response functions consequently cross on each side of zero.

The normalized medians cross at SUE values of $-0.1\%$ and $+0.1\%$, where both groups' forecasts equal the corresponding category default. The precise location partly reflects the normalization and is not itself an experimental estimand. The same steepening around zero is present in raw percentage forecasts, which do not use the elicited defaults (Table~\ref{tab:experiment_response_fits}). Our tests therefore focus on the change in sensitivity.

In line with our preregistration, we estimate the earnings response coefficient (ERC) over expanding SUE windows around zero:
\begin{equation}\label{eq:hc-linear}
    r_{ij} = \alpha + \beta \; \SUE_{ij} + \; \text{HC}_i \times \left(  \gamma + \delta \; \SUE_{ij} \right) + \varepsilon_{ij},
\end{equation}
where $r_{ij}$ is participant $i$'s normalized price forecast for firm $j$ and $\textit{HC}_i$ indicates assignment to \textit{High Constraints}. We estimate the regression over successively wider symmetric SUE windows \citep[cf.][]{freeman1992nonlinear,kinney2002earnings,burgstahler2017earnings}. The interaction $\delta$ measures how the treatment changes sensitivity to SUE; we divide it by the baseline slope $\beta$ to express this change proportionally. As preregistered, we use median regression because near-zero category defaults can yield extreme normalized forecasts.

Appendix Figure~\ref{fig:experiment_expanding_windows} plots $\hat\delta/\hat\beta$, while Appendix Table~\ref{tab:experiment_windows} reports the underlying regressions. In the smallest window, processing constraints increase sensitivity by about half. The estimate declines as the window expands and changes sign between $\pm0.1\%$ and $\pm0.5\%$. In the full sample, the treatment reduces sensitivity by about one quarter.  Processing constraints thus raise sensitivity to small surprises but lower it once larger surprises are included.

To summarize the $S$-shape predicted by our model through a single coefficient, we approximate it with a piecewise linear function. More precisely, we use the symmetric capped surprise function, which we call a \textit{hinge}:
\begin{equation}\label{eq:hinge}
h(s)\coloneqq
\begin{cases}
-k, & s<-k,\\
s, & -k\leq s\leq k,\\
k, & s>k.
\end{cases}
\end{equation}
Inside $[-k,k]$, it equals the surprise and moves one-for-one with it. Outside that interval, it is constant. Adding it in a regression next to the ordinary SUE term therefore estimates a continuous, piecewise linear response with a different slope near zero. Importantly, the ordinary SUE term remains: we do not impose a flat response to large surprises.

We estimate
\begin{equation}\label{eq:hc-interaction}
r_{ij}
=
\alpha+\beta\SUE_{ij}
+\beta_\Delta h(\SUE_{ij})
+\mathrm{HC}_i
\left[
\gamma+\delta\SUE_{ij}
+\delta_\Delta h(\SUE_{ij})
\right]
+ \varepsilon_{ij}.
\end{equation}
The coefficient $\beta$ is the baseline slope outside $[-k,k]$, while $\beta_\Delta$ is the additional slope in that interval, so the baseline slope in it is $\beta+\beta_\Delta$. The treatment interactions have the same interpretation: $\delta$ measures how \textit{High Constraints} changes the response to SUE for large surprises, $\delta+\delta_\Delta$ measures the change near the consensus, and $\delta_\Delta$ measures how much more the treatment changes sensitivity to small surprises than to large ones.

The threshold $k$ is selected to match the unconditional response pooling both treatments, which yields $k=0.02\%$, and then held fixed. This means around 20\% of responses lie in this central region. As in the expanding-window analysis, we use median regressions with standard errors clustered by respondent.

Table~\ref{tab:experiment_response_fits} tests the three model predictions in turn. Column (1) first establishes global attenuation: while in \textit{Baseline} a 1pp higher SUE increases the normalized return forecast by 2.37, \textit{High Constraints} \textit{reduces} this response by 0.55, or about one quarter. Column (2) then allows the sensitivity to differ between small and large surprises, and finds that the treatment \textit{increases} the sensitivity to each additional percentage point of SUE by 8.84 among small surprises – again about a quarter of the corresponding \textit{Baseline} slope of 40.00 – which is 9.65 more than the change for larger surprises. These three effects hence match the three predictions of Proposition~\ref{prop:uncertainty_effect}. Columns (3)--(4) show the same pattern in raw percentage forecasts.

\begin{table}[!htb]
    \centering
    \caption{Processing Constraints and the Sensitivity of the Experimental Response}
    \label{tab:experiment_response_fits}
    \fittable{\begin{tabular}{l*{4}{c}}
\toprule
& \multicolumn{4}{c}{\textit{Dependent variable: respondent return forecast}} \\
& \multicolumn{2}{c}{Normalized forecasts} & \multicolumn{2}{c}{Raw forecasts} \\
\cmidrule(lr){2-3}\cmidrule(lr){4-5}
& (1) & (2) & (3) & (4) \\
\midrule
SUE                 &        2.37\sym{***}&        1.22\sym{***}&       10.00\sym{***}&        6.67\sym{***}\\
                    &      (0.11)         &      (0.07)         &      (0.31)         &      (0.35)         \\
$\Delta$ Small SUE  &                     &       38.78\sym{***}&                     &      145.00\sym{***}\\
                    &                     &      (1.56)         &                     &      (6.12)         \\
High Constraints    &        0.12\sym{**} &        0.01         &        0.67\sym{***}&        0.33\sym{**} \\
                    &      (0.05)         &      (0.03)         &      (0.21)         &      (0.13)         \\
High Constraints $\times$ SUE&       -0.55\sym{***}&       -0.80\sym{***}&       -2.67\sym{***}&       -4.56\sym{***}\\
                    &      (0.14)         &      (0.11)         &      (0.49)         &      (0.55)         \\
High Constraints $\times$ $\Delta$ Small SUE&                     &        9.65\sym{***}&                     &       44.74\sym{***}\\
                    &                     &      (2.08)         &                     &     (11.33)         \\
\midrule
Small SUE           &                     &40.00\sym{***}         &                     &151.67\sym{***}         \\
                    &                     &      (1.53)         &                     &      (6.03)         \\
High Constraints $\times$ Small SUE&                     &8.84\sym{***}         &                     &40.18\sym{***}         \\
                    &                     &      (2.03)         &                     &     (11.09)         \\
\midrule
Hinge threshold \(k\)&                     &       0.020         &                     &       0.020         \\
Observations        &       4,480         &       4,480         &       4,480         &       4,480         \\
Respondents         &         896         &         896         &         896         &         896         \\
Pseudo \(R^{2}\)    &        0.22         &        0.34         &        0.28         &        0.42         \\
Regression          &      Median         &      Median         &      Median         &      Median         \\
\bottomrule
\end{tabular}
}
\caption*{\footnotesize \textit{Notes:} Conditional median regressions of the return forecast on the randomized surprise. Columns (1) and (3) estimate the linear specification~\eqref{eq:hc-linear} on the full sample, so that High Constraints $\times$ SUE is the change in the average sensitivity to SUE. Columns (2) and (4) estimate Equation~\eqref{eq:hc-interaction} with the hinge: SUE is the slope for surprises beyond $|\SUE|=k$, $\Delta$ Small SUE is the additional slope inside $[-k,k]$, and Small SUE is their sum, the slope at zero; the treatment interactions are defined analogously. Columns (1)--(2) divide each forecast by the absolute magnitude of the respondent's category default for a beat or miss; columns (3)--(4) use the raw percentage forecast. The scale $k$ is selected once from the pooled forecasts before the treatment indicator enters and then held fixed. No controls or fixed effects are included. Standard errors are clustered by respondent. The sample contains 4{,}480 observations from 896 respondents.}
\end{table}

\subsection{Robustness}\label{sec:experiment_robustness}

We show that the experimental result is robust to changing the measure of forecasted returns or the specification, to various sample restrictions and to using smooth functional forms for the $S$-shape.

The treatment comparison is not an artifact of dividing forecasts by elicited category defaults or of summarizing them by medians. Raw forecasts reproduce the preregistered expanding-window pattern (Appendix Figure~\ref{fig:experiment_excl_inconsistent}). Switching from median to least squares regressions, on normalized forecasts capped at $\pm3$, also recovers higher sensitivity in the narrowest window and lower sensitivity over the full range (Appendix Table~\ref{tab:experiment_windows_winsorized}). Importantly, it should be noted that while stable heterogeneity in respondents' response functions could contribute to the pooled $S$-shape, random assignment prevents it from systematically generating the treatment difference.

A more mechanical concern is that the $\pm15\%$ entry bound could mechanically flatten responses to large surprises. However, it binds for only $5.6\%$ of forecasts, slightly more often in \textit{Baseline} ($6.6\%$) than \textit{High Constraints} ($4.7\%$, Appendix Table~\ref{tab:experiment_censoring}). Removing bounded forecasts, or even more conservatively removing every respondent who ever reaches a bound, preserves the treatment-induced steepening (Appendix Figures~\ref{fig:experiment_censoring_uncensored}--\ref{fig:experiment_censoring_never}). The expanding-window pattern also remains after excluding timeouts, self-reported online searches, and sign-inconsistent forecasts (Appendix Figure~\ref{fig:experiment_excl_inconsistent}). Appendix Figure~\ref{fig:experiment_robustness_appendix} plots the corresponding response after excluding sign-inconsistent forecasts, which looks virtually identical to the full sample.

The piecewise-linear hinge used in~\eqref{eq:hinge} is continuous but has kinks at $\pm k$, where its slope changes abruptly. To show that our findings are not driven by our choice of functional form, we consider two smooth alternatives:
\[
h_A(s)=k\arctan(s/k),
\qquad \qquad \qquad
h_R(s)=\frac{ks}{\sqrt{s^2+k^2}}.
\]
The arctangent function has been used by \citet{freeman1992nonlinear} to model the nonlinear response of returns to earnings news, while the closely related rootsign function is motivated by the shape of the response function in our model. Just like the hinge function, both $h_A(s)$ and $h_R(s)$ have a slope of one at zero and flatten out for large positive or negative surprises. The regression coefficient multiplying the chosen function, $h_A(s)$ or $h_R(s)$, therefore measures how much more sensitive forecasts are near zero than for large surprises. In other words, this is essentially the same as in the hinge specification, but the slope changes gradually rather than abruptly. For each function, we choose $k$, the parameter that determines how quickly the slope flattens, to best fit the pooled response.

Using either smooth functional form doesn't change our key result: the \textit{High Constraints} treatment increases sensitivity more for small surprises than for large ones. This holds for both normalized and raw forecasts (Appendix Table~\ref{tab:experiment_response_fits_smooth}). Appendix Figure~\ref{fig:experiment_response_fits} shows that both forms fit the experimental responses closely, including the steeper response to small surprises under \textit{High Constraints}.

We argue the experiment provides causal evidence for the model's proposed mechanism. Specifically, in the experiment we are able to hold the news fixed and show that tighter processing constraints do not simply attenuate responses everywhere. Instead, they steepen the response function near the consensus while reducing it for larger surprises and on average. Having shown this causally in the laboratory, we next ask whether the signatures of multiplier uncertainty also appear in market reactions to earnings surprises.

\section{Field Evidence}\label{sec:setting}

We next ask whether, in the face of increased multiplier uncertainty, market prices display the globally attenuated response to earnings surprises, but increased sensitivity to surprises near zero predicted by our model. Using \citet{golubov2026accounting}'s measure of valuation uncertainty as an empirical proxy for multiplier uncertainty, we find that harder-to-value firms respond less to earnings news on average but more to small surprises relative to larger ones, mirroring our experimental results. We then show that this pattern is not specific to our choice of empirical proxy for multiplier uncertainty and is not sensitive to how we parameterize the shape of the earnings response function.  Moreover, we show that our findings are not driven by established explanations for the $S$-shaped response of prices to earnings news. 

\subsection{Setting and Data}

In this section, we analyze quarterly earnings announcements by publicly traded US firms. We combine announcement dates, reported EPS, and analyst forecasts from IBES with prices and returns from CRSP stock data and firm characteristics data from Compustat. The IBES release date and time identify the first day investors could trade on the announcement information, which we call the \textit{effective} earnings announcement date.  A day is labeled as an effective earnings announcement day if earnings are released before 4:00 pm eastern.  Earnings releases after 4:00 pm eastern and releases on weekends/holidays are assigned to the next valid trading day in CRSP. Reported earnings are IBES ``street'' EPS from the unadjusted detail file, which excludes many one-time items \citep{hillenbrand2024valuing}.\footnote{``Unadjusted'' refers to the fact that IBES does not itself apply split adjustments. We use the unadjusted file because the adjusted file rounds forecasts and earnings to the nearest cent, which significantly coarsens the measured earnings surprise \citep{payne2003implications}.}

For the consensus forecast, we take the mean EPS forecast in the final IBES statistical period before the announcement. Following \citet{dellavigna2009investor}, we define the standardized unexpected earnings (SUE) of firm $i$ at announcement $t$ as
\begin{equation}\label{eq:sue}
\SUE_{i,t}
= \frac{\text{EPS}_{i,t}-\Exp_{t-1} \left[\text{EPS}_{i,t}\right]}{P_{i,t-1}},
\end{equation}
where $\text{EPS}_{i,t}$ is reported earnings per share, $\Exp_{t-1}[\cdot]$ is the consensus forecast, and $P_{i,t-1}$ is the last closing price before the announcement. We winsorize $\SUE_{i,t}$ at the 1\% and 99\% percentiles. Importantly, the distribution of SUE is extremely concentrated around zero, with over 10\% of announcements having a SUE of exactly zero. Appendix Table~\ref{tab:field_summary} reports summary statistics on the distribution of SUE.

The primary outcome we focus on is market-adjusted returns, defined as the cumulative return on the stock minus the cumulative return on the CRSP value-weighted index over the same window \citep{campbell2001have}. For our main empirical tests, we focus on a post-earnings window which runs from the effective earnings announcement date to four trading days afterward. We use this 5-day window to allow for possible delays in price adjustment to earnings news, especially given that such delays may be correlated with multiplier uncertainty. Appendix Table~\ref{tab:field_event_windows} shows that shorter post-earnings announcement windows, and alternative definitions of abnormal returns, yield similar results to our baseline 5-day specification.

We proxy for multiplier uncertainty with the valuation uncertainty (VU) measure of \citet{golubov2026accounting}, described in Section~\ref{sec:VU} below. To identify our final sample of earnings announcements, we begin with CRSP ordinary common shares (share codes 10-11) traded on major exchanges (exchange codes 1-3).  We then require a match from CRSP stock data to IBES, non-missing reported and expected EPS, VU, pre-announcement price and market-adjusted returns.  We also require non-missing market capitalization, returns from months $t-12$ through $t-2$, book-to-market, institutional ownership and prior-year daily volatility, as these will all be included as control variables in our main regression specification. After applying these filters, we drop firm and year-quarter singletons because some of our specifications include year-quarter and firm fixed effects. To minimize the influence of extreme outliers in $\SUE$, we drop all observations with $|\SUE|>5\%$. This leaves $224{,}350$ announcements by more than $7{,}500$ firms between 1984 and 2023.

\begin{figure}[!htb]
\centering
\caption{Price Response to Earnings Surprises}
\label{fig:field_sshape_raw}
\includegraphics[width=\linewidth]{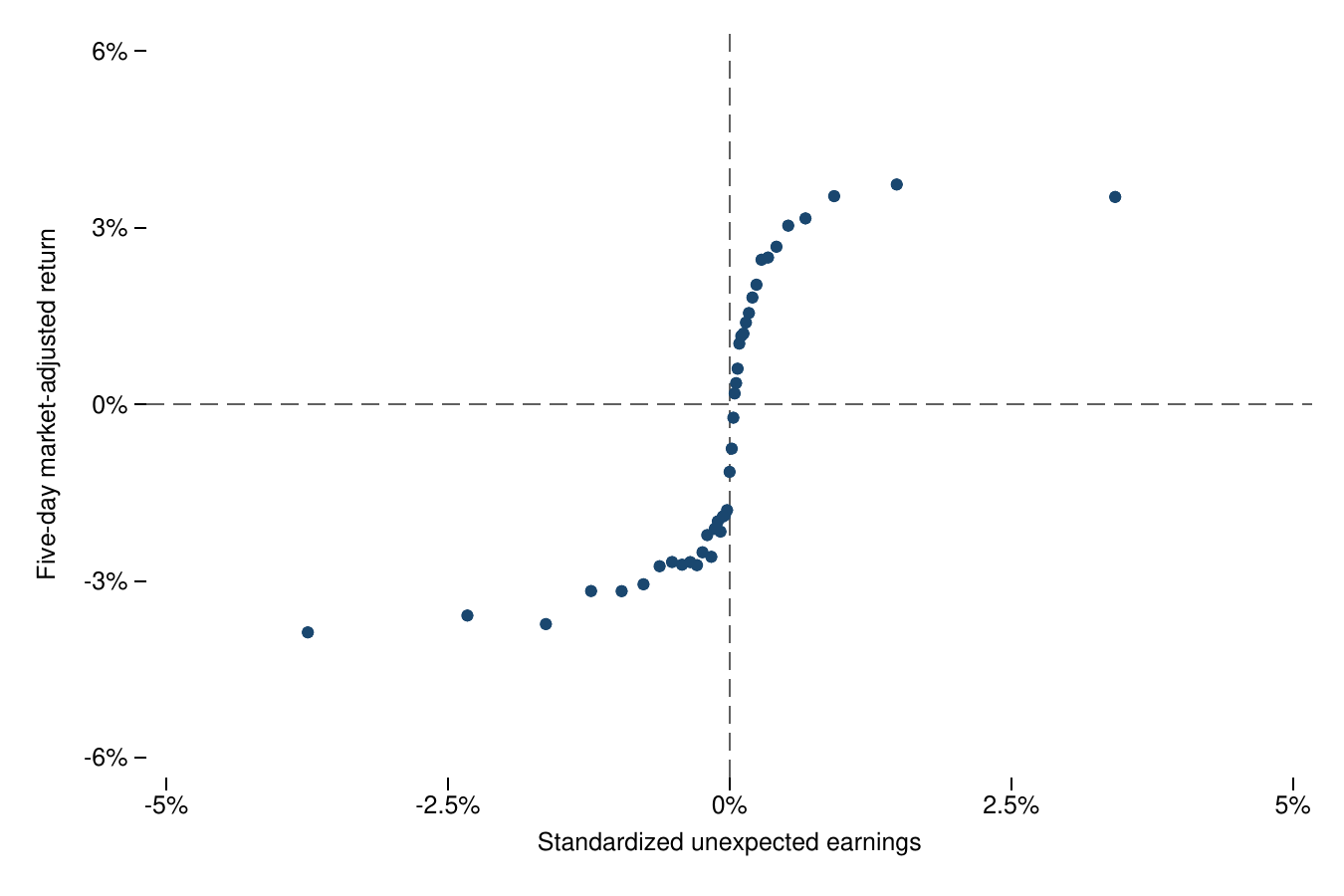}
\caption*{\footnotesize \textit{Notes:} Mean five-day market-adjusted returns in equal-count SUE bins over the full regression sample $|\SUE_{i,t}|\leq5\%$. There are twenty bins on each side of zero, while exact-zero surprises are put in a separate bin. Points show bin means. Both axes are in percentage points.}
\end{figure}

Figure~\ref{fig:field_sshape_raw} plots the average response of stock prices to $\SUE$. The empirical response function is steep as the surprise passes through zero and flattens quickly on either side of zero.  In terms of magnitudes, moving from the lowest plotted bin to the highest corresponds to a difference of about seven percentage points in market-adjusted returns, but almost all of this difference comes from the very steep slope near zero.\footnote{Mean returns are negative in some bins containing small positive surprises. These bins contain many sub-penny earnings beats, which the market treats less favorably than beats of at least one cent. Binned in one-cent increments of the dollar surprise, the first strictly positive bin has a positive mean return.} This unconditional $S$-shaped response to earnings news is well documented and has many proposed explanations \citep{freeman1992nonlinear,burgstahler1997earnings,skinner2002earnings,bhojraj2009making,stein1989efficient,huang2025strategic}. We return to these alternative explanations in Section~\ref{sec:alternative_explanations}.

\subsection{Valuation Uncertainty and the Earnings Response}\label{sec:VU}

In the model, the key driver of the \textit{shape} of the response function is multiplier uncertainty. To bring the model to the data, we use \citet{golubov2026accounting}'s valuation uncertainty (VU) measure as our empirical proxy of multiplier uncertainty.

Broadly, VU is meant to capture the range of plausible market values for a firm given its fundamentals and the valuations of similar firms at a specific point in time. To do this, \citet{golubov2026accounting} use a model that maps accounting inputs into market capitalization through a set of multipliers. Specifically, the model assigns a separate coefficient -- i.e., a  multiplier -- to book equity, earnings, an interaction term between earnings and an indicator for earnings being less than zero, leverage, and capitalized R\&D. To allow these multipliers to vary in both the cross-section and time-series, they estimate the model separately within each industry and year. To capture dispersion in valuations at a particular point in time, they use quantile regressions to estimate the multipliers at the 25th and 75th percentiles of the industry-year level valuation distribution.\footnote{To obtain a sensible residual scale, the model is estimated in logs, and predicted values are exponentiated back into levels before constructing VU.} The uncertainty around the multiplier mapping observables, specifically including earnings, to firm values is therefore the keystone of the construction of VU.

Their next step is to apply both sets of multipliers to each firm's inputs at a given point in time. This produces two implied equity values, one based on the 25th percentile multipliers and another based on the 75th percentile multipliers. VU is then defined as the difference in quantities divided by their midpoint. Variation in VU, therefore, has two main drivers. The first is that the multipliers -- both at the 25th percentile and at the 75th percentile -- vary across industries and over time as they are re-estimated. The second is that firms differ in their underlying inputs to the valuation model. For example, a firm with a large value for a particular input whose multiplier differs substantially across the 25th and 75th percentiles can have a wide range of implied valuations and thus higher VU. In this sense, VU is an input-weighted index of dispersion in valuation multipliers, making it a particularly well-suited measure to study multiplier uncertainty at the firm level. To avoid introducing a look-ahead bias, for each announcement, we use the latest VU estimate whose accounting inputs were public at the time earnings were released.

The two natural next questions are: (1) which of these two channels -- variation in multipliers or variation in inputs -- are the main drivers of variation in VU across firms and over time? and (2) how is VU different from other measures of processing constraints or valuation difficulty?  Appendix Table~\ref{tab:field_vu_variation} shows that, perhaps unsurprisingly because the quantile regressions are computed within each industry-year, a significant driver of variation in VU is across-industry differences.  For example, VU is substantially lower among utilities than among pharmaceutical and software firms. That being said, just 30 percent of the variation in VU is between industries, while about half is within industry-quarters.  There is still significant within-industry variation because firms may differ substantially in the inputs to the valuation model (e.g., book equity and earnings). As to the second question, in Section~\ref{sec:textual}, we re-run our empirical tests using text-based measures of uncertainty and complexity, which we show have little correlation with VU and yet yield similar results. Finally, in Section~\ref{sec:attenuation_benchmark} we compare VU with other measures which have been shown to shape the response function to earnings news such as analyst dispersion, Friday announcements, conflicting peer news, conflicting earnings surprise and guidance surprise, and prior return variance. 

\begin{figure}[!htb]
\centering
\caption{Earnings Response by Valuation Uncertainty}
\label{fig:field_response_fit_hinge}
\includegraphics[width=\linewidth]{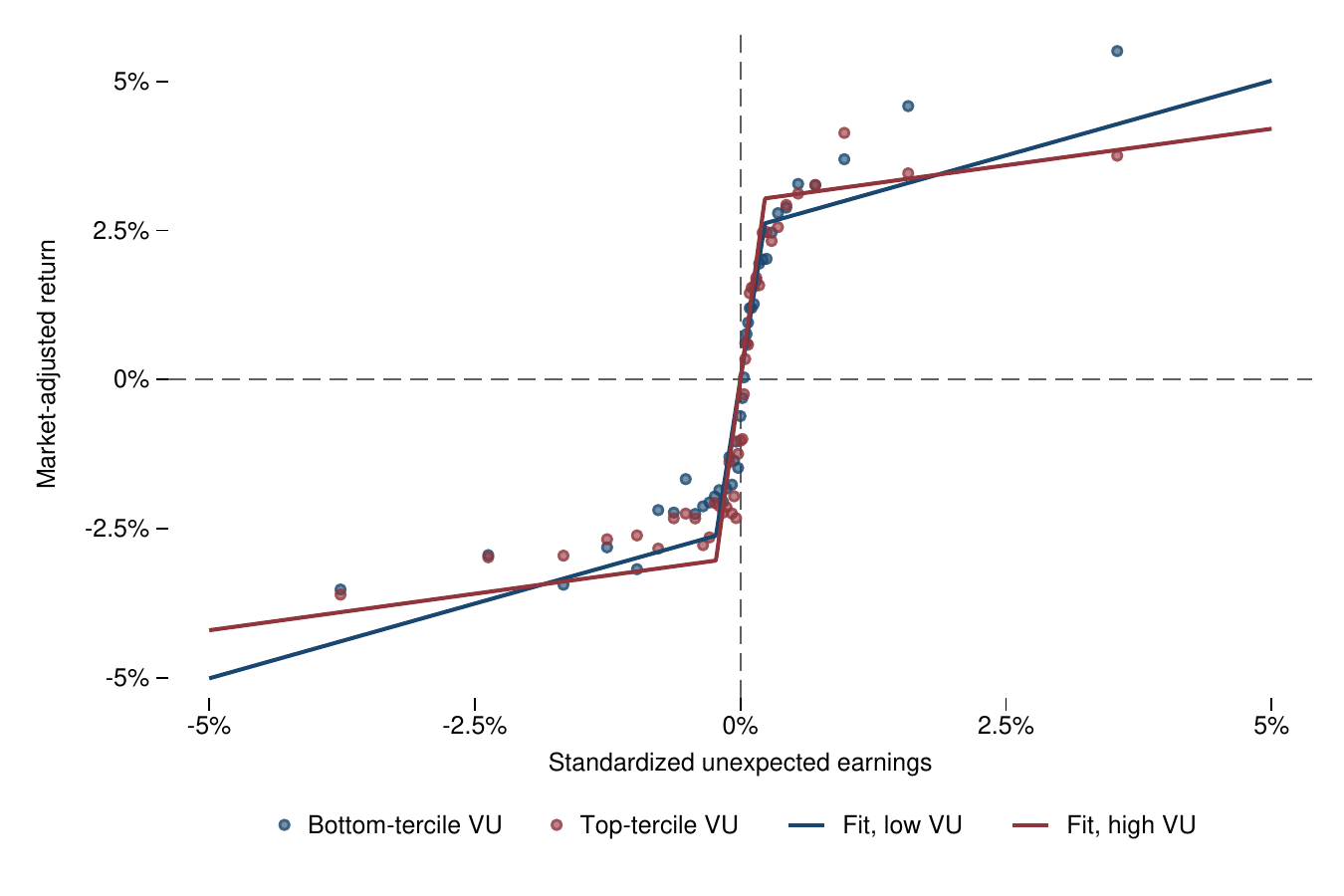}
\caption*{\footnotesize \textit{Notes:} Points are mean five-day market-adjusted returns in common SUE bins for announcements in the bottom and top VU terciles within year-quarter. The sample includes observations satisfying $|\SUE_{i,t}|\leq 5\%$. Lines show the fitted piecewise-linear responses from column (2) of Table~\ref{tab:field_vu_main}, evaluated at the mean VU of each tercile. Group-specific fitted levels are removed from the points and lines so that both are displayed on the same scale. Appendix Figure~\ref{fig:field_sshape_by_vu_unfitted} reports the binned means without fitted lines.}
\end{figure}

The model predicts that multiplier uncertainty changes the shape of the response function to earnings news. Specifically, multiplier uncertainty raises sensitivity near the prior mean while lowering sensitivity farther away from the mean. We first ask whether the model's predictions are visible in raw binned returns. To test this, Figure~\ref{fig:field_response_fit_hinge} splits announcements into the bottom and top VU terciles within each year-quarter.  As shown in Appendix Figure~\ref{fig:field_sue_histogram_by_vu}, the distribution of SUE differs somewhat across the two groups. So, to make the comparison clearer, we form common SUE bins using the full sample of earnings announcements and plot the average return for each VU group within the same SUE intervals, with exact-zero surprises placed in their own bin. Figure~\ref{fig:field_response_fit_hinge} shows that the high-VU response to SUE is steeper around zero but flatter for larger surprises, and the fitted response functions cross on both sides of zero, mirroring the results from the experiment in Figure~\ref{fig:experiment_median_forecasts}.

Next, we aim to quantify the shape of the response function to earnings news with two regressions. The first specification imposes one slope over the full SUE range:
\begin{equation}\label{eq:field_linear}
\begin{aligned}
r_{i,t}
&=\alpha+\beta\SUE_{i,t}
+\mathrm{VU}_{i,t-1}
\left(
\gamma+\delta\SUE_{i,t}
\right) \\
&\quad+\lambda'X_{i,t-1}
+\phi_t+\psi_{\mathrm{ind}}+\varepsilon_{i,t}.
\end{aligned}
\end{equation}
Given that we have normalized VU to have mean zero and standard deviation one, the interaction coefficient $\delta$ measures how a one-standard-deviation increase in VU changes the earnings response coefficient over the full SUE range. We use this as our test of average attenuation.

To investigate whether VU heightens sensitivity to small surprises, we use the same piecewise-linear specification as in the experiment. Specifically, we use the \textit{hinge} function $h(s)$ defined in Equation \eqref{eq:hinge}, which equals $s$ for $|s|\leq k$ and is capped at $-k$ and $k$ outside of this central interval. Including the hinge alongside SUE allows the slope of the response to differ between large and small surprises, while ensuring the fitted response is simply a symmetric, three-part piecewise-linear function, so that differences in coefficients remain interpretable.\footnote{Piecewise-linear response specifications date back at least to \citet{bacon1971estimating}. In economics and finance, two-threshold formulations with a flat middle have been used to identify bands of inaction where prices do not adjust via threshold cointegration or auto-regression \citep{balke1997threshold,obstfeld1997nonlinear}. Variants with differing slopes have been used in system identification \citep{hagenblad1999aspects,pupeikis2005identification} and under the name ``double-hinge'' in biology \citep{terrill2021threshold,neufeldt2026reaction}.} To reflect the symmetry of the optimal response function in the model around the prior mean, we assume that the two outer segments have the same slope.

Specifically, we augment the baseline regression with terms that measure the additional slope around small surprises:
\begin{equation}\label{eq:baseline}
\begin{aligned}
r_{i,t}
&=\alpha+\beta\SUE_{i,t}
+\beta_\Delta h(\SUE_{i,t}) \\
&\quad+\mathrm{VU}_{i,t-1}
\left(
\gamma+\delta\SUE_{i,t}
+\delta_\Delta h(\SUE_{i,t})
\right) \\
&\quad+\lambda'X_{i,t-1}
+\phi_t+\psi_{\mathrm{ind}}+\varepsilon_{i,t}.
\end{aligned}
\end{equation}

In this specification, the coefficient $\delta$ is the VU-related change in the response outside the hinge. The sum $\delta+\delta_\Delta$ is the VU-related change in the total response inside the hinge. The coefficient $\delta_\Delta$, therefore, measures how much more VU changes sensitivity inside the hinge than outside it.

When estimating Equation~\eqref{eq:baseline}, there is a natural question of how to choose $k$, and how robust our results are to our specific choice of $k$.  We choose the hinge threshold $k$ that best fits the unconditional return reaction to SUE plotted in Figure~\ref{fig:field_sshape_raw}. Specifically, as discussed in Appendix~\ref{app:scale_selection}, we select the $k$ that maximizes the $R^2$ in~\eqref{eq:baseline} before VU enters and without any fixed effects or controls, which yields $k=0.23\%$. Appendix Figure~\ref{fig:field_selection_profiles} shows that the corresponding $R^2$ profile has a well-behaved interior optimum. Finally, Appendix Table~\ref{tab:field_scale_selection} documents that this best-fitting hinge threshold is robust to adding fixed effects or controls to the regression. With this choice of $k$, 57\% of announcements sit in the central interval.

Table~\ref{tab:field_vu_main} reports the results. Our preferred specifications in columns (5) and (6) contain industry and year-quarter fixed effects, as well as a large set of controls. The fixed effects are motivated by the construction of VU, where the valuation model's coefficients are estimated at the industry-year level. We further control for firm characteristics which could drive earnings responses and could also be correlated with VU: time since listing (i.e., firm age), market capitalization, returns from months $t-12$ through $t-2$ (i.e., the returns used to form momentum portfolios), book-to-market, betas on the market, size, value and momentum factors, total institutional ownership, and prior-year daily volatility. All controls are measured using information available as of the month end before the earnings announcement. Standard errors are clustered by year-quarter.

Column (5) estimates Equation~\eqref{eq:field_linear} and shows that, on average, raising SUE by 0.1pp is associated with a 0.197pp higher five-day announcement return. Moreover, over the whole sample, a one-standard-deviation increase in VU \textit{reduces} this return difference by 0.029pp, or by about one seventh. We therefore find significant overall attenuation, matching the first prediction of Proposition~\ref{prop:uncertainty_effect}. Column (6) then separates small from large surprises. Among small surprises (defined here as $|\SUE|\leq0.23\%$), a one-standard-deviation increase in VU \textit{raises} the return difference associated with a 0.1pp increase in SUE by 0.088pp. This is the amplified sensitivity to small surprises predicted by Proposition~\ref{prop:uncertainty_effect}. Moreover, the effect of VU on this same 0.1pp increase in SUE is 0.101pp larger near the consensus than for large surprises, matching the third prediction of Proposition~\ref{prop:uncertainty_effect}.

To understand the magnitudes, this means that the difference in post-announcement returns between a 0pp and a 0.1pp surprise is 0.088pp larger when VU increases by one standard deviation. By contrast, the difference between a 1.00pp and a 1.10pp surprise is 0.013pp smaller. Thus, for the same 0.1pp increase in SUE, the effect of VU on return sensitivity is 0.101pp larger for small surprises than for large ones. To summarize, the evidence in Table~\ref{tab:field_vu_main} mirrors the evidence from the experiment: greater multiplier uncertainty weakens the response to news over the full distribution of surprises but increases it around zero.

\begin{table}[!htb]
\centering
\caption{Valuation Uncertainty and the Sensitivity of the Market Earnings Response}
\label{tab:field_vu_main}
\fittable{\begin{tabular}{l*{6}{Y}}
\toprule
& \multicolumn{6}{c}{\textit{Dependent variable: five-day market-adjusted return}} \\
& \multicolumn{2}{c}{No controls} & \multicolumn{2}{c}{Quarter \& industry FEs} & \multicolumn{2}{c}{All controls} \\
\cmidrule(lr){2-3}\cmidrule(lr){4-5}\cmidrule(lr){6-7}
& (1) & (2) & (3) & (4) & (5) & (6) \\
\midrule
SUE                 &        1.94\sym{***}&        0.37\sym{***}&        1.96\sym{***}&        0.38\sym{***}&        1.97\sym{***}&        0.38\sym{***}\\
                    &      (0.09)         &      (0.07)         &      (0.09)         &      (0.07)         &      (0.09)         &      (0.07)         \\
$\Delta$ Small SUE  &                     &       11.94\sym{***}&                     &       12.23\sym{***}&                     &       12.31\sym{***}\\
                    &                     &      (0.45)         &                     &      (0.45)         &                     &      (0.45)         \\
VU                  &       -0.24\sym{***}&       -0.28\sym{***}&       -0.30\sym{***}&       -0.25\sym{***}&       -0.14\sym{***}&       -0.12\sym{***}\\
                    &      (0.06)         &      (0.06)         &      (0.06)         &      (0.06)         &      (0.05)         &      (0.04)         \\
VU $\times$ SUE     &       -0.29\sym{***}&       -0.13\sym{**} &       -0.30\sym{***}&       -0.14\sym{**} &       -0.29\sym{***}&       -0.13\sym{**} \\
                    &      (0.07)         &      (0.06)         &      (0.07)         &      (0.06)         &      (0.06)         &      (0.06)         \\
VU $\times$ $\Delta$ Small SUE&                     &        1.07\sym{***}&                     &        1.02\sym{***}&                     &        1.01\sym{***}\\
                    &                     &      (0.26)         &                     &      (0.26)         &                     &      (0.26)         \\
\midrule
Small SUE           &                     &12.31\sym{***}         &                     &12.60\sym{***}         &                     &12.69\sym{***}         \\
                    &                     &      (0.44)         &                     &      (0.45)         &                     &      (0.45)         \\
VU $\times$ Small SUE&                     &0.94\sym{***}         &                     &0.88\sym{***}         &                     &0.88\sym{***}         \\
                    &                     &      (0.25)         &                     &      (0.24)         &                     &      (0.24)         \\
\midrule
Hinge threshold \(k\)&                     &        0.23         &                     &        0.23         &                     &        0.23         \\
Observations        &     224,350         &     224,350         &     224,350         &     224,350         &     224,350         &     224,350         \\
\(R^{2}\)           &        0.03         &        0.06         &        0.04         &        0.07         &        0.04         &        0.07         \\
Quarter \& industry FEs&                     &                     &$\checkmark$         &$\checkmark$         &$\checkmark$         &$\checkmark$         \\
All controls        &                     &                     &                     &                     &$\checkmark$         &$\checkmark$         \\
\bottomrule
\end{tabular}
}
\caption*{\footnotesize \textit{Notes:} 
The dependent variable is the cumulative stock return from the effective announcement day through four subsequent trading days, minus the cumulative value-weighted market return. Columns (1), (3), and (5) estimate Equation~\eqref{eq:field_linear}, so that VU $\times$ SUE measures the change in the overall earnings response coefficient associated with a one-standard-deviation increase in VU. Columns (2), (4), and (6) estimate Equation~\eqref{eq:baseline}, so that SUE is the response outside $[-k,k]$, $\Delta$ Small SUE the additional response inside that interval, and Small SUE the sum of the two coefficients. The VU interactions are defined analogously. The hinge threshold $k$ is selected from the unconditional response before VU enters and is held fixed across specifications. VU is standardized. The sample retains observations with $|\SUE_{i,t}|\leq 5\%$. Controls are listing age, market capitalization, returns from $t-12$ through $t-2$, book-to-market, four-factor loadings, institutional ownership, and prior-year daily volatility. Regressions are equal-weighted. Standard errors are clustered by year-quarter.}
\end{table}

\subsection{Robustness of Baseline Specification}\label{baselinerobustness}

We examine the robustness of our baseline results across four broad sets of sensitivity checks, where we consider alternative choices of regression specification, of sample, and of definition for abnormal returns and for earnings surprises.

First, Appendix Table~\ref{tab:field_specification_robustness} examines whether our empirical findings are stable across regression specifications and sample restrictions. Specifically, we find that replacing industry fixed effects with firm fixed effects, clustering standard errors by firm and quarter, using median regressions as in the experiment (instead of least-squares regressions), or measuring VU by its within-quarter percentile rank do not change our main findings. Our main results are also not sensitive to dropping zero surprises or earnings surprises smaller than one cent in absolute value. We find that further reducing the effect of outliers in $\SUE$ by restricting the sample to $|\SUE|\leq1\%$, $2\%$, or $3\%$ also has little qualitative effect on our results.

The next natural concern with our main results is that they are disproportionately driven by small firms, as our main regressions are equal-weighted. To test this, we re-run our baseline regression, but weight observations by market capitalization within each quarter, and then also by market capitalization rank within each quarter. We document that, in both cases, higher VU is now associated with an increase in the overall return sensitivity to earnings, but this is mainly driven by our central prediction of VU heightening sensitivity to small surprises, which becomes even stronger under value weighting.

Another possible concern is that our results are sensitive to when and how returns are measured. Appendix Table~\ref{tab:field_event_windows} shows that they are essentially unchanged when we instead consider returns just on the effective earnings announcement date or on the first 3 days post-announcement. Panel C of Appendix Table~\ref{tab:field_specification_robustness} further establishes that results are similar when, rather than using market-adjusted returns, we use residuals from a CAPM, a Fama--French three-factor model, a Carhart four-factor model, or the stock’s return relative to its industry.

Yet another concern is that our results are sensitive to the way we have defined SUE. Appendix Table~\ref{tab:field_sue_scalings} therefore repeats our main analysis with other definitions of SUE used in the literature: surprises relative to forecast dispersion \citep{latane1977standardized}, surprises relative to the absolute value of the consensus in dollars, and using sign-preserving percentile ranks of SUE \citep{shue2021can}. We find that, across all of these alternative definitions, VU increases the sensitivity of the return reaction to small earnings surprises relative to large ones.

We show that our results do not depend on the hinge specification by repeating the expanding-window design used in the experiment. Specifically, for each width $w$, we estimate Equation~\eqref{eq:field_linear} using only observations satisfying $|\SUE_{i,t}|\leq w$. Appendix Figure~\ref{fig:field_expanding_windows} reports the proportional change in sensitivity, $\widehat{\delta}/\widehat{\beta}$. The estimate is positive in narrow windows around zero and becomes negative as progressively larger surprises enter the sample. This confirms the result in Table~\ref{tab:field_vu_main}, and provides a directly comparable test to the experiment showing that the effect of VU on the response function does not depend on a particular hinge threshold or specification.

Finally, we document that, staying closer to the hinge specification but replacing it with smooth functional forms, or varying the calibration of the threshold $k$, does not overturn the results. Specifically, replacing the hinge with the arctangent response used by \citet{freeman1992nonlinear} or the rootsign response motivated by our model produces similarly good fits and identifies the same amplification of small surprises. As discussed in detail in Appendix~\ref{app:scale_selection}, the estimated hinge threshold itself is also very stable, moving only from $k=0.23\%$ without controls or fixed effects to $k=0.22\%$ when they are included.

\subsection{Heterogeneity and Sources of Variation}

Having established the robustness of the baseline result, a natural next question is what heterogeneity in the data actually drives the relationship between VU and the shape of the earnings response function. In this subsection, we hence explore whether our main empirical results hold across firm sizes, profitability and over time, before asking which components of VU drive the main results.

Firm size is a useful starting point, as large firms generally have greater analyst coverage and richer information environments than small firms. If our main results were driven by small stocks, they should weaken with firm size. In Section~\ref{baselinerobustness}, we already show that value weighting actually amplifies the effect of VU on surprises near zero.  As an additional check, Appendix Table~\ref{tab:field_size_and_period} shows that a one-standard-deviation increase in VU raises the return response to each additional percentage point of SUE by $0.81$pp \textit{less} for small surprises than for large ones among the smallest firms, but by $1.03$pp \textit{more} in the middle tercile and by $2.76$pp \textit{more} among the largest firms.

Another natural question is whether the effect of VU varies by firm profitability. This is especially relevant because previous research has established that the shape of the earnings reaction function is very different between profit- and loss-making firms \citep{hayn1995information,gipper2020public}. Also in our data, we find that the $S$-shaped response function which characterizes the amplification of small surprises is very prominent among announcements with positive earnings, but nearly absent among those posting losses (Appendix Figure~\ref{fig:field_sshape_by_profit_all}). Correspondingly, Panel B of Table~\ref{tab:field_size_and_period} shows that the effect of VU on the amplification of small surprises is strong among profitable firms, but not detectable among loss-making ones.

The information environment surrounding earnings announcements also changed substantially over our sample. Regulation FD, which became effective in 2000Q4, restricted selective disclosure and changed how information could reach investors before an earnings announcement. Sarbanes--Oxley followed in 2002Q3, strengthening financial reporting and audit requirements after the failures of Enron and WorldCom. Panel C of Table~\ref{tab:field_size_and_period} shows that while VU increased the return sensitivity to small earnings surprises relative to large ones before Regulation FD, afterwards the point estimate is no longer statistically distinguishable from zero. In Appendix Figure~\ref{fig:field_loss_share}, we show that this apparent change in our results is driven by a compositional shift in our sample.  Specifically, around 2000, there was a systematic increase in loss-making announcements, especially among high-VU firms.  And, as shown in Appendix Figure~\ref{fig:field_sshape_by_profit_all}, our results are weaker among loss-making firms.

Panel D of Appendix Table~\ref{tab:field_size_and_period}, therefore, further splits the pre- and post-Regulation FD analysis by profitability. It shows that, among profit-making firms, VU amplifies the sensitivity to small surprises relative to large ones both before and after 2000Q4. As an additional check, Appendix Figure~\ref{fig:field_coefficients_by_profit} estimates our main specification in rolling windows, separately for profitable and unprofitable firms, and finds that, among the former, the effect of VU on the sensitivity to small surprises is remarkably stable over time.

As discussed in Section~\ref{sec:VU}, VU combines, for each quarter, information from peer valuation multiples and the firm's fundamentals. To understand which part of VU drives our results, we separate variation across time and industries from variation within them. Decomposing VU by year-quarter produces positive local effects from both the between-quarter and within-quarter components, but Appendix Table~\ref{tab:field_variation_quarter} shows that the estimates become less statistically significant when we allow the components to enter separately. Between industries, a one-standard-deviation increase in VU raises the return response to each additional percentage point of SUE by 5.11pp more for small surprises than for large ones. The corresponding within-industry estimates are negative, at $-0.87$, and only marginally significant at the 10\% level. The strongest part of our empirical results, therefore, comes from comparisons across industries whose fundamentals map into market value with systematically different degrees of multiplier uncertainty.

\subsection{Textual Measures of Valuation Difficulty}\label{sec:textual}

We next ask whether our findings hold under alternative measures of valuation difficulty. In particular, because VU relies on industry-time-level valuation multipliers, it rests upon both accounting and price data. To identify potentially uncorrelated measures of valuation difficulty, we consider two proxies that use neither: the textual uncertainty (TU) and textual complexity (TC) measures from \citet{loughran2011liability,loughran2024measuring}, which compare the words in firms' latest 10-K and 10-Q filings with manually curated dictionaries. TU counts expressions of imprecision and contingencies, while TC captures language related to business and financial complexity, especially features that might complicate investors' assessment of future cash flows. TU and TC, therefore, offer alternate proxies for the difficulty of translating reported fundamentals into firm value. We find that their correlations with VU are only $0.09$ and $-0.05$, suggesting they provide a largely independent validation of our main results.

Table~\ref{tab:field_textual} repeats our baseline hinge regression with these measures of text-based uncertainty. We find that both textual measures are associated with greater sensitivity near $\SUE=0$. Column (2) shows that a one-standard-deviation increase in textual uncertainty raises the return response to earnings surprises by 1.75pp for each additional percentage point of SUE near the consensus, which is 1.94pp more than the corresponding increase for larger surprises. On the other hand, textual complexity raises the return response by 0.55pp for small surprises, which is 0.39pp more than for large surprises, although only the former is significant in column (4). Interestingly, TU does not produce significant average attenuation in column (1), while TC instead slightly raises the full-range ERC in column (3). Hence, both of these independent, text-based measures of valuation difficulty reproduce the central model prediction of heightened sensitivity to small surprises.

\begin{table}[!htb]
    \centering
    \caption{Textual Measures and the Sensitivity of the Market Earnings Response}
    \label{tab:field_textual}
    \fittable{\begin{tabular}{l*{4}{Y}}
\toprule
& \multicolumn{4}{c}{\textit{Dependent variable: five-day market-adjusted return}} \\
& \multicolumn{2}{c}{Textual uncertainty} & \multicolumn{2}{c}{Textual complexity} \\
\cmidrule(lr){2-3}\cmidrule(lr){4-5}
& (1) & (2) & (3) & (4) \\
\midrule
SUE                 &        2.25\sym{***}&        0.48\sym{***}&        2.19\sym{***}&        0.44\sym{***}\\
                    &      (0.08)         &      (0.07)         &      (0.09)         &      (0.07)         \\
$\Delta$ Small SUE  &                     &       14.95\sym{***}&                     &       14.71\sym{***}\\
                    &                     &      (0.44)         &                     &      (0.50)         \\
Textual measure     &       -0.20\sym{***}&       -0.28\sym{***}&       -0.01         &       -0.07\sym{**} \\
                    &      (0.04)         &      (0.04)         &      (0.03)         &      (0.03)         \\
Textual measure $\times$ SUE&        0.01         &       -0.19\sym{***}&        0.22\sym{***}&        0.17\sym{***}\\
                    &      (0.05)         &      (0.06)         &      (0.05)         &      (0.06)         \\
Textual measure $\times$ $\Delta$ Small SUE&                     &        1.94\sym{***}&                     &        0.39         \\
                    &                     &      (0.29)         &                     &      (0.27)         \\
\midrule
Small SUE           &                     &15.42\sym{***}         &                     &15.15\sym{***}         \\
                    &                     &      (0.42)         &                     &      (0.48)         \\
Textual measure $\times$ Small SUE&                     &1.75\sym{***}         &                     &0.55\sym{**}         \\
                    &                     &      (0.26)         &                     &      (0.23)         \\
\midrule
Corr. with VU       &        0.09         &        0.09         &       -0.05         &       -0.05         \\
Hinge threshold \(k\)&                     &        0.23         &                     &        0.23         \\
Observations        &     147,181         &     147,181         &     155,165         &     155,165         \\
\(R^{2}\)           &        0.05         &        0.09         &        0.05         &        0.08         \\
Quarter \& industry FEs&$\checkmark$         &$\checkmark$         &$\checkmark$         &$\checkmark$         \\
All controls        &$\checkmark$         &$\checkmark$         &$\checkmark$         &$\checkmark$         \\
\bottomrule
\end{tabular}
}
\caption*{\footnotesize \textit{Notes:} As Table~\ref{tab:field_vu_main}, replacing VU with textual uncertainty (TU) or textual complexity (TC). TU is the share of words in the firm's most recent 10-K or 10-Q filed during the preceding twelve months belonging to the uncertainty dictionary of \citet{loughran2011liability}. TC uses the complexity dictionary of \citet{loughran2024measuring}. Each measure is standardized. All specifications retain the baseline hinge threshold $k=0.23\%$. Samples include all announcements matched to a filing carrying the relevant measure and therefore slightly differ across measures.}
\end{table}

\subsection{Other Sources of Attenuation}\label{sec:attenuation_benchmark}

A weak average earnings response could arise through channels other than the difficulty of interpreting a firm's fundamentals. For example, investors may disagree about expected earnings, devote limited attention to the earnings release, or receive accompanying news that taxes their limited attention or contains conflicting information. Increased pre-earnings announcement volatility may further complicate an investor's optimal response to earnings news of a given magnitude. We examine all these channels by including each measure along with VU in our baseline hinge. Specifically, in Table~\ref{tab:field_joint_attenuation}, for each measure, we ask whether other known determinants of earnings responsiveness can account for the overall attenuation and amplification of small surprises  associated with VU.

Where VU measures the range of valuations consistent with a firm's fundamentals at a given point in time, analyst dispersion measures disagreement about the expected earnings number before it is released. Controlling for disagreement, therefore, allows us to distinguish between uncertainty about earnings and their valuation implications, as studied by \citet{subramanyam1996uncertain} and \citet{kinney2002earnings}. Column (1) shows that analyst dispersion lowers the overall return response to earnings. Column (2) estimates that a one-standard-deviation increase in analyst dispersion \textit{reduces} the return response to each additional percentage point of SUE by 0.95pp more for small surprises than for large ones. Further, even after controlling for dispersion, we find that VU still \textit{raises} the sensitivity to small surprises as in the baseline. As the two measures have opposite effects, analyst dispersion cannot explain the heightened sensitivity of the market reaction near the consensus associated with valuation uncertainty.

\citet{dellavigna2009investor} argue that Friday announcements are typically associated with weaker investor attention, and column (3) indeed shows that they lower overall sensitivity to earnings news. Allowing for a non-linear response, we find that they lower the sensitivity to a marginal percentage point of SUE by 1.10pp \textit{more} for small than for large surprises, while VU still increases it. Hence Friday announcements also attenuate the response to small surprises rather than amplifying it, so that they can similarly not account for the effect of VU.

Contemporaneous news released around the same time as a given earnings announcement can also influence how investors process that information \citep{hirshleifer2009driven}. We measure this using the dispersion of earnings surprises released by other firms in the same industry and calendar week.\footnote{It should be noted that, as the measure can include announcements later than the focal release, it describes the week's news environment rather than only information available on or before the announcement itself.} Like the previous measures, greater dispersion not only lowers full-range sensitivity, but furthermore reduces the return reaction to SUE by 0.77pp more for small than for large surprises, as can be seen in columns (5)--(6). Moreover, even after controlling for this dispersion, the small-surprise-enhancing effect of VU remains robust.

A muted reaction to an earnings beat could also reflect unfavorable guidance released alongside it, rather than a weak response to this signal \citep{rogers2013bundled,beaver2020increased}. To test this, we identify earnings announcements with contemporaneous management guidance and measure the guidance surprise as the difference between the annual EPS guidance and the median analyst forecast for the same fiscal period. We then define conflicting guidance as cases where the guidance surprise and $\SUE$ have opposite signs. We find that for this subset of earnings announcements, conflicting guidance sharply \textit{decreases} overall sensitivity and especially the sensitivity to small surprises.

Finally, we use prior return variance to assess whether the relationship between VU and the earnings response function is driven by the stock's broader volatility environment. Whereas VU summarizes the range of valuations consistent with given fundamentals, historical volatility measures realized price fluctuations, and we find that they indeed do not appear to tightly proxy the difficulty of mapping signals into returns. Indeed, columns (9)--(10) show that prior variance lowers the full-range ERC, and like the other measures \textit{lowers it more strongly} for small surprises than for large ones, so that it does not undo the effect of VU.

To summarize, we find that across all five specifications, VU continues to affect the earnings response as predicted by Proposition~\ref{prop:uncertainty_effect}, even after controlling for these possibly confounding sources of attenuation. In particular, VU amplifies the sensitivity to small surprises relative to large ones, while the other measures generally weaken it. This shows why it is important to empirically quantify the \textit{shape} of the earnings-reaction function: similar overall attenuation does not tell us whether or not a given measure amplifies or decreases sensitivity to small surprises. Our results suggest that only measures which directly speak to multiplier uncertainty appear to steepen the $S$-shape exhibited by the market return reaction.

\begin{table}[!htb]
    \centering
    \caption{Valuation Uncertainty and Other Sources of Attenuation}
    \label{tab:field_joint_attenuation}
    \fittable{\begin{tabular}{l*{10}{c}}
\toprule
& \multicolumn{10}{c}{\textit{Dependent variable: five-day market-adjusted return}} \\
& \multicolumn{2}{c}{\shortstack[c]{Analyst\\dispersion}} & \multicolumn{2}{c}{\shortstack[c]{Friday\\release}} & \multicolumn{2}{c}{\shortstack[c]{Peer-news\\dispersion}} & \multicolumn{2}{c}{\shortstack[c]{EPS-guidance\\conflict}} & \multicolumn{2}{c}{\shortstack[c]{Prior return\\variance}} \\
\cmidrule(lr){2-3}\cmidrule(lr){4-5}\cmidrule(lr){6-7}\cmidrule(lr){8-9}\cmidrule(lr){10-11}
& (1) & (2) & (3) & (4) & (5) & (6) & (7) & (8) & (9) & (10) \\
\midrule
SUE                 &        2.21\sym{***}&        0.36\sym{***}&        2.00\sym{***}&        0.38\sym{***}&        2.11\sym{***}&        0.45\sym{***}&        7.95\sym{***}&        1.88\sym{***}&        2.12\sym{***}&        0.44\sym{***}\\
                    &      (0.11)         &      (0.08)         &      (0.09)         &      (0.07)         &      (0.09)         &      (0.07)         &      (0.33)         &      (0.38)         &      (0.10)         &      (0.07)         \\
$\Delta$ Small SUE  &                     &       13.05\sym{***}&                     &       12.52\sym{***}&                     &       12.30\sym{***}&                     &       24.36\sym{***}&                     &       12.21\sym{***}\\
                    &                     &      (0.48)         &                     &      (0.44)         &                     &      (0.45)         &                     &      (1.08)         &                     &      (0.42)         \\
VU $\times$ SUE     &       -0.27\sym{***}&       -0.18\sym{***}&       -0.28\sym{***}&       -0.13\sym{**} &       -0.31\sym{***}&       -0.17\sym{***}&       -0.78\sym{***}&       -0.68\sym{***}&       -0.12\sym{*}  &       -0.15\sym{**} \\
                    &      (0.06)         &      (0.06)         &      (0.06)         &      (0.06)         &      (0.06)         &      (0.06)         &      (0.20)         &      (0.26)         &      (0.07)         &      (0.06)         \\
VU $\times$ $\Delta$ Small SUE&                     &        1.66\sym{***}&                     &        1.02\sym{***}&                     &        1.24\sym{***}&                     &        4.08\sym{***}&                     &        1.81\sym{***}\\
                    &                     &      (0.27)         &                     &      (0.26)         &                     &      (0.27)         &                     &      (0.68)         &                     &      (0.27)         \\
Candidate $\times$ SUE&       -0.07\sym{***}&        0.15\sym{***}&       -0.16\sym{**} &        0.03         &       -0.21\sym{***}&       -0.01         &       -6.44\sym{***}&       -2.03\sym{***}&       -0.28\sym{***}&        0.06         \\
                    &      (0.03)         &      (0.04)         &      (0.07)         &      (0.08)         &      (0.03)         &      (0.03)         &      (0.48)         &      (0.56)         &      (0.04)         &      (0.04)         \\
Candidate $\times$ $\Delta$ Small SUE&                     &       -0.95\sym{***}&                     &       -1.10\sym{**} &                     &       -0.77\sym{***}&                     &      -16.53\sym{***}&                     &       -1.88\sym{***}\\
                    &                     &      (0.24)         &                     &      (0.42)         &                     &      (0.23)         &                     &      (1.36)         &                     &      (0.29)         \\
\midrule
Small SUE           &                     &13.40\sym{***}         &                     &12.90\sym{***}         &                     &12.75\sym{***}         &                     &26.23\sym{***}         &                     &12.66\sym{***}         \\
                    &                     &      (0.47)         &                     &      (0.44)         &                     &      (0.44)         &                     &      (0.88)         &                     &      (0.42)         \\
VU $\times$ Small SUE&                     &1.48\sym{***}         &                     &0.89\sym{***}         &                     &1.07\sym{***}         &                     &3.40\sym{***}         &                     &1.65\sym{***}         \\
                    &                     &      (0.26)         &                     &      (0.24)         &                     &      (0.26)         &                     &      (0.54)         &                     &      (0.25)         \\
Candidate $\times$ Small SUE&                     &-0.80\sym{***}         &                     &-1.07\sym{***}         &                     &-0.78\sym{***}         &                     &-18.56\sym{***}         &                     &-1.82\sym{***}         \\
                    &                     &      (0.21)         &                     &      (0.39)         &                     &      (0.21)         &                     &      (1.05)         &                     &      (0.28)         \\
\midrule
Hinge threshold \(k\)&                     &        0.23         &                     &        0.23         &                     &        0.23         &                     &        0.23         &                     &        0.23         \\
Observations        &     176,323         &     176,323         &     224,350         &     224,350         &     193,779         &     193,779         &      23,089         &      23,089         &     224,350         &     224,350         \\
\(R^{2}\)           &        0.05         &        0.08         &        0.04         &        0.07         &        0.05         &        0.07         &        0.11         &        0.17         &        0.05         &        0.07         \\
Quarter \& industry FEs&$\checkmark$         &$\checkmark$         &$\checkmark$         &$\checkmark$         &$\checkmark$         &$\checkmark$         &$\checkmark$         &$\checkmark$         &$\checkmark$         &$\checkmark$         \\
All controls        &$\checkmark$         &$\checkmark$         &$\checkmark$         &$\checkmark$         &$\checkmark$         &$\checkmark$         &$\checkmark$         &$\checkmark$         &$\checkmark$         &$\checkmark$         \\
\bottomrule
\end{tabular}
}
\caption*{\footnotesize \textit{Notes:} Each column pair enters VU jointly with the candidate measure named in the header. The first column presents the estimates from estimating a linear earnings response on the full range of SUEs, while the second column estimates Equation~\eqref{eq:baseline}. In the hinge columns, the coefficient on SUE represents the ERC outside $[-k,k]$, and $\Delta$ Small SUE is the additional ERC inside that interval, while Small SUE is the sum of these two coefficients. All the interaction terms are defined in the same way. Any continuous measures included in the regression are winsorized at the first and ninety-ninth percentiles and standardized. Friday and EPS-guidance conflict are dummies. Peer-news dispersion is the leave-one-out standard deviation of SUE among other firms in the same FF49 industry and calendar week. Prior return variance sums squared daily returns over the previous twelve calendar months, lagged by one month, and is removed from the control vector in columns (9)-(10). Standard errors are clustered by year-quarter.}
\end{table}

\subsection{Relationship to Alternative Explanations}\label{sec:alternative_explanations}

As we model in Appendix~\ref{app:model_nonlinear}, multiplier uncertainty acts as a complement to other forces that may drive the $S$-shaped response to news. In this subsection, we ask whether the heightened sensitivity to small surprises induced by VU is explained by other previously proposed explanations for the $S$-shape in the literature: earnings persistence, reporting incentives around the consensus, information incorporation before the earnings announcement, or bundled guidance. We would like to highlight, however, that while these forces may affect the \textit{unconditional} shape of the earnings response function, they threaten our results only if they differentially steepen the response near zero for high-VU relative to low-VU firms.

Early work on the $S$-shaped response of prices to earnings news argued that large surprises had weaker marginal responses because the resulting earnings growth was less persistent \citep{freeman1992nonlinear,das1994nonlinearity,lipe1998nonlinearity}. For this to create issues for our empirical design, however, this would have to apply more strongly to high- than low-VU announcements. To test this, Appendix Table~\ref{tab:field_future_fundamentals} augments our baseline regression with future EPS growth and an interaction term between future EPS growth and VU. We find that controlling for future EPS growth has little effect on our baseline results. If anything, we find that it increases the effect of VU on the relative amplification of small surprises from $1.086$ to $1.135$, and to $1.178$ when the future-SUE and EPS-growth interactions enter jointly.

Incentives for managers to meet or beat earnings expectations may lead the market to interpret a small beat differently from a small miss \citep{burgstahler2006management,bhojraj2009making}. Specifically, given that the market understands that management has a strong incentive to engineer at least a small earnings beat, observing a miss suggests that the firm has serious problems. This could affect our results if these incentives, or the ability to engineer small beats, differs across high- and low-VU announcements. To test this, Appendix Figure~\ref{fig:field_sue_histogram_by_vu} compares the distributions of earnings surprises in the top and bottom VU terciles. Both distributions are concentrated around the consensus, but high-VU firms do not exhibit greater bunching: if anything, the central mass is larger among low-VU firms. This is evidence against a simple differential-bunching account of our findings.

Another explanation for our results is that high-VU firms have a different share of their earnings information incorporated into prices before it is formally released. If less information has been incorporated beforehand, more is left to be learned which could lead to a larger return response \citep{atiase1985predisclosure,weller2018algorithmic}.  As mentioned above, however, this would only create a problem for our results if it differentially affected the incorporation of information for small versus large surprises for higher-VU announcements. To test this, Appendix Table~\ref{tab:field_preannouncement} conditions on two measures of pre-earnings announcement information incorporation. We find that controlling for the price-jump measure of \citet{weller2018algorithmic} has little effect on our baseline results. Instead using volatility-normalized absolute announcement returns -- which can be computed for all earnings announcements -- leaves the results qualitatively unchanged but statistically insignificant. However, part of the statistical weakness is unavoidable, since these controls are themselves based on the announcement return and hence absorb part of the variation our specification is designed to explain. Overall, the evidence in Appendix Table~\ref{tab:field_preannouncement}  suggests that differences in pre-announcement information incorporation across high- and low-VU firms are not driving our results. 

One final concern is that other information may be released at the same time as earnings news, e.g., management guidance \citep{rogers2013bundled,beaver2020increased}. In Section~\ref{sec:attenuation_benchmark}, we showed that conflicting guidance attenuates responses but does not overturn the amplification of small surprises by VU. Out of an abundance of caution, we now simply exclude all announcements with contemporaneous guidance and re-estimate the baseline. Appendix Table~\ref{tab:field_future_fundamentals}, Panel B, shows this does not materially change the results. The amplification of small surprises is therefore present even when the earnings announcement contains no identified EPS guidance.

\subsection{Valuation Uncertainty and Return Predictability}

One final question is whether the amplified response to earnings surprises near zero for high-VU firms reflects overreaction. Importantly, if it were overreaction, we would expect to observe reversals for these events after the announcement. On the other hand, one might be concerned that the weaker response to large surprises could reflect underreaction. In turn, if this were the case, we would expect to observe post-event continuation. In our model in Section~\ref{sec:model}, however, neither part of the $S$-shaped response to earnings news is a ``mistake'', i.e., an under- or overreaction. In fact, both are part of the optimal response when investors are uncertain about how earnings news maps into value, and therefore neither should generate predictable returns after the announcement. Strictly, this prediction also requires that beliefs are calibrated and are reflected in prices, so we read the following tests as consistency checks rather than as proof of optimality.

We test this prediction in two steps. First, we verify that the familiar post-earnings-announcement drift (PEAD) is present in our sample across several horizons \citep{ball1968empirical,bernard1989post}. Second, we test whether VU predicts differential reversal or continuation following the announcement, i.e., differences in PEAD.

Table~\ref{tab:field_pead_horizons} shows that over our sample, there is strong evidence of unconditional PEAD. In terms of magnitudes, we find that over the entire sample, a 0.1pp higher SUE is associated with a 0.052pp higher return over days~5--21 and a 0.112pp higher return over days~5--59. However, we detect neither a reversal of the stronger short-term response near zero over the following days, nor a continuation of the weaker response in the tails. Over days~5--59, a one-standard-deviation increase in VU raises the subsequent return response to each additional percentage point of SUE by only $0.07$pp more for small surprises than for large ones, compared with $1.01$pp during the announcement window. The post-announcement point estimate is economically small, but is also not estimated precisely: its 95\% confidence interval is $[-1.07, 1.21]$, so it does not rule out a reversal comparable in magnitude to the announcement-window effect of $1.01$. Appendix Figure~\ref{fig:field_response_pead} illustrates the same absence of a systematic pattern in subsequent returns by SUE bin and VU tercile.

\begin{table}[!htb]
\centering
\caption{Valuation Uncertainty and Subsequent Returns}
\label{tab:field_pead_horizons}
\fittable{\begin{tabular}{l*{6}{Y}}
\toprule
& \multicolumn{6}{c}{\textit{Dependent variable: post-announcement market-adjusted return}} \\
& \multicolumn{2}{c}{Days 5--21} & \multicolumn{2}{c}{Days 5--43} & \multicolumn{2}{c}{Days 5--59} \\
\cmidrule(lr){2-3}\cmidrule(lr){4-5}\cmidrule(lr){6-7}
& (1) & (2) & (3) & (4) & (5) & (6) \\
\midrule
SUE                 &        0.52\sym{***}&        0.39\sym{***}&        0.87\sym{***}&        0.53\sym{***}&        1.12\sym{***}&        0.61\sym{***}\\
                    &      (0.04)         &      (0.06)         &      (0.07)         &      (0.09)         &      (0.09)         &      (0.12)         \\
$\Delta$ Small SUE  &                     &        1.06\sym{***}&                     &        2.61\sym{***}&                     &        3.89\sym{***}\\
                    &                     &      (0.30)         &                     &      (0.56)         &                     &      (0.69)         \\
VU                  &        0.05         &        0.05         &        0.14         &        0.14         &        0.07         &        0.08         \\
                    &      (0.08)         &      (0.08)         &      (0.16)         &      (0.16)         &      (0.21)         &      (0.21)         \\
VU $\times$ SUE     &       -0.06         &       -0.06         &       -0.12\sym{*}  &       -0.07         &       -0.12         &       -0.04         \\
                    &      (0.05)         &      (0.06)         &      (0.07)         &      (0.09)         &      (0.10)         &      (0.12)         \\
VU $\times$ $\Delta$ Small SUE&                     &        0.18         &                     &        0.04         &                     &        0.07         \\
                    &                     &      (0.24)         &                     &      (0.41)         &                     &      (0.58)         \\
\midrule
Small SUE           &                     &1.44\sym{***}         &                     &3.15\sym{***}         &                     &4.50\sym{***}         \\
                    &                     &      (0.26)         &                     &      (0.49)         &                     &      (0.62)         \\
VU $\times$ Small SUE&                     &        0.12         &                     &       -0.02         &                     &        0.03         \\
                    &                     &      (0.21)         &                     &      (0.36)         &                     &      (0.51)         \\
\midrule
Hinge threshold \(k\)&                     &        0.23         &                     &        0.23         &                     &        0.23         \\
Observations        &     222,536         &     222,536         &     222,536         &     222,536         &     222,536         &     222,536         \\
\(R^{2}\)           &        0.04         &        0.04         &        0.06         &        0.06         &        0.07         &        0.07         \\
Quarter \& industry FEs&$\checkmark$         &$\checkmark$         &$\checkmark$         &$\checkmark$         &$\checkmark$         &$\checkmark$         \\
All controls        &$\checkmark$         &$\checkmark$         &$\checkmark$         &$\checkmark$         &$\checkmark$         &$\checkmark$         \\
\bottomrule
\end{tabular}
}
\caption*{\footnotesize \textit{Notes:} As Table~\ref{tab:field_vu_main}, using cumulative market-adjusted returns over the post-announcement window indicated in each column pair, winsorized at the first and ninety-ninth percentiles. All columns include the controls and industry and year-quarter fixed effects. Standard errors are clustered by year-quarter.}
\end{table}

The fact that we do not find a statistically significant relationship between valuation uncertainty and PEAD does not rule out that such a relationship actually exists. As we look at progressively longer return windows, cumulative returns incorporate more volatility unrelated to the earnings announcement, potentially increasing residual variance and thus the standard errors. We therefore also examine whether the VU-related steepening of the initial earnings response around zero persists as we extend the post-earnings announcement return window. Specifically, Appendix Table~\ref{tab:field_cumulative_returns} and Appendix Figure~\ref{fig:field_cumulative_returns} examine cumulative returns from the announcement day through progressively longer horizons. This differs from Table~\ref{tab:field_pead_horizons}, which focuses only on the post-announcement window and excludes the initial earnings response whose persistence we wish to assess.

We find that the relationship between VU and the amplification of small surprises is very stable across horizons. The coefficient measuring how much more VU raises sensitivity to small than to large surprises is $1.02$ over days~0--4 and $1.00$ over days~0--59, i.e., it remains nearly unchanged as the window approaches the next earnings announcement. Perhaps unsurprisingly, standard errors increase with the horizon: despite these similar point estimates, our baseline finding of steepening near zero for high-VU earnings announcements remains statistically significant through day~29 but is no longer significant at day~59.

These results are consistent with the model's prediction that the VU-related amplification of the initial response to small earnings surprises is not expected to subsequently reverse. We also observe unconditional post-earnings-announcement drift in our sample, but find no statistically significant evidence that VU changes continuation differently for small and large surprises. Together, we view these results as all pointing to the same conclusion: there is no clear indication of VU-related underreaction or overreaction.

\FloatBarrier

\section{Conclusion}\label{sec:conclusion}

In this paper, we study the problem of translating a signal into an optimal response when the mapping between the two is unknown to the decision maker. Our specific application is to earnings announcements, where investors must determine how much an earnings surprise should move a stock's price but cannot directly observe the multiplier that translates earnings news into post-announcement returns. We develop a model which captures investors' simultaneous updating about the magnitude and importance of the news, where inference about the multiplier is driven by investors' prior beliefs about how much stock prices typically move around earnings announcements. This mechanism produces an $S$-shaped response of returns to news. The key comparative static is that greater \textit{uncertainty} about the multiplier attenuates responses on average but steepens the slope of the response function near the consensus, which distinguishes our model from results in the literature in which decision difficulty attenuates responses everywhere \citep{gabaix2019behavioral,woodford2020modeling,enke2023cognitive,enke2025behavioral}.

We test this prediction by exploring two sources of multiplier uncertainty. In our experiment, we randomly tighten processing constraints for participants forecasting returns for real firms about to announce earnings, while holding the earnings news fixed. In our field evidence, we use \cite{golubov2026accounting}'s measure of valuation uncertainty, along with two text-based measures of valuation difficulty. In both settings, greater multiplier uncertainty is associated with weaker responses to earnings news on average but stronger responses near the consensus.

One might also ask whether these patterns reflect mistakes by investors. In our model, investors update their beliefs using Bayes' rule, i.e., they update rationally given their uncertainty about the multiplier. Although we observe post-earnings-announcement drift in our sample, we find no evidence of systematic continuation or reversal associated with valuation uncertainty, even conditional on the magnitude of the earnings surprise. We view this as consistent with the model's prediction that stronger responses to small surprises and weaker responses to large ones need not reflect mistakes but can instead arise from investors learning about both cash flows and the multiplier from the same earnings news.

More broadly, our findings suggest that making a decision harder can change which news people respond to most strongly. This may matter in other settings where investors can judge the direction of news more easily than its implications for value, such as deciding how to price an IPO when roadshow demand exceeds the number of shares offered or inferring a fund manager's skill from returns. We believe that future research could examine how investors learn about the strength of these mappings and how that learning affects their responses to news.

\clearpage
\FloatBarrier

\begin{spacing}{0.9}
\bibliographystyle{aer}
\bibliography{references}   
\end{spacing}

\clearpage

\pagebreak
\setcounter{page}{1}

\appendix

\renewcommand{\thetable}{\thesection\arabic{table}}
\renewcommand{\thefigure}{\thesection\arabic{figure}}
\renewcommand{\theHtable}{\thesection.\arabic{table}}
\renewcommand{\theHfigure}{\thesection.\arabic{figure}}

\setcounter{table}{0} \setcounter{figure}{0}

\begin{center}

{\Large Appendix for \\[.5cm]
\textsc{``Huge If True:\\[.2cm] Learning from the Size of the Surprise''}}

\vspace{1cm}

{\large
\begin{tabular}{@{}c@{\hspace{2.2cm}}c@{}}
Thomas Graeber & Christopher Roth \\[0.35em]
Marco Sammon & Constantin Schesch
\end{tabular}
}

\end{center}

This appendix provides supplementary theoretical and empirical material. Appendix~\ref{app:model} contains proofs and extensions of the model. Appendix~\ref{app:exper} reports additional results and robustness analyses for the experiment. Appendix~\ref{app:field} provides additional field evidence and robustness checks.

\section{Model Proofs and Extensions}\label{app:model}

\subsection{Proofs}

\begin{proof}[Proof of Proposition~\ref{prop:rootsign_posterior_mean}]
Since $s=r/\beta$, the likelihood is $s\mid\beta\sim\mathcal N(0,\sigma_r^2/\beta^2)$. Normal--Gamma conjugacy gives
\[
\beta^2\mid s
\sim
\mathrm{Gamma}\left(
\kappa^{-1}+\tfrac12,
\frac{1}{\kappa\bar\beta^2}+\frac{s^2}{2\sigma_r^2}
\right).
\]
For $X\sim\mathrm{Gamma}(a,d)$, $\Exp[X^p]=d^{-p}\Gamma(a+p)/\Gamma(a)$ whenever $a+p>0$. Applying this identity with $p=1/2$ gives the posterior multiplier stated below Proposition~\ref{prop:rootsign_posterior_mean}, and  multiplying by $s$ yields~\eqref{eq:response}. The response is odd and $C^\infty$, and calculation gives
\begin{equation}\label{eq:local_sensitivity}
\hat r'(s)
=
A(\kappa)\bar\beta
\left(
1+\frac{\kappa\bar\beta^2s^2}{2\sigma_r^2}
\right)^{-3/2}
>0.
\end{equation}
\end{proof}

\begin{proof}[Proof of Proposition~\ref{prop:uncertainty_effect}]
For $0<\kappa<1$, iterated expectations and independence give
\[
\Exp[s\hat r(s)]
=
\Exp[sr]
=
\sigma_r^2\Exp[\beta^{-1}],
\qquad
\Exp[s^2]
=
\sigma_r^2\Exp[\beta^{-2}].
\]
The Gamma moment formula and recurrence therefore imply
\[
b_{\mathrm{OLS}}
=
\frac{\Exp[\beta^{-1}]}{\Exp[\beta^{-2}]}
=
\sqrt{\kappa}\bar\beta
\frac{\Gamma(\kappa^{-1}-\tfrac12)}
{\Gamma(\kappa^{-1}-1)}
=
\frac{1-\kappa}{1-\kappa/2}
\frac{\bar\beta}{A(\kappa)}.
\]

To sign the comparative statics, we use the integral representation of the digamma function, which yields
\begin{equation}\label{eq:amplification_bounds}
\begin{aligned}
\frac{A'(\kappa)}{A(\kappa)}
&=
\frac{1}{2\kappa}
-
\frac{\psi(\kappa^{-1}+1)-\psi(\kappa^{-1}+\tfrac12)}{\kappa^2}\\
&=
\frac{1}{2\kappa}
\int_0^\infty e^{-t}\tanh\left(\frac{\kappa t}{4}\right)\,dt
\in
\left(0,\min\left\{\frac{1}{2\kappa},\frac18\right\}\right).
\end{aligned}
\end{equation}
The bounds then follow from $0<\tanh x<\min\{x,1\}$ for $x>0$. Thus $A$ is strictly increasing. The Gamma-ratio asymptotic gives $A(\kappa)\to1$ as $\kappa\to0$, so $A(\kappa)>1$. Since $(1-\kappa)/(1-\kappa/2)$ is positive, strictly decreasing, and below one on $(0,1)$, $b_{\mathrm{OLS}}$ is strictly decreasing and below $\bar\beta$. Equation~\eqref{eq:local_sensitivity} gives $\hat r'(0)=A(\kappa)\bar\beta>\bar\beta$ and the stated slope ratio, which strictly increases with $\kappa$ for fixed $s\neq0$. Finally,~\eqref{eq:response} and~\eqref{eq:local_sensitivity} give the certainty limits.
\end{proof}

\begin{proof}[Proof of Corollary~\ref{cor:categorical_limit}]
As $\kappa\to\infty$, $\Gamma(\kappa^{-1}+1)\to\Gamma(1)=1$ and $\Gamma(\kappa^{-1}+\tfrac12)\to\Gamma(\tfrac12)=\sqrt\pi$, so $A(\kappa)/\sqrt{\kappa}\to1/\sqrt{\pi}$. For fixed $s\neq0$, $\sqrt{1+\tfrac{\kappa\bar\beta^2}{2\sigma_r^2}s^2}\sim\sqrt{\kappa}\,\bar\beta|s|/(\sqrt2\,\sigma_r)$, so that
\[
\hat r(s)
=A(\kappa)\,\bar\beta\,\frac{s}{\sqrt{1+\tfrac{\kappa\bar\beta^2}{2\sigma_r^2}s^2}}
\; \to \;
\frac{\sqrt{\kappa}}{\sqrt{\pi}}\,\bar\beta\,\frac{\sqrt2\,\sigma_r\,s}{\sqrt{\kappa}\,\bar\beta\,|s|}
=\sqrt{\tfrac{2}{\pi}}\,\sigma_r\,\sign(s).
\]
Since $\beta>0$, $\sign(s)=\sign(r)$, and the half-normal mean of $r\sim\mathcal N(0,\sigma_r^2)$ gives $\Exp[r\mid r>0]=\sigma_r\sqrt{2/\pi}$, so the limit equals $\Exp[r\mid\sign(r)=\sign(s)]$.
\end{proof}

\subsection{Additional results}

\begin{corollary}[Difference in local sensitivities]
\label{cor:slope_difference}
For every $\kappa>0$ and fixed $s\neq0$,
\begin{equation}\label{eq:slope_difference}
\hat r'(0)-\hat r'(s)
=
A(\kappa)\bar\beta
\left[
1-
\left(
1+\frac{\kappa\bar\beta^2s^2}{2\sigma_r^2}
\right)^{-3/2}
\right],
\qquad
\frac{\partial}{\partial\kappa}
\left[\hat r'(0)-\hat r'(s)\right]>0.
\end{equation}
\end{corollary}

\begin{proof}
The identity follows from~\eqref{eq:local_sensitivity}. For fixed $s\neq0$, both $A(\kappa)$ and the expression in brackets are positive and have strictly positive derivatives with respect to $\kappa$.
\end{proof}

\begin{corollary}[Single crossing in local sensitivity]
\label{corollary:slope_single_crossing}
For every $\kappa>0$, there exists a unique cutoff $s^*(\kappa)>0$ such that
\[
\frac{\partial\hat r'(s)}{\partial\kappa}
\begin{cases}
>0, & |s|<s^*(\kappa),\\
=0, & |s|=s^*(\kappa),\\
<0, & |s|>s^*(\kappa),
\end{cases}
\qquad
s^*(\kappa)^2
=
\frac{2\sigma_r^2}{\bar\beta^2}
\frac{A'(\kappa)/A(\kappa)}
{\tfrac32-\kappa A'(\kappa)/A(\kappa)}.
\]
For any $\kappa_H>\kappa_L>0$, the response functions cross once on each side of zero.
\end{corollary}

\begin{proof}
Equation~\eqref{eq:local_sensitivity} implies
\[
\frac{\partial}{\partial\kappa}\log\hat r'(s)
=
\frac{A'(\kappa)}{A(\kappa)}
-
\frac32
\frac{\bar\beta^2s^2/(2\sigma_r^2)}
{1+\kappa\bar\beta^2s^2/(2\sigma_r^2)}.
\]
This expression strictly decreases with $|s|$, from $A'/A>0$ to $A'/A-3/(2\kappa)<0$ by~\eqref{eq:amplification_bounds}. Solving its unique zero gives the cutoff. (Note that, for the parametrization used in Figure~\ref{fig:model_sensitivities}, the same bound gives $\partial_\kappa\log\hat r'(1)<1/8-3/[2(2+\kappa)]<0$ on $(0,1)$.)

For $s>0$, the ratio of response levels is
\[
\frac{\hat r(s;\kappa_H)}{\hat r(s;\kappa_L)}
=
\frac{A(\kappa_H)}{A(\kappa_L)}
\left(
\frac{1+\kappa_L\bar\beta^2s^2/(2\sigma_r^2)}
{1+\kappa_H\bar\beta^2s^2/(2\sigma_r^2)}
\right)^{1/2}.
\]
It strictly decreases from a limit above one to a limit below one, because $A$ increases while $A(\kappa)/\sqrt{\kappa}$ decreases by~\eqref{eq:amplification_bounds}. Oddness gives the negative-side crossing. The crossings of response levels need not coincide with those of marginal sensitivities.
\end{proof}

\begin{corollary}[Uniform attenuation]\label{cor:uniform_attenuation}
Fix $\kappa>0$ and $0<c<1$. Let $\hat r_c(s)$ denote the posterior response after replacing $\bar\beta$ by $c\bar\beta$ and $\sigma_r$ by $c\sigma_r$. Then
\[
\hat r_c(s)=c\hat r(s),
\qquad
\hat r_c'(s)=c\hat r'(s)
\]
for every $s$. Relative sensitivities are unchanged, and for $0<\kappa<1$, $b_{\mathrm{OLS},c}=c\,b_{\mathrm{OLS}}$.
\end{corollary}

\begin{proof}
The joint scaling leaves $\kappa\bar\beta^2/(2\sigma_r^2)$ unchanged and multiplies the numerator of~\eqref{eq:response} by $c$. Hence $\hat r_c=c\hat r$ and $\hat r_c'=c\hat r'$. It also leaves the Student scale $\sigma_r/\bar\beta$ unchanged, so $b_{\mathrm{OLS},c}=c\,b_{\mathrm{OLS}}$ whenever $0<\kappa<1$.
\end{proof}

\subsection{Nonlinear full-information response}\label{app:model_nonlinear}

For clarity of exposition, our baseline model assumes the full-information response $r=\beta \cdot s$ is linear in $s$. We now consider the case where the full-information response is not linear. In particular, it might already have an $S$-like shape, as implied by other explanations proposed in the literature, such as earnings persistence. Multiplier uncertainty can then steepen that shape without being its sole source.

We denote the full-information reaction $r^*$. A first possible generalization lets the multiplier act on the value of a known non-linear response, $r=\beta r^*(s)$. If $r^*$ is an odd, strictly increasing bijection, the baseline posterior formula applies up to a change of variables, i.e., after replacing $s$ by $r^*(s)$. (Note that the associated OLS result then concerns a regression on the transformed signal $r^*(s)$, not the original surprise.) This case is therefore relatively straightforward.

We also explore the alternative which instead lets uncertainty concern the fundamental news entering the non-linear response. Let
\begin{equation}\label{eq:nonlinear_index}
z=\beta s,
\qquad
z\sim\mathcal N(0,\sigma_z^2),
\qquad
r=r^*(z),
\end{equation}
where $\beta^2\sim\mathrm{Gamma}(\kappa^{-1},1/(\kappa\bar\beta^2))$ is independent of $z$. The Gaussian assumption now applies to fundamental news, while returns inherit their distribution from $r^*$.

\begin{proposition}[Average attenuation and local amplification with non-linear responses]
\label{prop:nonlinear_steepening}
Suppose $r^*$ is odd, strictly increasing, and of at most polynomial growth. For $0<\kappa<1$,
\[
b_{\mathrm{OLS}}
=
\frac{1-\kappa}{1-\kappa/2}
\frac{\bar\beta}{A(\kappa)}
\frac{\Exp[zr^*(z)]}{\sigma_z^2},
\qquad
\frac{\partial b_{\mathrm{OLS}}}{\partial\kappa}<0.
\]

If $r^*$ is continuously differentiable with a polynomially bounded derivative and $r^{*\prime}(0)>0$, then for every $\kappa>0$,
\[
\hat r'(0)=A(\kappa)\bar\beta r^{*\prime}(0),
\qquad
\frac{\partial\hat r'(0)}{\partial\kappa}>0.
\]

If, additionally, $r^*$ is three times continuously differentiable with polynomially bounded derivatives and $r^{*\prime\prime\prime}(0)\leq0$, then as $s\to0$,
\begin{equation}\label{eq:nonlinear_local_steepening}
\frac{\hat r'(0)}{\hat r'(s)}
=
1+\frac{\bar\beta^2s^2}{2}
\left[
\frac{3\kappa}{2\sigma_z^2}
-
(1+\kappa)
\frac{r^{*\prime\prime\prime}(0)}{r^{*\prime}(0)}
\right]
+o(s^2).
\end{equation}
The second-order coefficient is strictly increasing in $\kappa$, as is the second-order coefficient of $\hat r'(0)-\hat r'(s)$.
\end{proposition}

\begin{proof}
For average sensitivity, iterated expectations and independence give
\[
\Exp[s\hat r(s)]
=
\Exp[sr]
=
\Exp[zr^*(z)]\Exp[\beta^{-1}],
\qquad
\Exp[s^2]
=
\sigma_z^2\Exp[\beta^{-2}].
\]
Hence
\[
b_{\mathrm{OLS}}
=
\frac{\Exp[zr^*(z)]}{\sigma_z^2}
\frac{\Exp[\beta^{-1}]}{\Exp[\beta^{-2}]}.
\]
The Gamma moments give
\[
\frac{\Exp[\beta^{-1}]}{\Exp[\beta^{-2}]}
=
\frac{1-\kappa}{1-\kappa/2}
\frac{\bar\beta}{A(\kappa)},
\]
so the stated expression follows. Since $\Exp[zr^*(z)]>0$ and does not depend on $\kappa$, the comparative static follows directly from Proposition~\ref{prop:uncertainty_effect}.

For the remaining results, conjugacy gives
\begin{equation}\label{eq:nonlinear_posterior}
\beta^2\mid s
\sim
\mathrm{Gamma}\left(
\kappa^{-1}+\tfrac12,
\frac{1}{\kappa\bar\beta^2}+\frac{s^2}{2\sigma_z^2}
\right).
\end{equation}
Thus, for any $q>0$,
\begin{equation}\label{eq:nonlinear_posterior_moment}
\Exp[\beta^q\mid s]
=
(\kappa\bar\beta^2)^{q/2}
\frac{\Gamma(\kappa^{-1}+\tfrac12+\tfrac q2)}
{\Gamma(\kappa^{-1}+\tfrac12)}
\left(
1+\frac{\kappa\bar\beta^2s^2}{2\sigma_z^2}
\right)^{-q/2}.
\end{equation}

The posterior distribution depends on $s$ only through $s^2$, so its first-order change at $s=0$ is zero. Differentiating $\hat r(s)=\Exp[r^*(\beta s)\mid s]$ at zero and setting $q=1$ in \eqref{eq:nonlinear_posterior_moment} therefore gives
\[
\hat r'(0)
=
r^{*\prime}(0)\Exp[\beta\mid s=0]
=
A(\kappa)\bar\beta r^{*\prime}(0).
\]
Since $A(\kappa)$ is strictly increasing, local sensitivity is strictly increasing in $\kappa$.

For local steepening, \eqref{eq:nonlinear_posterior_moment} gives
\[
\Exp[\beta\mid s]
=
A(\kappa)\bar\beta
\left[
1-\frac{\kappa\bar\beta^2}{4\sigma_z^2}s^2
\right]
+o(s^2),
\qquad
\Exp[\beta^3\mid s=0]
=
(1+\kappa)A(\kappa)\bar\beta^3.
\]
Since $r^*$ is odd and three times continuously differentiable, Taylor expansion and the polynomial bounds on its derivatives yield
\[
\begin{aligned}
\hat r(s)
&=
A(\kappa)\bar\beta r^{*\prime}(0)s 
+
A(\kappa)\bar\beta^3
\left[
\frac{1+\kappa}{6}r^{*\prime\prime\prime}(0)
-
\frac{\kappa}{4\sigma_z^2}r^{*\prime}(0)
\right]s^3
+
o(s^3).
\end{aligned}
\]
Thus
\[
\hat r'''(0)
=
A(\kappa)\bar\beta^3
\left[
(1+\kappa)r^{*\prime\prime\prime}(0)
-
\frac{3\kappa}{2\sigma_z^2}r^{*\prime}(0)
\right].
\]
Because $\hat r$ is odd,
\[
\frac{\hat r'(0)}{\hat r'(s)}
=
1-
\frac{\hat r'''(0)}{2\hat r'(0)}s^2
+
o(s^2),
\]
which gives \eqref{eq:nonlinear_local_steepening}. Under $r^{*\prime\prime\prime}(0)\leq0$, its second-order coefficient is positive and strictly increasing in $\kappa$. The second-order coefficient of $\hat r'(0)-\hat r'(s)$ is the product of this coefficient and $\hat r'(0)$, and is therefore likewise strictly increasing.
\end{proof}

Average attenuation and amplification at zero thus survive a general non-linear full-information response. The additional shape condition gives local steepening around zero, but the generality of the response function naturally does not allow us to establish global comparative statics. We can obtain explicit solutions and hence sharper characterizations in two useful special cases: if the full-information response is a fractional polynomial, or if it is already rootsign.

The posterior remains tractable for fractional-polynomial full-information responses, which form an extremely broad class and can approximate many non-linear functions in the limit.

For the following results, define
\[
\rho_\kappa(s)
\coloneqq
\frac{\bar\beta s}
{\sqrt{1+\kappa\bar\beta^2s^2/(2\sigma_z^2)}}.
\]

\begin{proposition}
\label{prop:nonlinear_fractional_polynomial}
Let
\[
r^*(z)=\sum_{j=1}^J c_j\sign(z)|z|^{p_j},
\qquad p_j>0.
\]
Then
\[
\hat r(s)
=
\sum_{j=1}^J c_j A_{p_j}(\kappa)\sign(s)
|\rho_\kappa(s)|^{p_j},
\qquad
A_p(\kappa)
\coloneqq
\kappa^{p/2}
\frac{\Gamma(\kappa^{-1}+\tfrac12+\tfrac p2)}
{\Gamma(\kappa^{-1}+\tfrac12)}.
\]
The response is bounded and admits finite limits as $s\to\pm\infty$. Moreover, $A_1=A$, and each $A_p$ strictly increases with $\kappa$ from its certainty limit of one.

For the pure power $r^*(z)=\sign(z)|z|^p$, multiplier uncertainty increases sensitivity near zero relative to sensitivity farther away, as, for $0<|s_1|<|s_2|$,
\[
\frac{\hat r'(s_1)}{\hat r'(s_2)}
=
\left|\frac{s_1}{s_2}\right|^{p-1}
\left(
\frac{1+\kappa\bar\beta^2s_2^2/(2\sigma_z^2)}
{1+\kappa\bar\beta^2s_1^2/(2\sigma_z^2)}
\right)^{(p+2)/2},
\]
which strictly increases with $\kappa$. Taking $s_1\to0$ clarifies the connection to the main result: for $0<p<1$, both the primitive and the posterior response have an infinite slope at zero; for $p>1$, both have zero slope at zero; only for the linear case $p=1$ is $\hat r'(0)$ finite and nonzero, with its ratio to $\hat r'(s)$ at any $s\neq0$ rising strictly with $\kappa$.
\end{proposition}

\begin{proof}
By \eqref{eq:nonlinear_posterior_moment} and the definition of $A_p(\kappa)$,
\[
\sign(s)|s|^p\Exp[\beta^p\mid s]
=
A_p(\kappa)\sign(s)|\rho_\kappa(s)|^p.
\]
Applying this term by term gives the stated response. Boundedness and finite limits follow from
$|\rho_\kappa(s)|\to\sqrt{2}\sigma_z/\sqrt{\kappa}$ as $|s|\to\infty$.

To establish monotonicity of $A_p$, the digamma integral gives
\[
\psi\left(\kappa^{-1}+\tfrac12+\tfrac p2\right)
-
\psi\left(\kappa^{-1}+\tfrac12\right)
=
\int_0^\infty e^{-t/\kappa}
\frac{1-e^{-pt/2}}{2\sinh(t/2)}\,dt
<
\frac{p\kappa}{2}.
\]
Therefore,
\[
\frac{\partial\log A_p(\kappa)}{\partial\kappa}
=
\frac{p}{2\kappa}
-
\frac{1}{\kappa^2}
\left[
\psi\left(\kappa^{-1}+\tfrac12+\tfrac p2\right)
-
\psi\left(\kappa^{-1}+\tfrac12\right)
\right]
>0.
\]
The Gamma-ratio asymptotic gives $A_p(\kappa)\to1$ as $\kappa\to0$, and $A_1=A$.

For a pure power,
\[
\hat r'(s)
=
p A_p(\kappa)\bar\beta^p|s|^{p-1}
\left(
1+\frac{\kappa\bar\beta^2s^2}{2\sigma_z^2}
\right)^{-(p+2)/2},
\qquad s\neq0,
\]
which yields the stated ratio. For $0<|s_1|<|s_2|$, its second factor is strictly increasing in $\kappa$, proving the comparative static. Finally,
$\hat r(s)\sim A_p(\kappa)\bar\beta^p\sign(s)|s|^p$ as $s\to0$, so $\hat r'(0)=+\infty$ for $0<p<1$, $\hat r'(0)=0$ for $p>1$, and $\hat r'(0)=A(\kappa)\bar\beta$ for $p=1$.
\end{proof}

In particular, the fractional-polynomial case includes $r^*(z)=\sign(z)\sqrt{|z|}$. Here the full-information response already has an infinite slope at zero, so that amplification is best described by the leading coefficient:
\[
\lim_{s\to0}
\frac{\hat r(s)}{r^*(\bar\beta s)}
=
A_{1/2}(\kappa)>1.
\]
Greater uncertainty raises this coefficient and increases sensitivity to smaller nonzero surprises relative to larger ones. The population OLS coefficient still falls for $0<\kappa<1$, by the first part of Proposition~\ref{prop:nonlinear_steepening}.

More broadly, this proposition and its proof show that signed power functions are eigenfunctions of the model's multiplicative averaging operator, with eigenvalues $A_p(\kappa)$. This reflects the underlying Mellin-transform structure of the problem: random rescaling acts separately on each power component, changing only its coefficient.

Although it is extremely broad, the fractional-polynomial case does not include full-information responses that reach a finite limit at $\pm \infty$. We show that if the full-information response is already a rootsign function, which saturates, the reaction under multiplier uncertainty can still be solved for, and that increasing multiplier uncertainty has the usual effects on its shape.

\begin{proposition}
\label{prop:rootsign_in_rootsign}
Let
\[
r^*(z)=\frac{z}{\sqrt{1+z^2/\tau^2}},
\qquad \tau>0.
\]
For $s\neq0$, define $z_\kappa(s)=\tau^2/[\kappa\rho_\kappa(s)^2]$. Then
\[
\hat r(s)
=
\tau\sign(s)\,
\frac{A(\kappa)}{\sqrt{\kappa}}\,
z_\kappa(s)^{\kappa^{-1}+\tfrac12}
U\left(
\kappa^{-1}+1,\kappa^{-1}+\tfrac32,z_\kappa(s)
\right),
\]
where $U$ denotes Tricomi's confluent hypergeometric function. With $\hat r(0)=0$, the response is odd, $C^\infty$, strictly increasing, concave on $(0,\infty)$, and bounded by $\tau$ in absolute value.

Average sensitivity falls for $0<\kappa<1$, while $\hat r'(0)=A(\kappa)\bar\beta$ rises for every $\kappa>0$. Near zero,
\[
\frac{\hat r'(0)}{\hat r'(s)}
=
1+\frac32\bar\beta^2
\left(
\frac{\kappa}{2\sigma_z^2}
+
\frac{1+\kappa}{\tau^2}
\right)s^2+o(s^2).
\]
Thus the relative and absolute slope gaps have increasing second-order coefficients.
\end{proposition}

\begin{proof}
Let
\[
B_\kappa(s)
\coloneqq
\frac{1}{\kappa\bar\beta^2}
+
\frac{s^2}{2\sigma_z^2},
\qquad
Y\coloneqq B_\kappa(s)\beta^2.
\]
By \eqref{eq:nonlinear_posterior},
\[
Y\mid s
\sim
\mathrm{Gamma}\left(
\kappa^{-1}+\tfrac12,1
\right).
\]
Moreover,
\[
\beta s
=
\sqrt{\kappa Y}\,\rho_\kappa(s).
\]
Hence, for $s\neq0$,
\[
\hat r(s)
=
\tau\sign(s)
\Exp\left[
\left(
\frac{Y}{Y+z_\kappa(s)}
\right)^{1/2}
\right].
\]
For $z>0$, the integral representation of $U$ gives
\[
\begin{aligned}
\Exp\left[\left(\frac{Y}{Y+z}\right)^{1/2}\right]
&=
\frac{1}{\Gamma(\kappa^{-1}+\tfrac12)}
\int_0^\infty
y^{\kappa^{-1}}(y+z)^{-1/2}e^{-y}\,dy \\
&=
z^{\kappa^{-1}+\tfrac12}
\frac{\Gamma(\kappa^{-1}+1)}
{\Gamma(\kappa^{-1}+\tfrac12)}
U\left(
\kappa^{-1}+1,
\kappa^{-1}+\tfrac32,
z
\right).
\end{aligned}
\]
Substituting $z=z_\kappa(s)$ and using $\Gamma(\kappa^{-1}+1)/\Gamma(\kappa^{-1}+\tfrac12) =A(\kappa)/\sqrt{\kappa}$ proves the stated formula.

The same representation gives the shape properties. For each $Y>0$, $r^*(\sqrt{\kappa Y}\rho_\kappa(s))$ is odd and strictly increasing, and is concave on $(0,\infty)$ because both $\rho_\kappa$ and $r^*$ are increasing and concave there. Taking expectations preserves these properties. Smoothness follows by dominated differentiation, since $Y$ has moments of all orders, and boundedness follows from $|r^*(z)|<\tau$.

Finally, $r^{*\prime}(0)=1$ and $r^{*\prime\prime\prime}(0)=-3/\tau^2$. Proposition~\ref{prop:nonlinear_steepening} therefore gives the statements about average sensitivity, sensitivity at zero, and the local relative and absolute slope gaps.
\end{proof}

\begin{figure}[!htb]
\caption{Multiplier Uncertainty with a Nonlinear Response}
    \centering
    \includegraphics[width=\linewidth,height=0.855\textheight]{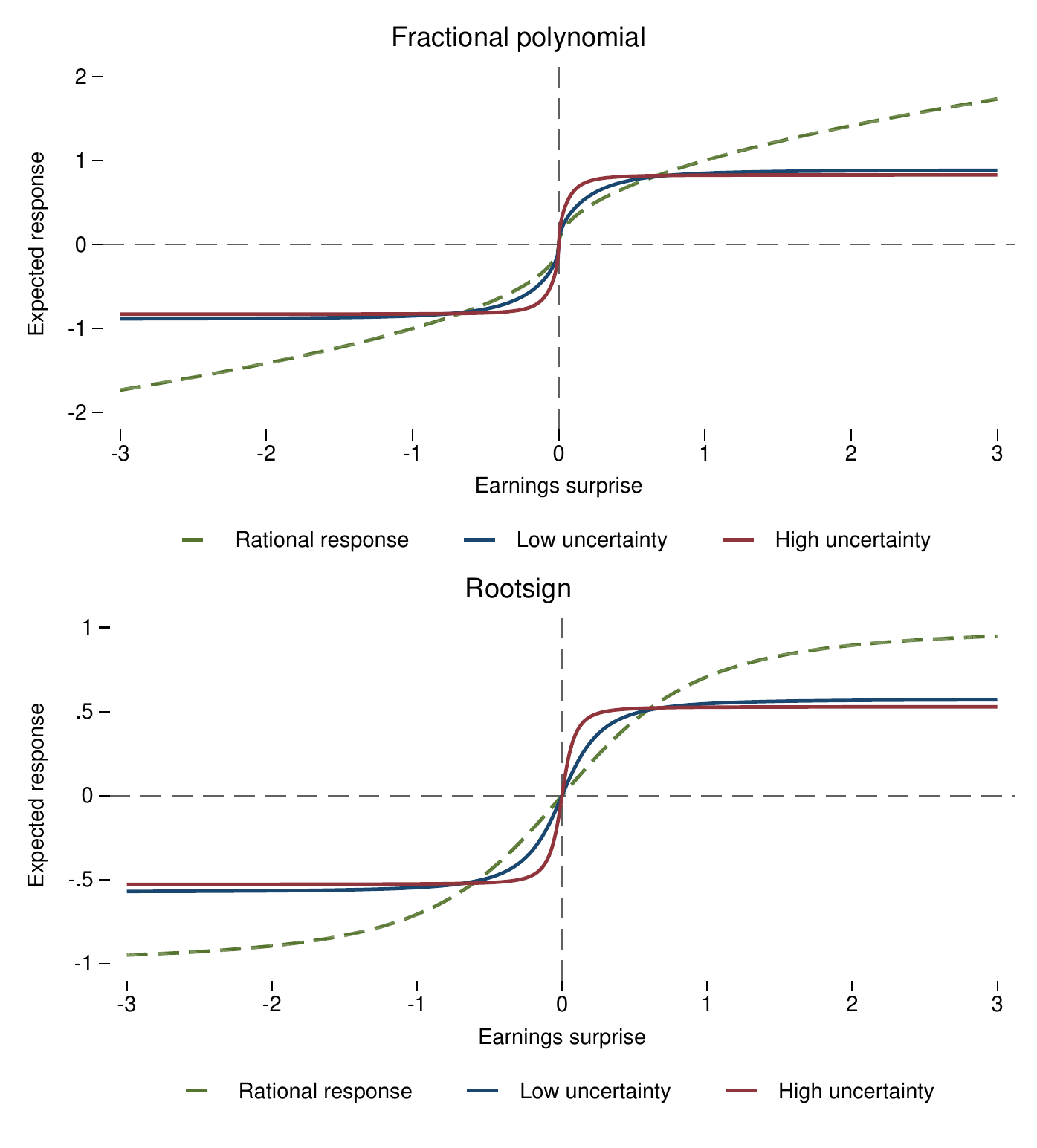}
\caption*{\footnotesize \textit{Notes:} The top panel plots $\hat r(s)$ for the full-information response $r^*(z)=\sign(z)\sqrt{|z|}$ in Proposition~\ref{prop:nonlinear_fractional_polynomial}. The bottom panel uses $r^*(z)=z/\sqrt{1+z^2}$ in Proposition~\ref{prop:rootsign_in_rootsign}, with $\tau=1$. We set $\bar\beta=\sigma_z=1$. Low and high multiplier uncertainty correspond to $\kappa=10$ and $\kappa=100$.}
    \label{fig:model_nonlinear_rational}
\end{figure}

Appendix Figure~\ref{fig:model_nonlinear_rational} plots the posterior responses when $r^*(z)=\sign(z)\sqrt{|z|}$ and when $r^*(z)=z/\sqrt{1+z^2}$, as in Propositions~\ref{prop:nonlinear_fractional_polynomial} and~\ref{prop:rootsign_in_rootsign}. In both examples, the full-information response is already non-linear but multiplier uncertainty nonetheless amplifies the reaction to small signals. The mechanism is therefore a complement to other forces behind the $S$-shape.

\subsection{Asymmetric reactions under multiplier uncertainty}\label{app:model_asymmetric}

Because firms have incentives to report positive earnings surprises, one may expect asymmetric market reactions to earnings news. Our model accommodates such asymmetries by letting the multiplier differ between good and bad news, with a separate prior on each branch. The DM may thus assess the meaning of positive and negative surprises differently. This yields a posterior that has the same closed form as before, but with different coefficients on each side of zero, and the same comparative statics, now also separately on each side of zero.

\begin{proposition}
\label{prop:rootsign_asymmetry}
Retain $r\sim\mathcal N(0,\sigma_r^2)$ and let
\[
r=\beta_+s\quad\text{if }s\geq0,
\qquad
r=\beta_-s\quad\text{if }s<0,
\]
where the multipliers are independent of $r$ and of each other, with
\[
\beta_\pm^2
\sim
\mathrm{Gamma}\left(
\frac{1}{\kappa_\pm},
\frac{1}{\kappa_\pm\bar\beta_\pm^2}
\right).
\]
On each side of zero,
\[
\hat r(s)
=
A(\kappa_\pm)\bar\beta_\pm
\frac{s}
{\sqrt{1+\tfrac{\kappa_\pm\bar\beta_\pm^2s^2}{2\sigma_r^2}}},
\qquad
\hat r'(0^\pm)=A(\kappa_\pm)\bar\beta_\pm.
\]
For $0<\kappa_\pm<1$, the population OLS slope within the corresponding sign subsample, allowing for an intercept, is
\[
b_{\mathrm{OLS}}^\pm
=
\frac{(1-2/\pi)\Exp[\beta_\pm^{-1}]}
{\Exp[\beta_\pm^{-2}]-(2/\pi)\Exp[\beta_\pm^{-1}]^2}
<
\bar\beta_\pm
<
\hat r'(0^\pm).
\]
Increasing $\kappa_\pm$ lowers $b_{\mathrm{OLS}}^\pm$ and raises $\hat r'(0^\pm)$. For every nonzero $s$ on each side of zero,
\[
\frac{\hat r'(0^\pm)}{\hat r'(s)}
=
\left(
1+\frac{\kappa_\pm\bar\beta_\pm^2s^2}{2\sigma_r^2}
\right)^{3/2},
\]
which strictly increases with $\kappa_\pm$.
\end{proposition}

\begin{proof}
The sign of $s$ reveals the relevant branch. The half-normal likelihood differs from the baseline likelihood only by a factor independent of the multiplier, so conjugacy and the posterior formula apply separately on each side.

For one branch, write $m_j=\Exp[\beta_\pm^{-j}]$. The half-normal moments give
\[
\operatorname{Cov}(s,r\mid\pm)
=
\sigma_r^2(1-2/\pi)m_1,
\qquad
\Var(s\mid\pm)
=
\sigma_r^2\left(m_2-\frac{2}{\pi}m_1^2\right).
\]
Their ratio is the stated coefficient. The ratio $m_1/m_2$ strictly decreases with $\kappa_\pm$ by Proposition~\ref{prop:uncertainty_effect}, and
\[
\frac{m_1^2}{m_2}
=
\frac{1-\kappa_\pm}
{(1-\kappa_\pm/2)^2A(\kappa_\pm)^2}
\]
also strictly decreases. Hence
\[
b_{\mathrm{OLS}}^\pm
=
\frac{(1-2/\pi)(m_1/m_2)}
{1-(2/\pi)(m_1^2/m_2)}
\]
strictly decreases from its certainty limit $\bar\beta_\pm$. Local comparisons follow from this posterior.
\end{proof}

Hence higher multiplier uncertainty drives overall attenuation, amplification of small surprises, and steepening relative to large surprises separately for beats and misses, even when the two response functions differ.

\begin{figure}[t]
\caption{Asymmetric Responses under Multiplier Uncertainty}
    \centering
    \includegraphics[width=\linewidth]{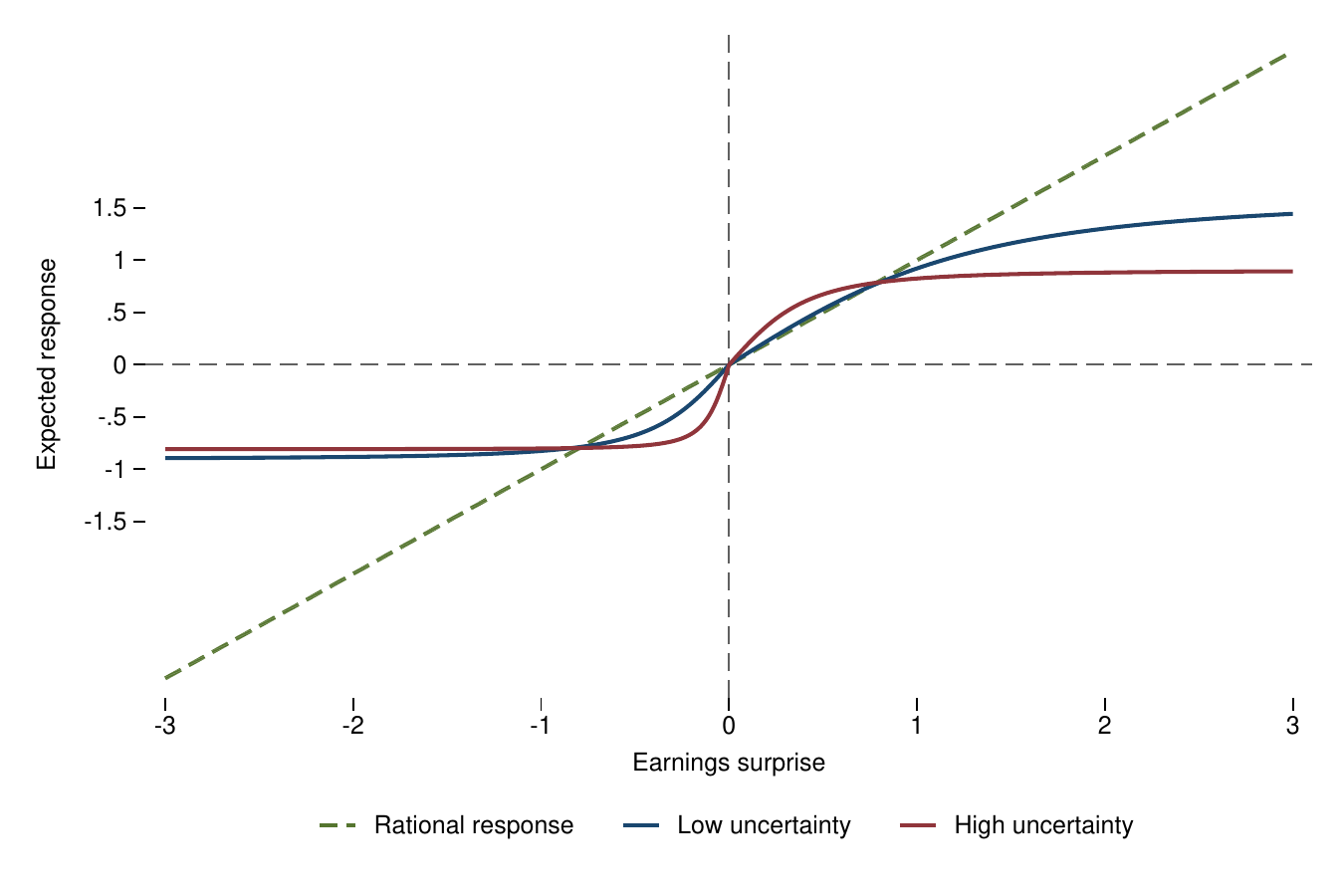}
\caption*{\footnotesize \textit{Notes:} This figure plots the asymmetric response in Proposition~\ref{prop:rootsign_asymmetry}. We set $\sigma_r=1$ and $\bar\beta_+=\bar\beta_-=1$. Low uncertainty uses $\kappa_+=1$ and $\kappa_-=10$, while high uncertainty uses $\kappa_+=10$ and $\kappa_-=100$. Multiplier uncertainty is therefore ten times higher for negative surprises.}
    \label{fig:model_asymmetric_reaction}
\end{figure}

\subsection{Uncertainty about return scale}\label{app:model_return_uncertainty}

In the baseline, the DM knows the return distribution but is uncertain about the multiplier. We now allow uncertainty about the return scale as well. The likelihood of the surprise depends on the multiplier and return scale only through the surprise precision
\[
\Lambda\coloneqq\frac{\beta^2}{\sigma_r^2}.
\]
We hence place the Gamma prior directly on $\Lambda$, independently of the return scale. Thus, $\kappa$ now measures dispersion in the mapping-to-scale ratio, which is the relevant multiplier in the updating problem.

\begin{proposition}
\label{prop:unknown_sigma_rootsign}
Suppose
\[
\Lambda
\sim
\mathrm{Gamma}\left(
\frac{1}{\kappa},
\frac{1}{\kappa\bar\Lambda^2}
\right),
\qquad
s\mid \Lambda,\sigma_r \sim\mathcal N(0,\Lambda^{-1}),
\qquad
r=\sigma_r\sqrt{\Lambda}\,s,
\]
where $\sigma_r>0$ is independent of $\Lambda$ and $\Exp[\sigma_r^2]<\infty$. Then
\[
\hat r(s)
=
\Exp[\sigma_r]\,A(\kappa)\bar\Lambda
\frac{s}{\sqrt{1+\tfrac{\kappa\bar\Lambda^2s^2}{2}}}.
\]
For $0<\kappa<1$,
\[
b_{\mathrm{OLS}}
=
\Exp[\sigma_r]\,
\frac{1-\kappa}{1-\kappa/2}
\frac{\bar\Lambda}{A(\kappa)},
\]
while for every $\kappa>0$,
\[
\hat r'(0)=\Exp[\sigma_r]A(\kappa)\bar\Lambda,
\qquad
\frac{\hat r'(0)}{\hat r'(s)}
=
\left(1+\frac{\kappa\bar\Lambda^2s^2}{2}\right)^{3/2}.
\]
Holding $\bar\Lambda$ and the return-scale distribution fixed, greater $\kappa$ lowers average sensitivity, raises sensitivity at zero, and raises the relative sensitivity to small surprises. Its effect on the absolute sensitivity gap and the single-crossing property are also unchanged.
\end{proposition}

\begin{proof}
The likelihood does not depend on $\sigma_r$, so its posterior equals its prior and remains independent of $\Lambda\mid s$. Conjugacy gives
\[
\Lambda\mid s
\sim
\mathrm{Gamma}\left(
\kappa^{-1}+\tfrac12,
\frac{1}{\kappa\bar\Lambda^2}+\frac{s^2}{2}
\right).
\]
Consequently,
\[
\hat r(s)
=
s\Exp[\sigma_r]\Exp[\sqrt{\Lambda}\mid s],
\]
which yields the posterior formula. Moreover,
\[
\Exp[sr]
=
\Exp[\sigma_r]\Exp[\Lambda^{-1/2}],
\qquad
\Exp[s^2]=\Exp[\Lambda^{-1}].
\]
The Gamma moment formula gives $b_{\mathrm{OLS}}$. The remaining conclusions follow from Proposition~\ref{prop:uncertainty_effect} and Corollaries~\ref{cor:slope_difference} and~\ref{corollary:slope_single_crossing}, setting $\bar\beta=\bar\Lambda$ and $\sigma_r=1$ and multiplying by $\Exp[\sigma_r]$.
\end{proof}

This extension, while allowing us to generalize the hypotheses of the model, is therefore also useful to understand its mechanics. The return-scale distribution affects the posterior mean only through $\Exp[\sigma_r]$: changes in its dispersion that hold this mean fixed do not alter the response, while changes in its mean scale the response uniformly up or down. This is precisely the uniform-attenuation benchmark we contrast our model with in the main text. The analogy identified in Corollary~\ref{cor:uniform_attenuation} hence generalizes to settings where $\sigma_r$ is potentially random and hence unknown to the decision maker. Conversely, it is uncertainty about the multiplier relative to the scale of returns that drives the non-linearity of the response function, and which potentially steepens the reaction to small surprises.

\FloatBarrier
\clearpage
\setcounter{figure}{0}\setcounter{table}{0}
\section{Additional Experimental Evidence}\label{app:exper}

\subsection{Additional Response Estimates}\label{app:exper_additional}

\begin{figure}[!htbp]
\centering
\caption{Processing Constraints over Expanding SUE Windows}
\label{fig:experiment_expanding_windows}
\includegraphics[width=\linewidth]{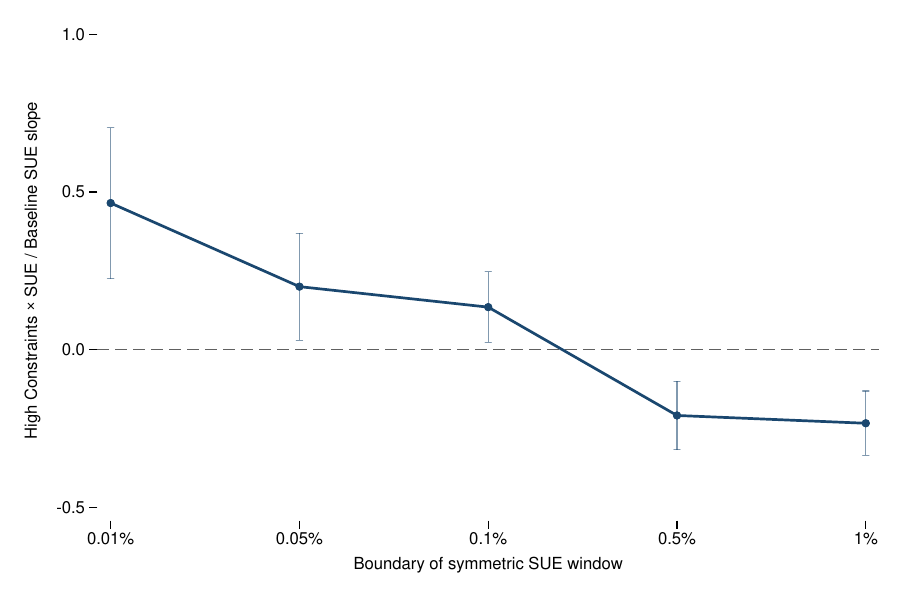}
\caption*{\footnotesize \textit{Notes:} $\hat\delta/\hat\beta$ from Appendix Table~\ref{tab:experiment_windows}, where $\delta$ is the treatment interaction and $\beta$ the baseline SUE slope. Whiskers are pointwise 95\% confidence intervals from delta-method standard errors clustered by respondent.}
\end{figure}

\begin{figure}[!htb]
\centering
\caption{Processing Constraints and the Shape of the Experimental Response}
\label{fig:experiment_response_fits}
\includegraphics[height=0.9\textheight]{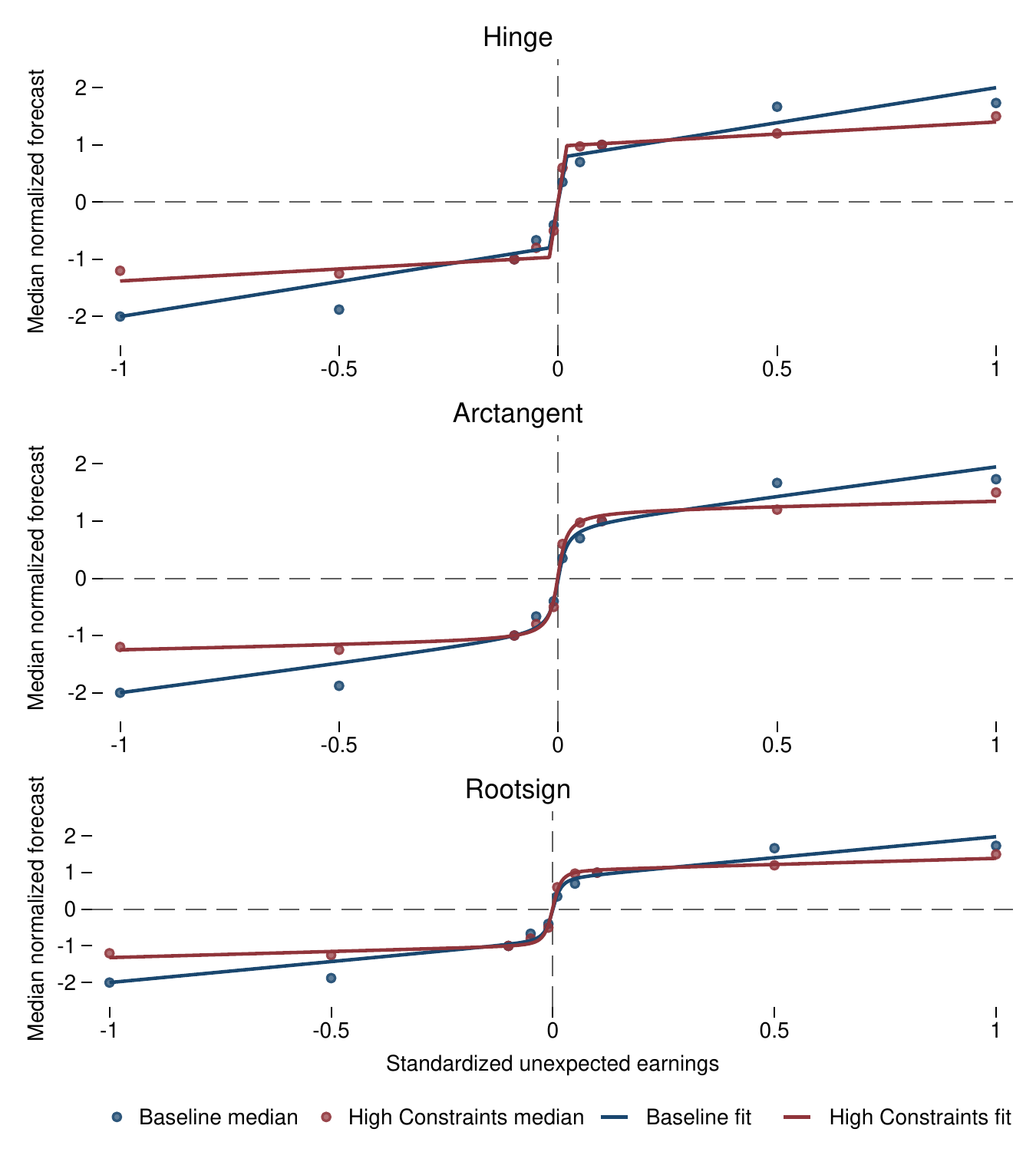}
\caption*{\footnotesize \textit{Notes:} Points are treatment-cell medians of normalized forecasts and are identical across panels. Lines are the fitted conditional medians corresponding to Table~\ref{tab:experiment_response_fits} (hinge) and Appendix Table~\ref{tab:experiment_response_fits_smooth} (arctangent and rootsign), using the response family indicated in each panel.}
\end{figure}

Appendix Table~\ref{tab:experiment_windows} and Appendix Figure~\ref{fig:experiment_expanding_windows} report the preregistered expanding-window estimates. Appendix Table~\ref{tab:experiment_response_fits_smooth} and Appendix Figure~\ref{fig:experiment_response_fits} report the smooth-response specifications described in Section~\ref{sec:experiment_robustness}. Finally, we explain in detail the selection of the response scale and its robustness in Appendix~\ref{app:scale_selection}.

\begin{table}[!htbp]
\centering
\caption{Processing Constraints over Expanding SUE Windows}
\label{tab:experiment_windows}
\fitpage{\begin{tabular}{l*{5}{c}}
\toprule
& \multicolumn{5}{c}{\textit{Dependent variable: normalized respondent return forecast}} \\
 & (1) & (2) & (3) & (4) & (5) \\
 & $|\text{SUE}|\leq 0.01\%$ & $|\text{SUE}|\leq 0.05\%$ & $|\text{SUE}|\leq 0.1\%$ & $|\text{SUE}|\leq 0.5\%$ & $|\text{SUE}|\leq 1\%$ \\
\midrule

SUE             &    37.54\sym{***}&    15.28\sym{***}&    10.28\sym{***}&     4.00\sym{***}&     2.37\sym{***}\\
                &   (2.24)         &   (0.83)         &   (0.35)         &   (0.16)         &   (0.11)         \\
SUE $\times$ HC &    17.46\sym{***}&     3.06\sym{**} &     1.39\sym{**} &    -0.83\sym{***}&    -0.55\sym{***}\\
                &   (3.90)         &   (1.20)         &   (0.56)         &   (0.24)         &   (0.14)         \\
HC              &     0.07\sym{**} &     0.07\sym{*}  &     0.10\sym{***}&     0.08         &     0.12\sym{**} \\
                &   (0.04)         &   (0.04)         &   (0.03)         &   (0.05)         &   (0.05)         \\
\midrule
Observations    &      899         &    1,812         &    2,675         &    3,554         &    4,480         \\
Respondents     &      607         &      831         &      887         &      896         &      896         \\
Pseudo \(R^{2}\)&     0.24         &     0.26         &     0.28         &     0.22         &     0.22         \\
Regression      &   Median         &   Median         &   Median         &   Median         &   Median         \\
\bottomrule
\end{tabular}
}
\caption*{\footnotesize \textit{Notes:} Conditional median estimates of the expanding-window specification~\eqref{eq:hc-linear}. Each column uses the symmetric SUE window indicated. Standard errors are clustered by respondent.}
\end{table}

\FloatBarrier

\begin{table}[!htbp]
\centering
\caption{Processing Constraints with Smooth Functional Forms}
\label{tab:experiment_response_fits_smooth}
\fittable{\begin{tabular}{l*{4}{c}}
\toprule
& \multicolumn{4}{c}{\textit{Dependent variable: respondent return forecast}} \\
& \multicolumn{2}{c}{Normalized forecasts} & \multicolumn{2}{c}{Raw forecasts} \\
\cmidrule(lr){2-3}\cmidrule(lr){4-5}
& Arctangent & Rootsign & Arctangent & Rootsign \\
& (1) & (2) & (3) & (4) \\
\midrule
SUE                 &        1.02\sym{***}&        1.14\sym{***}&        5.89\sym{***}&        6.38\sym{***}\\
                    &      (0.07)         &      (0.07)         &      (0.35)         &      (0.36)         \\
$\Delta$ Small SUE  &       40.88\sym{***}&       42.33\sym{***}&      215.02\sym{***}&      203.50\sym{***}\\
                    &      (1.38)         &      (1.53)         &      (7.64)         &      (7.25)         \\
High Constraints    &        0.07\sym{***}&        0.04\sym{*}  &        0.42\sym{***}&        0.42\sym{***}\\
                    &      (0.03)         &      (0.03)         &      (0.13)         &      (0.13)         \\
High Constraints $\times$ SUE&       -0.85\sym{***}&       -0.81\sym{***}&       -4.75\sym{***}&       -4.52\sym{***}\\
                    &      (0.14)         &      (0.12)         &      (0.53)         &      (0.47)         \\
High Constraints $\times$ $\Delta$ Small SUE&        7.46\sym{***}&        8.74\sym{***}&       59.75\sym{***}&       56.04\sym{***}\\
                    &      (1.92)         &      (2.08)         &     (15.39)         &     (14.78)         \\
\midrule
Small SUE           &41.90\sym{***}         &43.48\sym{***}         &220.91\sym{***}         &209.88\sym{***}         \\
                    &      (1.35)         &      (1.50)         &      (7.51)         &      (7.12)         \\
High Constraints $\times$ Small SUE&6.61\sym{***}         &7.93\sym{***}         &55.00\sym{***}         &51.51\sym{***}         \\
                    &      (1.86)         &      (2.04)         &     (15.12)         &     (14.55)         \\
\midrule
Scale \(k\)&       0.015         &       0.020         &       0.010         &       0.015         \\
Observations        &       4,480         &       4,480         &       4,480         &       4,480         \\
Respondents         &         896         &         896         &         896         &         896         \\
Pseudo \(R^{2}\)    &        0.35         &        0.35         &        0.42         &        0.42         \\
Regression          &      Median         &      Median         &      Median         &      Median         \\
\bottomrule
\end{tabular}
}
\caption*{\footnotesize \textit{Notes:} As the hinge columns of Table~\ref{tab:experiment_response_fits}, replacing the hinge in Equation~\eqref{eq:hc-interaction} by the arctangent term $h_A(\SUE)$ or rootsign term $h_R(\SUE)$. $\Delta$ Small SUE is the bounded term, whose coefficient is again the additional slope at zero. For each family and outcome, the scale $k$ is selected once from the pooled forecasts before the treatment indicator enters and then held fixed. Standard errors are clustered by respondent.}
\end{table}

\FloatBarrier
\clearpage

\subsection{Robustness}\label{app:exper_robustness}

\begin{figure}[!htbp]
\centering
\caption{Experimental Response, Excluding Capped Forecasts}
\label{fig:experiment_censoring_uncensored}
\includegraphics[width=.8\linewidth]{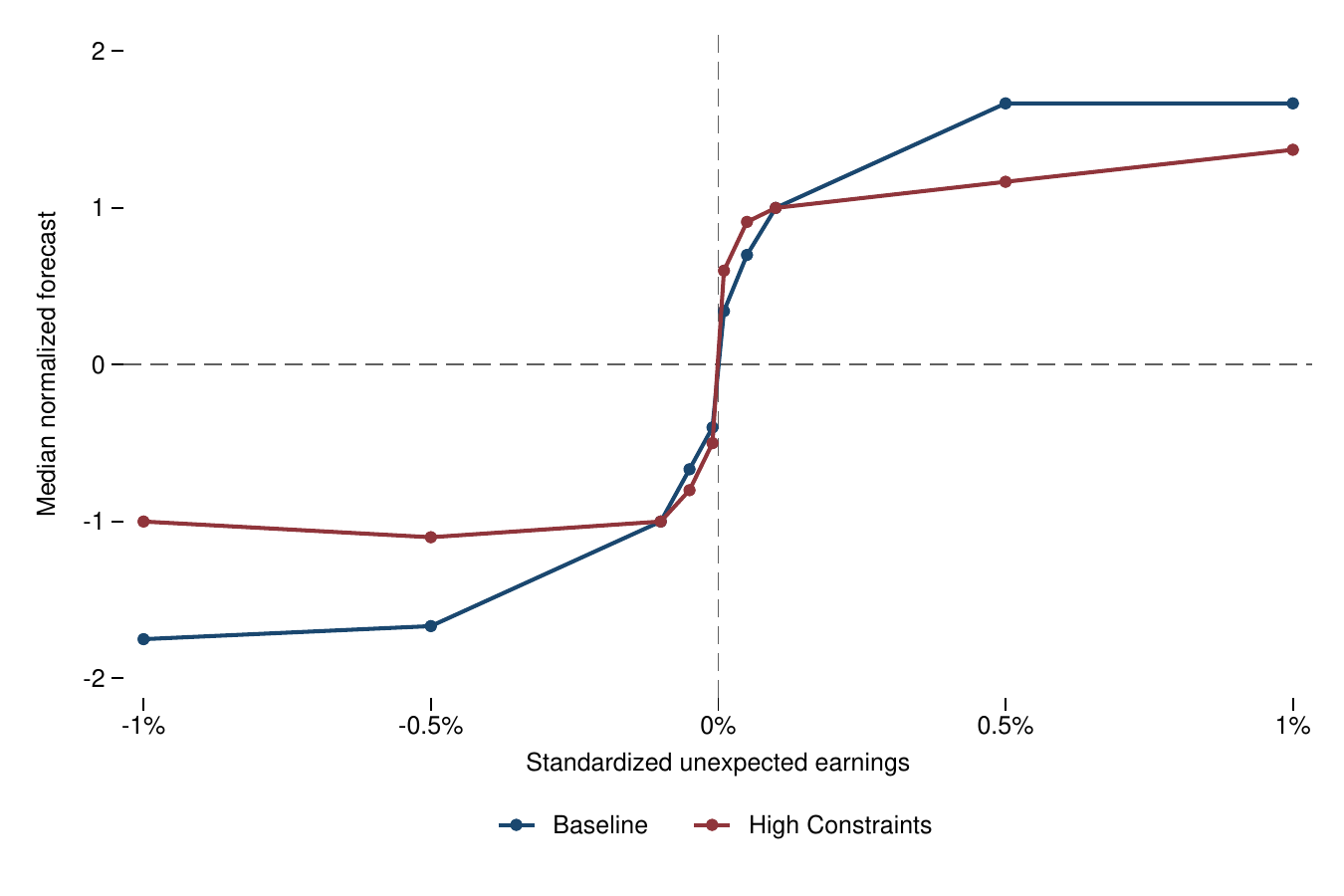}
\caption*{\footnotesize \textit{Notes:} As Figure~\ref{fig:experiment_median_forecasts}, dropping forecasts that reach either end of the $\pm15\%$ entry range.}
\par\medskip
\caption{Experimental Response for Never-Capped Respondents}
\label{fig:experiment_censoring_never}
\includegraphics[width=.8\linewidth]{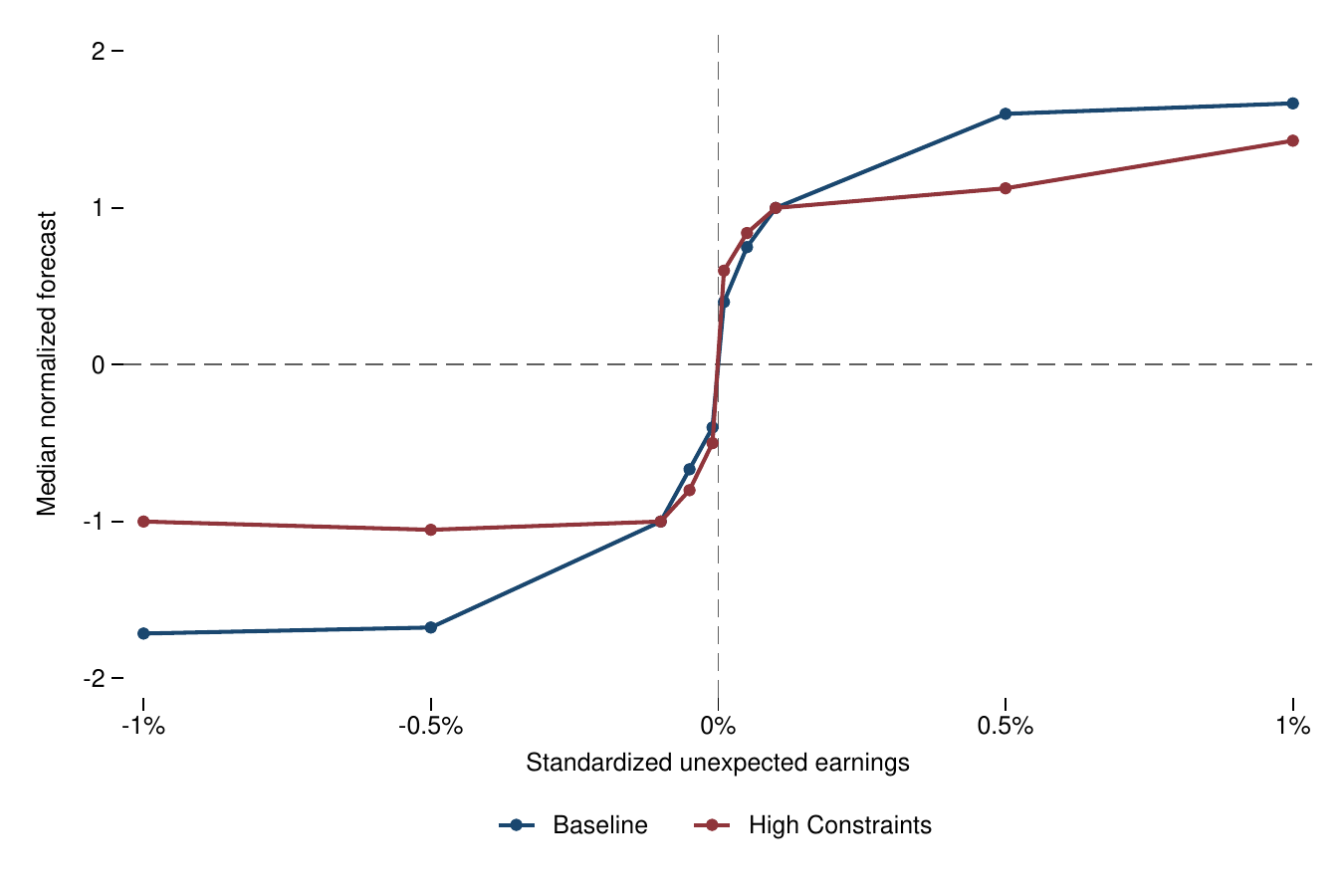}
\caption*{\footnotesize \textit{Notes:} As Figure~\ref{fig:experiment_median_forecasts}, dropping respondents who reach either entry bound in any round.}
\end{figure}

We report expanding-window interactions under alternative outcomes and sample restrictions in Appendix Figure~\ref{fig:experiment_excl_inconsistent}. Appendix Table~\ref{tab:experiment_windows_winsorized} gives the corresponding conditional-mean estimates. In turn, Appendix Figure~\ref{fig:experiment_robustness_appendix} shows the response shape after excluding sign-inconsistent forecasts. Finally, Appendix Table~\ref{tab:experiment_censoring} and the corresponding Appendix Figures~\ref{fig:experiment_censoring_uncensored}--\ref{fig:experiment_censoring_never} assess the raw forecast-entry bounds.

\begin{figure}[!htbp]
\caption{Experimental Expanding-Window Robustness}
\label{fig:experiment_excl_inconsistent}
    \centering
       \begin{subfigure}[b]{0.48\textwidth}
        \includegraphics[width=\textwidth]{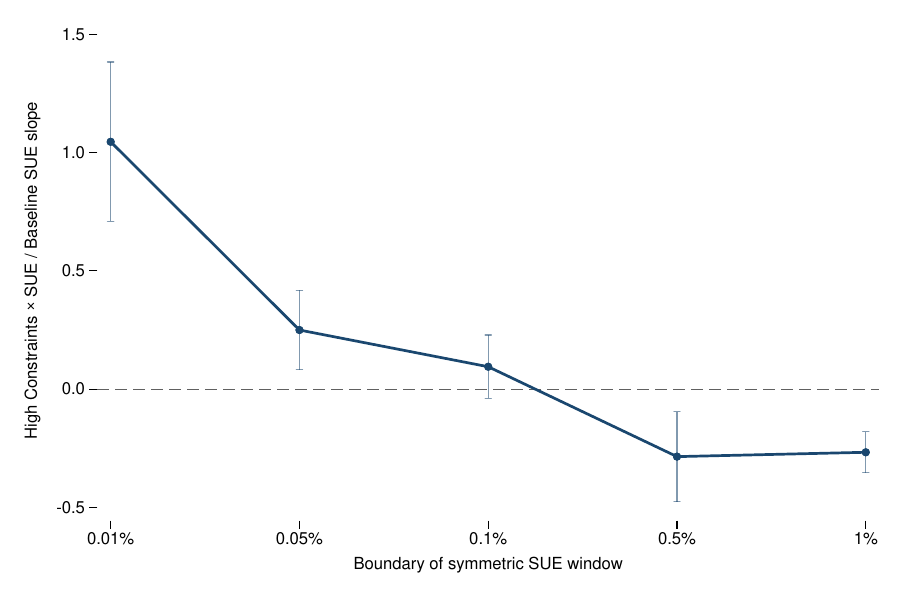}
        \caption{Raw forecasts}
        \label{fig:robust_check4}
    \end{subfigure}
\hfill
        \begin{subfigure}[b]{0.48\textwidth}
        \includegraphics[width=\textwidth]{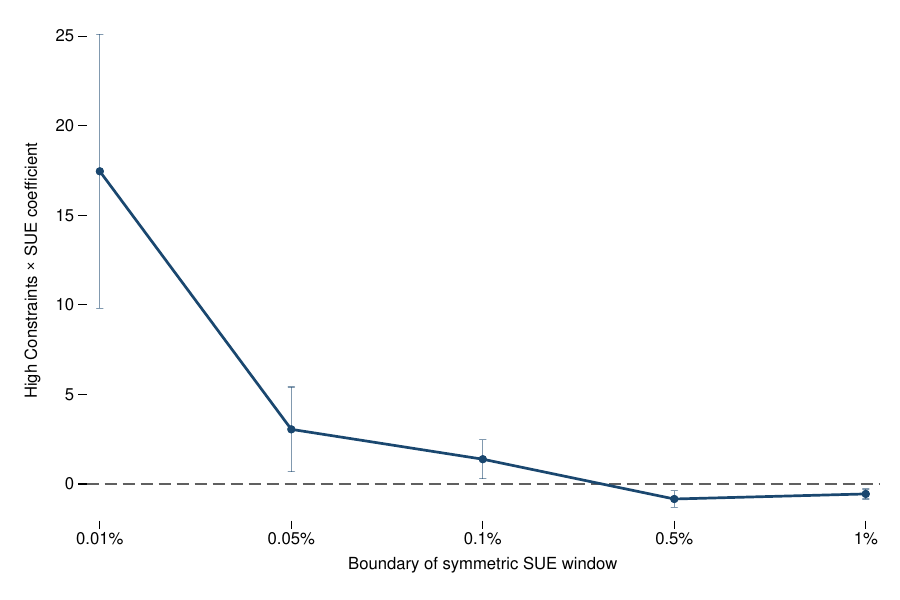}
        \caption{Unscaled interactions}
        \label{fig:robust_check5}
    \end{subfigure}
    \vspace{0.25cm}
    \begin{subfigure}[b]{0.48\textwidth}
        \includegraphics[width=\textwidth]{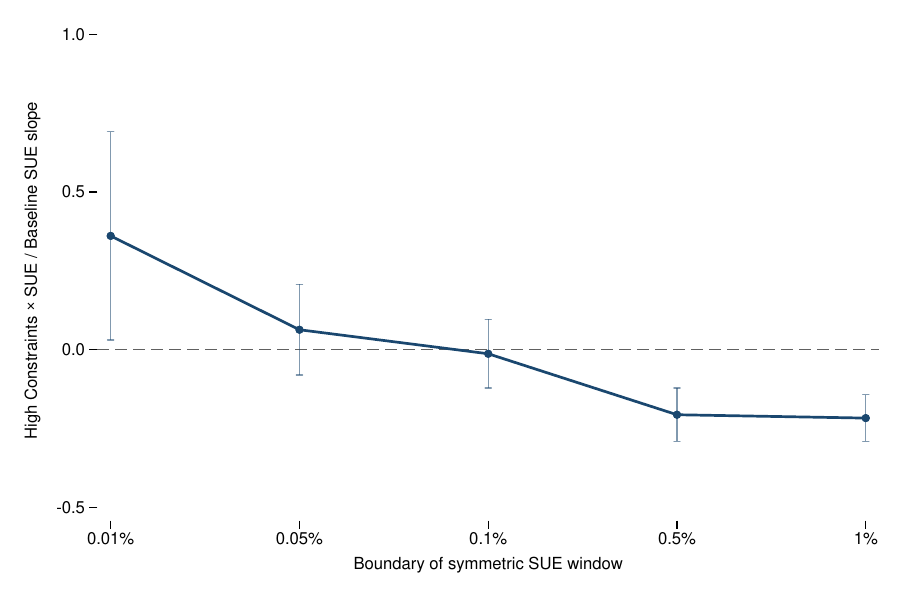}
        \caption{Conditional means}
        \label{fig:robust_check6}
    \end{subfigure}
\hfill
    \begin{subfigure}[b]{0.48\textwidth}
        \includegraphics[width=\textwidth]{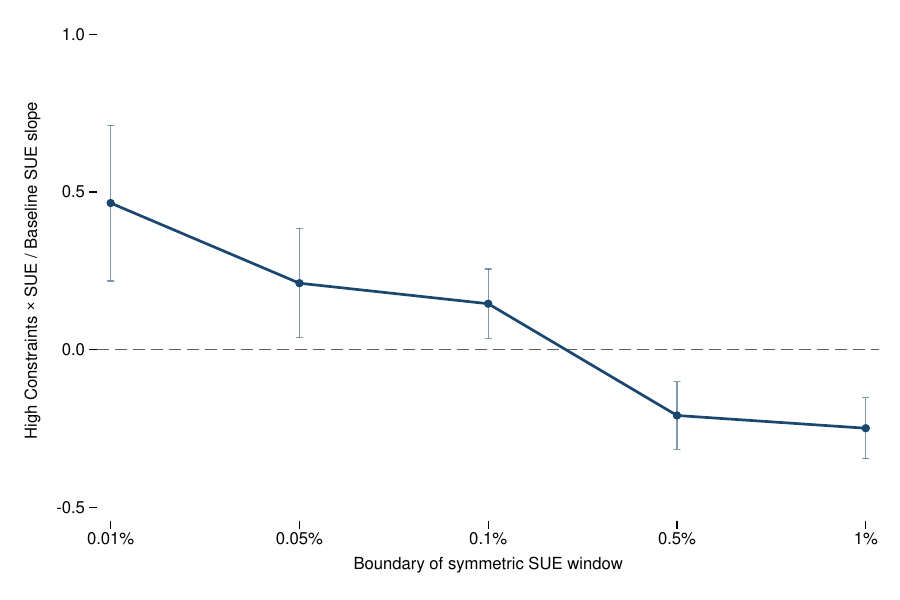}
        \caption{Excluding timeouts}
        \label{fig:robust_check2}
    \end{subfigure}
    \vspace{0.25cm}
    \begin{subfigure}[b]{0.48\textwidth}
        \includegraphics[width=\textwidth]{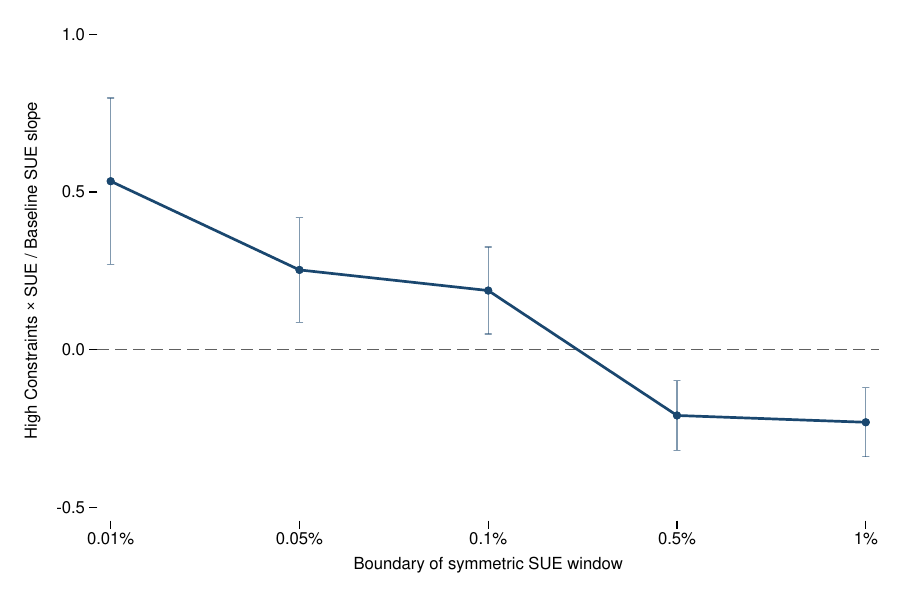}
        \caption{Excluding online searches}
        \label{fig:robust_check3}
    \end{subfigure}
\hfill
    \begin{subfigure}[b]{0.48\textwidth}
        \includegraphics[width=\textwidth]{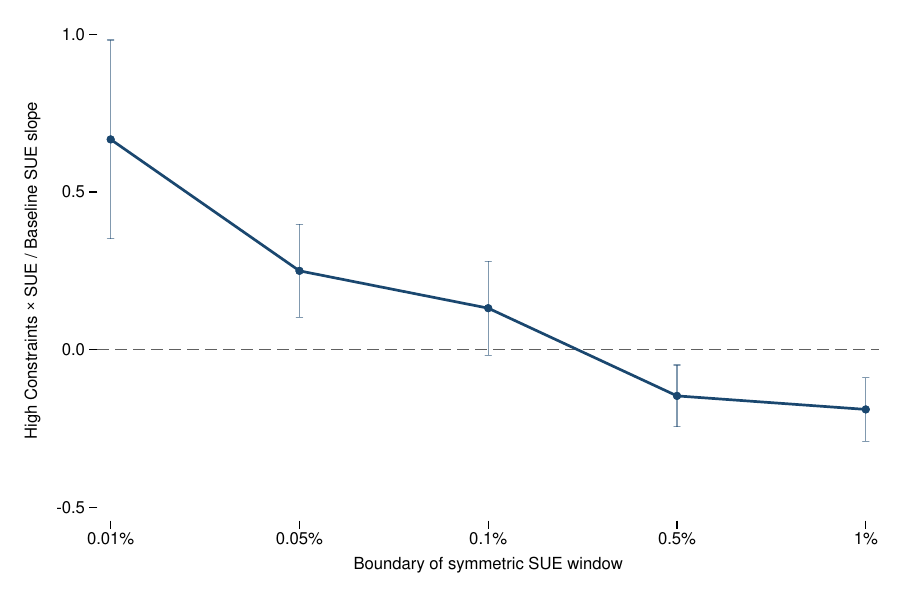}
        \caption{Excluding sign-inconsistent forecasts}
        \label{fig:robust_check1}
    \end{subfigure}
\caption*{\footnotesize \textit{Notes:} Panels modify Appendix Figure~\ref{fig:experiment_expanding_windows}. Panel (a) uses raw forecasts. Panel (b) plots $\hat\delta$ rather than $\hat\delta/\hat\beta$. Panel (c) uses OLS with normalized forecasts capped at $\pm3$. Panels (d) and (f) drop individual responses that timed out or had the wrong sign. Panel (e) drops all responses from respondents reporting an online search. A timeout is a response exceeding the 40-second bonus window. Whiskers are pointwise 95\% intervals, using respondent-clustered standard errors for $\hat\delta$ and delta-method standard errors for $\hat\delta/\hat\beta$.}
\end{figure}

\begin{table}[!htbp]
\caption{Expanding SUE Windows Using Conditional Means}
\label{tab:experiment_windows_winsorized}
\fittable{
\begin{tabular}{l*{5}{c}}
\toprule
& \multicolumn{5}{c}{\textit{Dependent variable: normalized respondent return forecast}} \\
 & (1) & (2) & (3) & (4) & (5) \\
 & $|\text{SUE}|\leq 0.01\%$ & $|\text{SUE}|\leq 0.05\%$ & $|\text{SUE}|\leq 0.1\%$ & $|\text{SUE}|\leq 0.5\%$ & $|\text{SUE}|\leq 1\%$ \\
\midrule

SUE             &    44.80\sym{***}&    17.37\sym{***}&    12.49\sym{***}&     3.83\sym{***}&     2.19\sym{***}\\
                &   (3.77)         &   (0.84)         &   (0.49)         &   (0.12)         &   (0.06)         \\
SUE $\times$ HC &    16.16\sym{**} &     1.10         &    -0.16         &    -0.79\sym{***}&    -0.47\sym{***}\\
                &   (6.69)         &   (1.23)         &   (0.70)         &   (0.18)         &   (0.09)         \\
HC              &     0.13\sym{**} &     0.11\sym{**} &     0.12\sym{***}&     0.06         &     0.09\sym{**} \\
                &   (0.06)         &   (0.05)         &   (0.04)         &   (0.04)         &   (0.04)         \\
\midrule
Observations    &      899         &    1,812         &    2,675         &    3,554         &    4,480         \\
Respondents     &      607         &      831         &      887         &      896         &      896         \\
\(R^{2}\)       &     0.28         &     0.34         &     0.41         &     0.38         &     0.41         \\
Regression      &      OLS         &      OLS         &      OLS         &      OLS         &      OLS         \\
\bottomrule
\end{tabular}

}
\caption*{\footnotesize \textit{Notes:} As Appendix Table~\ref{tab:experiment_windows}, except regressions are estimated by OLS with normalized forecasts capped at $\pm3$ (three times the respondent's category-default magnitude in raw-return units).}
\end{table}

\begin{figure}[!htbp]
\centering
\caption{Experimental Response, Excluding Sign-Inconsistent Forecasts}
\label{fig:experiment_robustness_appendix}
\includegraphics[width=.8\linewidth]{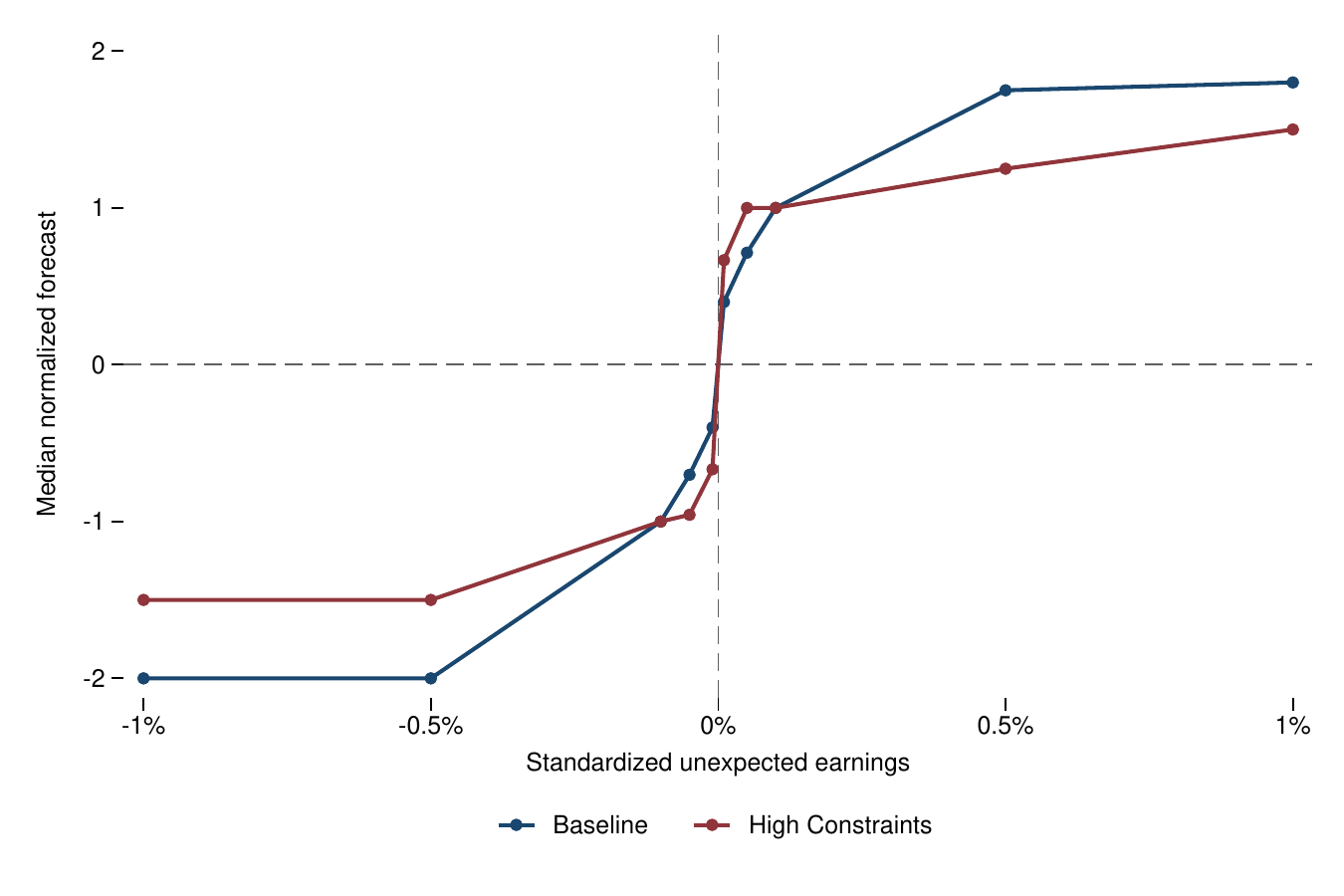}
\caption*{\footnotesize \textit{Notes:} As Figure~\ref{fig:experiment_median_forecasts}, excluding individual forecasts whose sign is opposite to that of the earnings surprise. This response-shape plot is distinct from the expanding-window interaction in Appendix Figure~\ref{fig:experiment_excl_inconsistent}(f).}
\end{figure}

\FloatBarrier

\begin{table}[!htbp]
\centering
\caption{Incidence of Forecasts at the Entry Bound}
\label{tab:experiment_censoring}
\fitpage{\begin{tabular}{lccc}
\toprule
$|\mathrm{SUE}|$ & Baseline & High Constraints & All \\
\midrule
$0.01\%$ & 0.5\% & 1.3\% & 0.9\% \\
$0.05\%$ & 0.0\% & 0.9\% & 0.4\% \\
$0.1\%$ & 1.3\% & 3.0\% & 2.2\% \\
$0.5\%$ & 12.2\% & 8.7\% & 10.5\% \\
$1\%$ & 18.2\% & 9.6\% & 13.9\% \\
\midrule
All responses & 6.6\% & 4.7\% & 5.6\% \\
Respondents ever at the bound & 22.4\% & 18.1\% & 20.2\% \\
\bottomrule
\end{tabular}
}
\caption*{\footnotesize \textit{Notes:} Percentage of forecasts that reach the end of the $\pm15\%$ entry range, by the absolute size of the surprise and by treatment. A response counts as capped if it lies within a thousandth of a percentage point of either end. The final row reports the share of respondents who reach a bound in at least one round.}
\end{table}

\FloatBarrier
\clearpage

\setcounter{figure}{0}\setcounter{table}{0}
\section{Additional Field Evidence}\label{app:field}

\subsection{Sample and Earnings-Surprise Distributions}\label{app:field_distribution}

\begin{figure}[!htbp]
\centering
\caption{Distribution of Earnings Surprises}
\label{fig:field_sue_histogram}
\includegraphics[width=\linewidth]{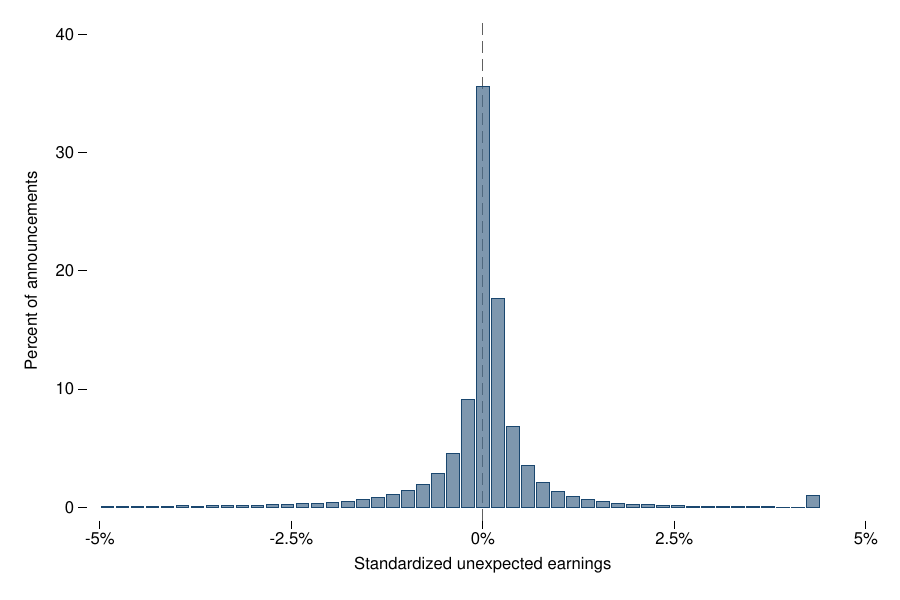}
\caption*{\footnotesize \textit{Notes:} SUE (defined in~\eqref{eq:sue}) in 51 equal-width bins from $-5\%$ to $+5\%$.}
\end{figure}

\begin{figure}[!htb]
\centering
\caption{Earnings Surprises by Valuation Uncertainty}
\label{fig:field_sue_histogram_by_vu}
\includegraphics[width=\linewidth]{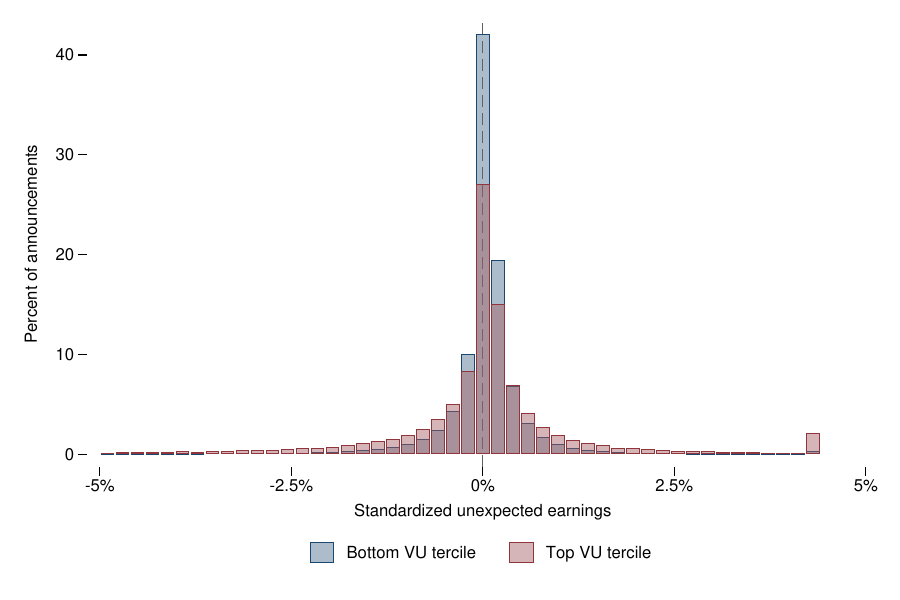}
\caption*{\footnotesize \textit{Notes:} As Appendix Figure~\ref{fig:field_sue_histogram}, comparing the bottom and top VU terciles formed within year-quarter.}
\end{figure}

Appendix Table~\ref{tab:field_summary} describes the baseline five-day return regression sample, while Appendix Figure~\ref{fig:field_sue_histogram} shows the distribution of SUEs, which Appendix Figure~\ref{fig:field_sue_histogram_by_vu} further splits by VU.

\begin{table}[!htbp]
\centering
\caption{Summary Statistics for the Field Sample}
\label{tab:field_summary}
\fittable{\begin{tabular}{lrrrrrr}
\toprule
& Obs. & Mean & SD & P25 & Median & P75 \\
\midrule
Standardized unexpected earnings &   224,350 &     0.011 &     0.936 &    -0.125 &     0.033 &     0.207 \\
Reported earnings per share &   224,350 &     0.365 &     1.069 &     0.050 &     0.240 &     0.520 \\
Five-day market-adjusted return &   224,350 &    -0.068 &     9.479 &    -4.965 &    -0.246 &     4.597 \\
Valuation uncertainty &   224,350 &     0.769 &     0.238 &     0.622 &     0.767 &     0.920 \\
Analyst forecast dispersion &   176,323 &     0.423 &     0.431 &     0.188 &     0.269 &     0.444 \\
\bottomrule
\end{tabular}
}
\caption*{\footnotesize \textit{Notes:} The sample contains the observations used in the main five-day return specification. SUE and five-day market-adjusted returns are both in units of percentage points. Reported earnings per share (i.e., EPS) are ``street'' earnings from IBES and are in units of dollars per share. Here, we report the raw VU index instead of the standardized version we use in our main empirical tests. Analyst forecast dispersion is defined as the standard deviation of analysts' forecasts and is therefore available only when there are at least three analysts covering the stock.}
\end{table}

\FloatBarrier
\clearpage

\subsection{Variation in Valuation Uncertainty}\label{app:fieldadditional}

\begin{table}[!htb]
\centering
\caption{Where Valuation Uncertainty Varies}
\label{tab:field_vu_variation}
\fittable{\begin{minipage}{\linewidth}
\centering
\begin{tabular}{@{}lccc@{}}
\multicolumn{4}{c}{\textit{Panel A: Aggregate variation over time}} \\
\toprule
Year & Mean VU & Median VU & Observations \\
\midrule
1985 & $-1.02$ & $-0.97$ &     2,811 \\
2000 & $ 0.76$ & $ 0.86$ &     7,831 \\
2020 & $ 0.64$ & $ 0.64$ &     4,329 \\
\bottomrule
\end{tabular}

\vspace{1.25em}

\begin{tabular}{@{}lccc@{}}
\multicolumn{4}{c}{\textit{Panel B: Cross-industry variation}} \\
\toprule
Industry & Mean VU & Median VU & Observations \\
\midrule
Utilities & $-1.76$ & $-1.90$ &    12,187 \\
Pharmaceutical Products & $ 0.48$ & $ 0.53$ &    12,018 \\
Computer Software & $ 0.62$ & $ 0.55$ &    17,421 \\
\bottomrule
\end{tabular}

\vspace{1.25em}

\begin{tabular}{@{}lccc@{}}
\multicolumn{4}{c}{\textit{Panel C: Within computer software over time}} \\
\toprule
Year & Mean VU & Median VU & Observations \\
\midrule
2005 & $-0.13$ & $-0.08$ &       766 \\
2018 & $ 0.81$ & $ 0.73$ &       279 \\
\bottomrule
\end{tabular}

\vspace{1.25em}

\begin{tabular}{@{}lccccc@{}}
\multicolumn{6}{c}{\textit{Panel D: Costco within retail over time}} \\
\toprule
Announcement date & VU & \shortstack[c]{Overall\\percentile} & \shortstack[c]{Quarter\\percentile} & \shortstack[c]{Industry-quarter\\percentile} & Reported EPS \\
\midrule
March 3, 2016 & $-1.94$ &   3 &   3 &   1 & $ 1.24$ \\
June 1, 2018 & $ 0.45$ &  69 &  73 &  85 & $ 1.70$ \\
March 4, 2022 & $ 1.06$ &  87 &  73 &  95 & $ 2.92$ \\
\bottomrule
\end{tabular}
\end{minipage}
}
\caption*{\footnotesize \textit{Notes:} All summaries use the baseline complete-case regression sample, counting each announcement once. Here, VU is standardized to have mean zero and standard deviation one on that sample. Panels A-C report means and medians for the indicated calendar years or FF49 industries, where each announcement is given an equal weight. Panel D selects Costco's lowest and highest industry-quarter VU percentiles and its June 1, 2018 announcement from its modal FF49 industry in the sample (retail). Percentile ranks use all sample announcements, all announcements in the same quarter, or all announcements in the same industry-quarter, respectively. A Shapley decomposition of raw VU variance across announcements attributes 28.2\% to industry effects and 15.6\% to calendar-quarter effects, leaving 56.2\% unexplained; when firm effects are included, the respective shares are 14.4\%, 13.5\%, and 40.0\%, with 32.1\% unexplained.}
\end{table}

\FloatBarrier
\clearpage

\subsection{Response-Scale Selection}\label{app:scale_selection}

For each response family, we select the scale $k$ that achieves the highest $R^2$ in fitting the unconditional response, i.e., in the regression in~\eqref{eq:baseline} before adding VU or any fixed effects and controls.

In the experiment, we use median regression of forecasts on an intercept, SUE, and $h(\SUE;k)$, pooling over both treatments. We then maximize the pseudo-$R^2$ across a grid of candidate values $k=0.010,0.015,\ldots,0.990$ for the hinge and $k=0.010,0.015,\ldots,2.000$ for the smooth families. We do this separately for raw and normalized forecasts, since they have very different scales. The scale is then held fixed in all subsequent experimental applications.

In the field, the corresponding five-day return regression maximizes the ordinary $R^2$ over $k=0.02,0.03,\ldots,1.00$. We express scales in percentage points of SUE. This selection excludes controls, fixed effects, VU, and their interactions. Subsequent field hinge specifications also all inherit this numerical threshold. For comparability across measures, alternative SUE measures are rescaled to the standard deviation of baseline SUE without changing the inherited threshold. On the other hand, smooth response families are naturally given their own optimal scales.

\begin{figure}[!htbp]
\centering
\caption{Unconditional Fit across Response Scales}
\label{fig:field_selection_profiles}
\includegraphics[width=\linewidth]{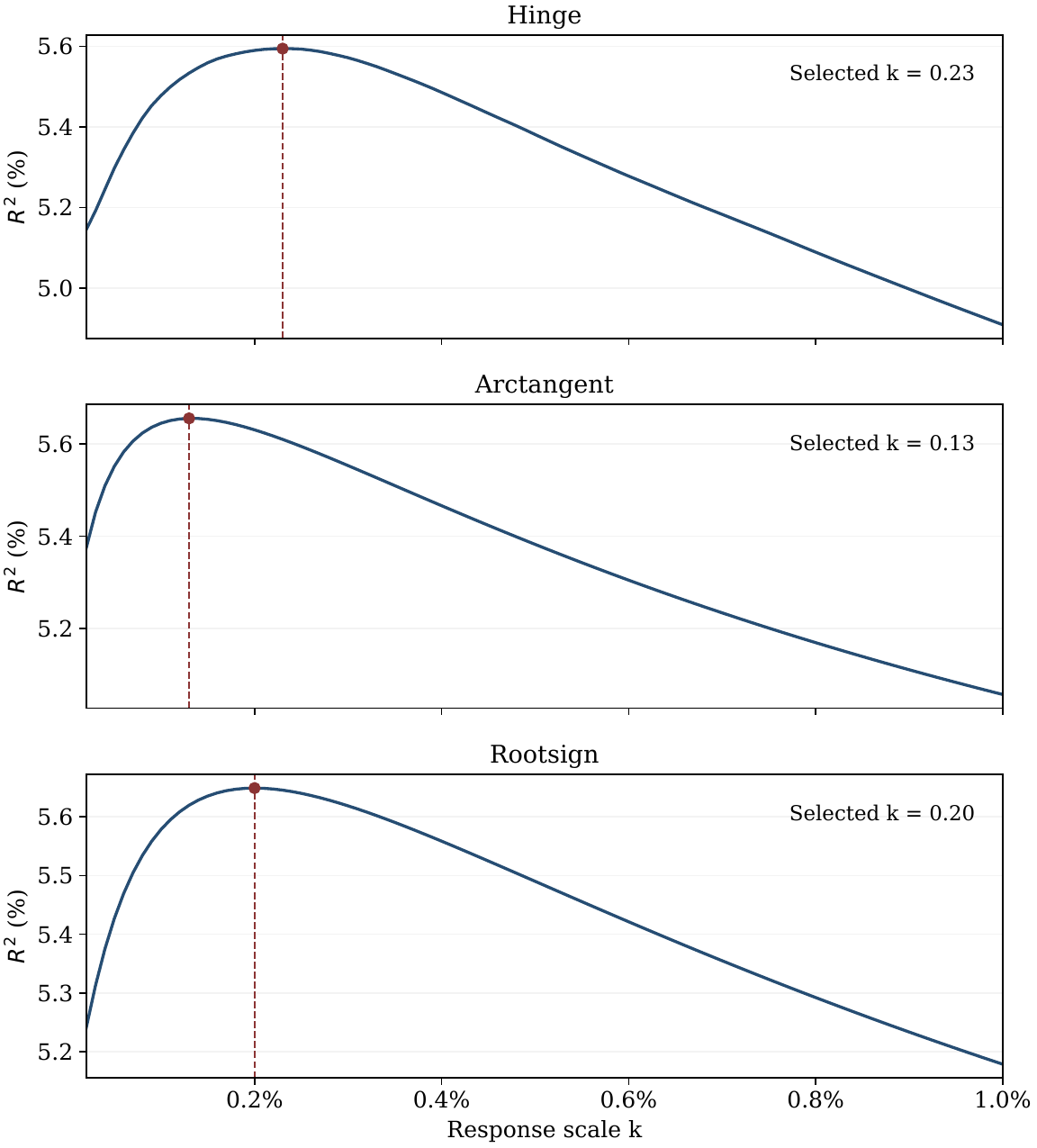}
\caption*{\footnotesize \textit{Notes:} Ordinary $R^2$ from the unconditional five-day return regression at each candidate scale, on the common VU sample. Each regression includes an intercept, SUE, and the indicated response term, without controls, fixed effects, or VU. The marked maximum sets the scale for that response specification.}
\end{figure}

We first investigate whether this optimization is brittle. For that, Appendix Figure~\ref{fig:field_selection_profiles} plots the $R^2$'s from the corresponding regressions in the baseline field specification, showing that the optimal hinge threshold indeed corresponds to a smooth global maximum. This also holds true for the best-fitting arctangent and rootsign scales.

We then explore how sensitive this scale selection is in the field. Appendix Table~\ref{tab:field_scale_selection} compares alternative selection regressions on the same 224,350 field observations: adding controls, adding industry and quarter fixed effects, or adding both barely moves the optimal hinge threshold. Furthermore, it also barely moves the optimal rootsign and arctangent scales.

It should be stressed that we view the threshold of the hinge function as a useful parametrization, but not directly as a parameter of interest to be estimated. We hence consider it given in all regressions that use the hinge function. This means that reported standard errors condition on the selected scale and do not account for its estimation. This also applies to all regressions using the smooth alternatives.

\begin{table}[!htbp]
\centering
\caption{Response-Scale Selection with Controls and Fixed Effects}
\label{tab:field_scale_selection}
\fittable{\begin{tabular*}{\linewidth}{@{\extracolsep{\fill}}lrrr@{}}
\toprule
Selection regression & Hinge & Rootsign & Arctangent \\
\midrule
No controls or fixed effects & 0.23 & 0.20 & 0.13 \\
Controls & 0.22 & 0.19 & 0.13 \\
Industry and quarter fixed effects & 0.22 & 0.19 & 0.12 \\
Controls and fixed effects & 0.22 & 0.19 & 0.12 \\
\bottomrule
\end{tabular*}
}
\caption*{\footnotesize \textit{Notes:} Selected scales $k$, in percentage points of SUE, maximize the ordinary $R^2$ of the five-day return regression on an intercept, SUE, and the indicated response term. All rows use the same 224,350 observations and the grid $0.02,0.03,\ldots,1.00$. Controls are those of Table~\ref{tab:field_vu_main} and fixed effects are industry and year-quarter effects. VU and its interactions are excluded throughout. The first row gives the paper's selection rule.}
\end{table}

\FloatBarrier

\subsection{Additional Response Estimates}\label{app:field_full_range}

\begin{figure}[!htbp]
\centering
\caption{Binned Earnings Responses by Valuation Uncertainty}
\label{fig:field_sshape_by_vu_unfitted}
\includegraphics[width=\linewidth]{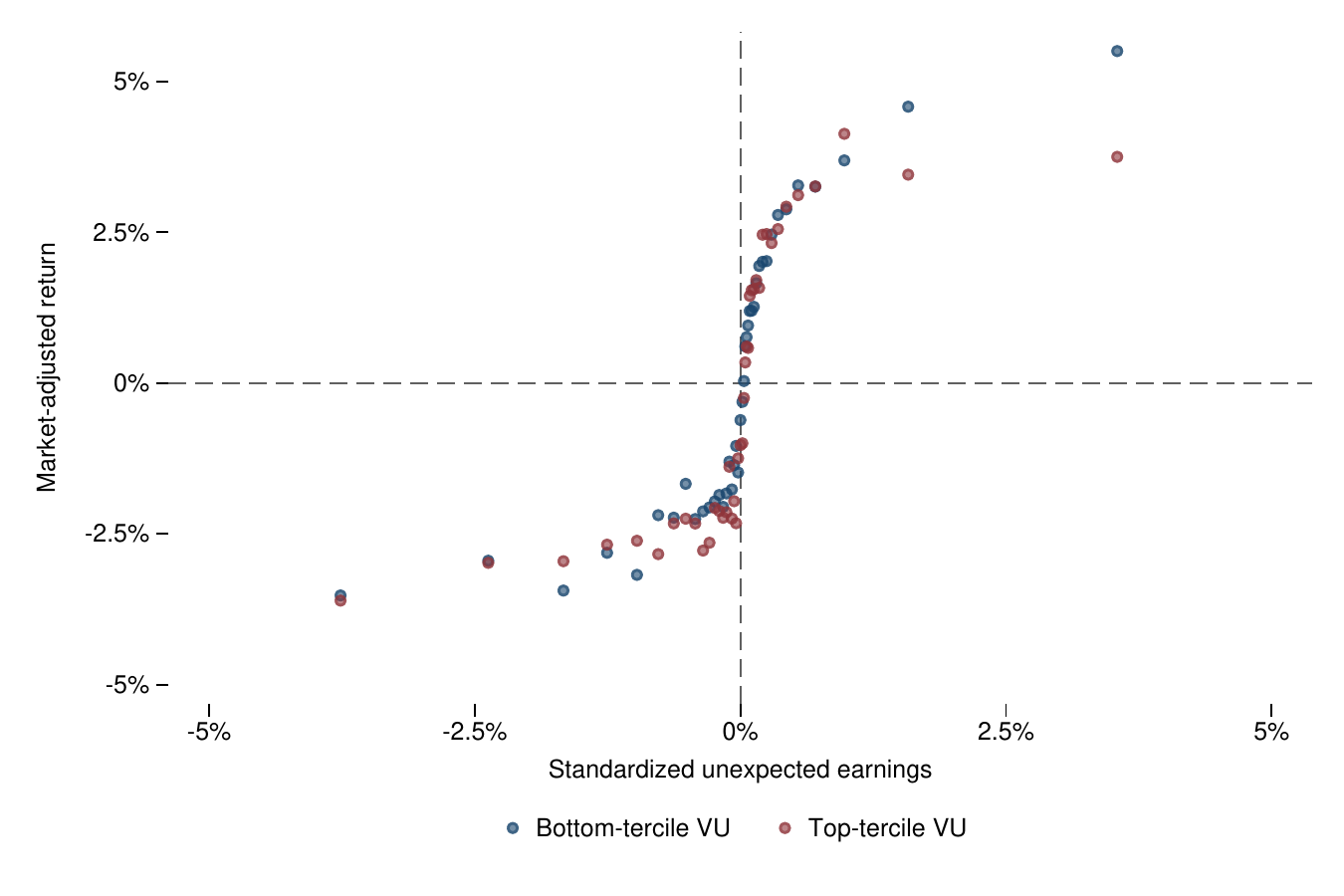}
\caption*{\footnotesize \textit{Notes:} Announcements are split into bottom and top VU terciles within year-quarter, then placed in 20 common equal-count SUE bins on each side of zero over the $-5\%$ to $+5\%$ regression range, while exact-zero surprises are put in a separate bin. Group-specific fitted levels from the no-controls hinge regression are subtracted from returns before plotting.}
\end{figure}

\begin{figure}[!htbp]
\centering
\caption{Valuation Uncertainty over Expanding SUE Windows}
\label{fig:field_expanding_windows}
\includegraphics[width=\linewidth]{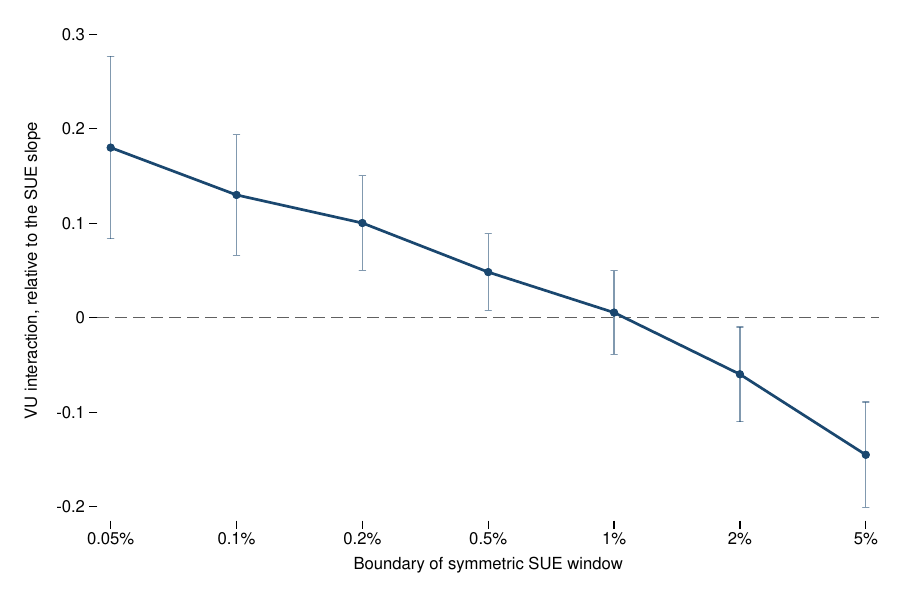}
\centering
\caption*{\footnotesize \textit{Notes:} Each point is $\hat\delta/\hat\beta$ from Equation~\eqref{eq:field_linear}, estimated over the symmetric SUE window $w$ shown on the horizontal axis. Regressions include the paper's controls and industry and year-quarter fixed effects. Whiskers are 95\% confidence intervals based on delta-method standard errors clustered by year-quarter.}
\end{figure}

\begin{figure}[!htb]
\centering
\caption{Valuation Uncertainty and the Shape of the Market Earnings Response}
\label{fig:field_response_fits}
\includegraphics[width=\linewidth,height=0.61\textheight,keepaspectratio]{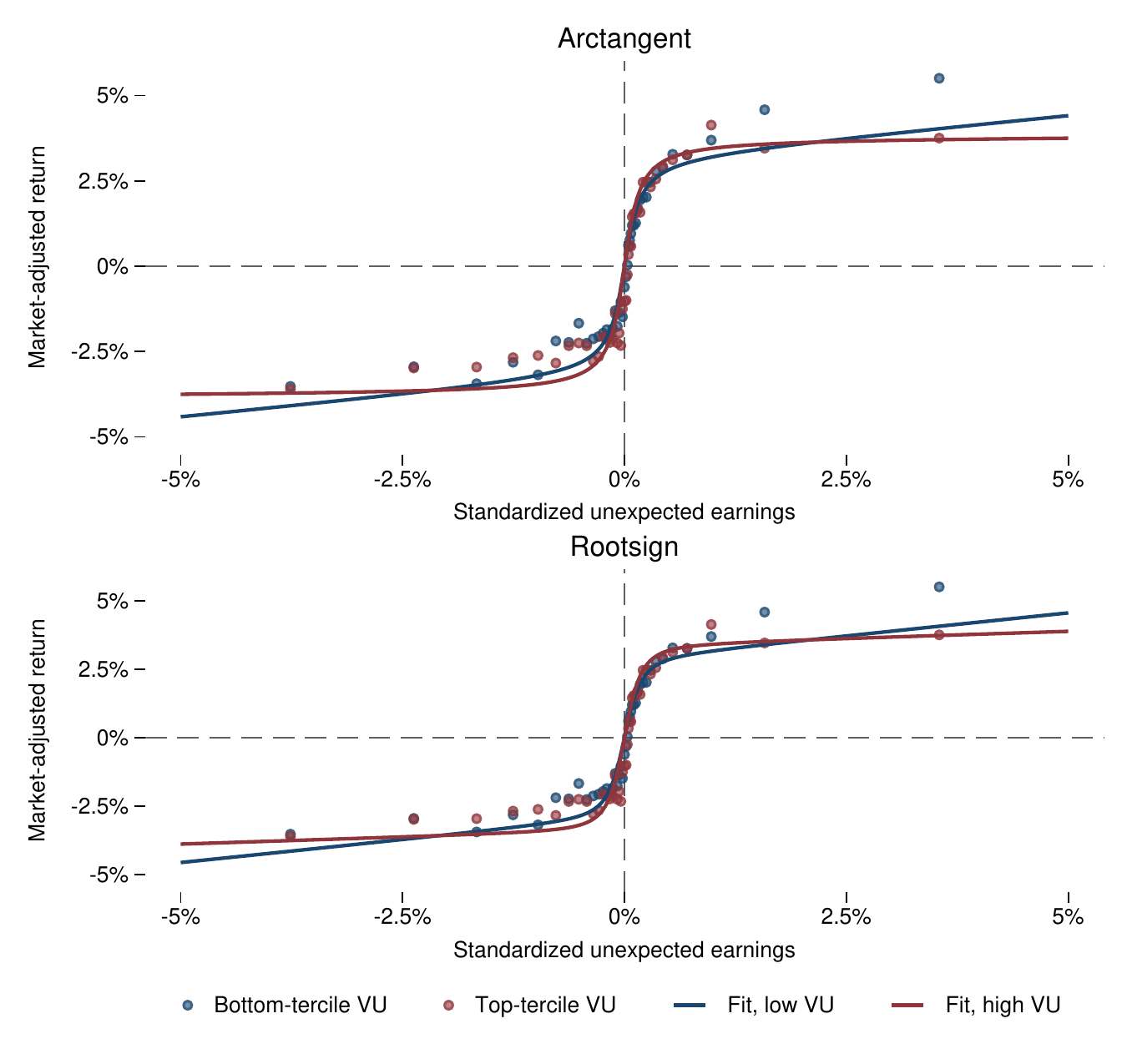}
\caption*{\footnotesize \textit{Notes:} Points are those in Appendix Figure~\ref{fig:field_sshape_by_vu_unfitted} and are identical across panels over the full $-5\%$ to $+5\%$ regression range. Lines are the fitted values corresponding to Appendix Table~\ref{tab:field_vu_smooth}, using the response family of each panel. Group-specific fitted levels are removed from both points and lines. The hinge fit appears in Figure~\ref{fig:field_response_fit_hinge}.}
\end{figure}

Appendix Figure~\ref{fig:field_sshape_by_vu_unfitted} shows the binned response function without any fit, while Appendix Figure~\ref{fig:field_expanding_windows} reproduces the expanding-window estimate pre-registered for the experiment. In turn, Appendix Table~\ref{tab:field_vu_smooth} and Appendix Figure~\ref{fig:field_response_fits} show arctangent and rootsign fits.

\begin{table}[!htbp]
\centering
\caption{Valuation Uncertainty under Smooth Functional Forms}\label{tab:field_vu_smooth}
\fittable{\begin{tabular}{l*{2}{Y}}
\toprule
\multicolumn{3}{c}{\textit{Dependent variable: five-day market-adjusted return}} \\
& Arctangent & Rootsign \\
& (1) & (2) \\
\midrule
SUE                 &        0.13\sym{*}  &        0.22\sym{***}\\
                    &      (0.07)         &      (0.07)         \\
$\Delta$ Small SUE  &       17.69\sym{***}&       16.20\sym{***}\\
                    &      (0.65)         &      (0.60)         \\
VU                  &       -0.11\sym{**} &       -0.11\sym{**} \\
                    &      (0.04)         &      (0.04)         \\
VU $\times$ SUE     &       -0.12\sym{**} &       -0.12\sym{**} \\
                    &      (0.06)         &      (0.06)         \\
VU $\times$ $\Delta$ Small SUE&        1.25\sym{***}&        1.17\sym{***}\\
                    &      (0.37)         &      (0.34)         \\
\midrule
Small SUE           &17.82\sym{***}         &16.42\sym{***}         \\
                    &      (0.64)         &      (0.59)         \\
VU $\times$ Small SUE&1.13\sym{***}         &1.06\sym{***}         \\
                    &      (0.35)         &      (0.32)         \\
\midrule
Scale \(k\)         &        0.13         &        0.20         \\
Observations        &     224,350         &     224,350         \\
\(R^{2}\)           &        0.07         &        0.07         \\
Quarter \& industry FEs&$\checkmark$         &$\checkmark$         \\
All controls        &$\checkmark$         &$\checkmark$         \\
\bottomrule
\end{tabular}
}
\caption*{\footnotesize \textit{Notes:} As the hinge columns of Table~\ref{tab:field_vu_main}, replacing the hinge in Equation~\eqref{eq:baseline} by the arctangent term $h_A(\SUE)$ or rootsign term $h_R(\SUE)$. $\Delta$ Small SUE is the bounded term, whose coefficient is again the additional slope at zero. For each family, the scale $k$ is selected once from the unconditional response before VU enters and then held fixed. Both columns include the baseline controls and industry and year-quarter fixed effects.}
\end{table}

\FloatBarrier
\clearpage

\subsection{Robustness}\label{app:field_robustness}

Appendix Table~\ref{tab:field_specification_robustness} repeats the main specification from Table~\ref{tab:field_vu_main} while varying estimation method, weighting, fixed effects, sample restrictions, return benchmark, and VU dating.

\begin{table}[!htb]
\centering
\caption{Robustness Checks for Table~\ref{tab:field_vu_main}}\label{tab:field_specification_robustness}
\fittable{\begin{tabular}{l r@{}l@{\hspace{0.3em}}l r@{}l@{\hspace{0.3em}}l r@{}l@{\hspace{0.3em}}l r}
\toprule
& \multicolumn{10}{c}{\textit{Dependent variable: five-day abnormal return}} \\
 & \multicolumn{3}{c}{Global} & \multicolumn{3}{c}{Local} & \multicolumn{3}{c}{Relative} & \\
 & \multicolumn{3}{c}{VU \(\times\) SUE} & \multicolumn{3}{c}{VU \(\times\) Small SUE} & \multicolumn{3}{c}{VU \(\times\) \(\Delta\) Small SUE} & Obs. \\
\midrule
\quad Baseline (Table \ref{tab:field_vu_main}, columns 5--6) & -0.285 & \sym{***} & (0.065) & 0.880 & \sym{***} & (0.245) & 1.011 & \sym{***} & (0.257) & 224,350 \\
\multicolumn{11}{l}{\textit{Panel A: specification}} \\
\quad Two-way clustered standard errors & -0.285 & \sym{***} & (0.066) & 0.880 & \sym{***} & (0.265) & 1.011 & \sym{***} & (0.278) & 224,350 \\
\quad Firm fixed effects & -0.223 & \sym{***} & (0.063) & 1.052 & \sym{***} & (0.238) & 1.127 & \sym{***} & (0.243) & 224,217 \\
\quad Firm fixed effects, two-way clustered & -0.223 & \sym{***} & (0.065) & 1.052 & \sym{***} & (0.258) & 1.127 & \sym{***} & (0.264) & 224,217 \\
\quad Market-cap weights within quarter & 0.425 & \sym{***} & (0.087) & 1.996 & \sym{***} & (0.371) & 1.720 & \sym{***} & (0.431) & 224,350 \\
\quad Market-cap-rank weights within quarter & 0.206 & \sym{***} & (0.073) & 2.221 & \sym{***} & (0.249) & 2.110 & \sym{***} & (0.270) & 224,350 \\
\quad Median regression & -0.120 & \sym{***} & (0.021) & 1.627 & \sym{***} & (0.113) & 1.618 & \sym{***} & (0.130) & 224,350 \\
\quad VU as within-quarter percentile rank & -0.426 & \sym{***} & (0.041) & 0.479 & \sym{***} & (0.147) & 0.707 & \sym{***} & (0.164) & 224,350 \\
\addlinespace
\multicolumn{11}{l}{\textit{Panel B: sample}} \\
\quad Estimation window $|\text{SUE}|\leq1\%$ & 0.032 &  & (0.128) & 0.883 & \sym{***} & (0.240) & 1.045 & \sym{***} & (0.332) & 195,152 \\
\quad Estimation window $|\text{SUE}|\leq2\%$ & -0.224 & \sym{**} & (0.101) & 0.899 & \sym{***} & (0.234) & 1.106 & \sym{***} & (0.265) & 211,619 \\
\quad Estimation window $|\text{SUE}|\leq3\%$ & -0.303 & \sym{***} & (0.084) & 0.915 & \sym{***} & (0.236) & 1.116 & \sym{***} & (0.253) & 217,613 \\
\quad Excluding exact-zero surprises & -0.282 & \sym{***} & (0.065) & 0.853 & \sym{***} & (0.248) & 0.982 & \sym{***} & (0.261) & 197,978 \\
\quad Dollar surprise at least half a cent & -0.282 & \sym{***} & (0.065) & 0.850 & \sym{***} & (0.248) & 0.979 & \sym{***} & (0.261) & 197,590 \\
\quad Dollar surprise at least one cent & -0.280 & \sym{***} & (0.065) & 1.024 & \sym{***} & (0.254) & 1.167 & \sym{***} & (0.268) & 186,568 \\
\addlinespace
\multicolumn{11}{l}{\textit{Panel C: return definition}} \\
\quad CAPM-adjusted return & -0.286 & \sym{***} & (0.065) & 0.894 & \sym{***} & (0.236) & 1.029 & \sym{***} & (0.246) & 224,350 \\
\quad Fama--French three-factor return & -0.285 & \sym{***} & (0.064) & 0.931 & \sym{***} & (0.232) & 1.069 & \sym{***} & (0.240) & 224,350 \\
\quad Carhart four-factor return & -0.291 & \sym{***} & (0.063) & 0.936 & \sym{***} & (0.234) & 1.081 & \sym{***} & (0.240) & 224,350 \\
\quad Industry-adjusted return & -0.264 & \sym{***} & (0.062) & 0.975 & \sym{***} & (0.228) & 1.108 & \sym{***} & (0.238) & 224,350 \\
\addlinespace
\multicolumn{11}{l}{\textit{Panel D: timing}} \\
\quad VU dated by accounting-data & -0.322 & \sym{***} & (0.066) & 0.800 & \sym{***} & (0.246) & 0.943 & \sym{***} & (0.256) & 224,350 \\
\quad VU delayed until January & -0.235 & \sym{***} & (0.066) & 1.322 & \sym{***} & (0.223) & 1.482 & \sym{***} & (0.236) & 172,251 \\
\quad Excluding non-market days & -0.285 & \sym{***} & (0.065) & 0.879 & \sym{***} & (0.245) & 1.011 & \sym{***} & (0.257) & 224,345 \\
\bottomrule
\end{tabular}
}
\caption*{\footnotesize \textit{Notes:} Each row modifies the linear and hinge specifications with full fixed effects and controls from Table~\ref{tab:field_vu_main}, with the baseline row combining columns (5) and (6). The global attenuation measure ``VU $\times$ SUE'' comes from the linear specification, while the local measure ``VU $\times$ Small SUE'' and the relative measure ``VU $\times\Delta$ Small SUE'' comes from the hinge specification. In each row, the observations and controls included are identical, and the hinge threshold is held fixed at $k=0.23\%$. Further, in each row, we include the same fixed effects and cluster by year-quarter unless the row states otherwise.  Specifically, the median regression has no fixed effects and uses heteroskedasticity-robust errors, while two-way clustering additionally clusters by firm. Market-capitalization and size-rank weights are normalized to sum to one within year-quarter. Industry-adjusted returns subtract the FF49 industry-year-quarter mean return from each firm's return. Other return definitions use standard CAPM, Fama--French three-factor, or Carhart four-factor residuals, and are winsorized at the first and ninety-ninth percentiles. The VU-rank specification uses within-quarter percentile ranks of VU instead of VU itself. Future earnings growth is defined as the EPS growth over the next four quarters divided by the price. The alternative VU dates either ignore the June peer-model formation date or conservatively delay each new vintage until the following January.}
\end{table}

\FloatBarrier
\clearpage

\paragraph{Announcement windows.}
Appendix Table~\ref{tab:field_event_windows} uses returns over day~0, days~0--2 and 0--4, where day~0 is the first tradable announcement day. All use the common sample and baseline hinge threshold.

\begin{table}[!htbp]
\centering
\caption{Announcement-Window Robustness}
\label{tab:field_event_windows}
\fittable{\begin{tabular}{l*{6}{Y}}
\toprule
& \multicolumn{6}{c}{\textit{Dependent variable: announcement market-adjusted return}} \\
& \multicolumn{2}{c}{Day 0} & \multicolumn{2}{c}{Days 0--2} & \multicolumn{2}{c}{Days 0--4} \\
\cmidrule(lr){2-3}\cmidrule(lr){4-5}\cmidrule(lr){6-7}
& (1) & (2) & (3) & (4) & (5) & (6) \\
\midrule
SUE                 &        1.58\sym{***}&        0.25\sym{***}&        1.90\sym{***}&        0.34\sym{***}&        1.98\sym{***}&        0.39\sym{***}\\
                    &      (0.08)         &      (0.05)         &      (0.09)         &      (0.06)         &      (0.09)         &      (0.07)         \\
$\Delta$ Small SUE  &                     &       10.37\sym{***}&                     &       12.14\sym{***}&                     &       12.39\sym{***}\\
                    &                     &      (0.44)         &                     &      (0.44)         &                     &      (0.46)         \\
VU                  &       -0.03         &       -0.02         &       -0.10\sym{***}&       -0.08\sym{**} &       -0.14\sym{***}&       -0.12\sym{***}\\
                    &      (0.02)         &      (0.02)         &      (0.04)         &      (0.04)         &      (0.05)         &      (0.04)         \\
VU $\times$ SUE     &       -0.20\sym{***}&       -0.08\sym{**} &       -0.25\sym{***}&       -0.10\sym{**} &       -0.27\sym{***}&       -0.12\sym{**} \\
                    &      (0.05)         &      (0.03)         &      (0.06)         &      (0.05)         &      (0.06)         &      (0.06)         \\
VU $\times$ $\Delta$ Small SUE&                     &        0.98\sym{***}&                     &        1.00\sym{***}&                     &        1.02\sym{***}\\
                    &                     &      (0.20)         &                     &      (0.23)         &                     &      (0.25)         \\
\midrule
Small SUE           &                     &10.62\sym{***}         &                     &12.48\sym{***}         &                     &12.77\sym{***}         \\
                    &                     &      (0.44)         &                     &      (0.43)         &                     &      (0.45)         \\
VU $\times$ Small SUE&                     &0.89\sym{***}         &                     &0.90\sym{***}         &                     &0.90\sym{***}         \\
                    &                     &      (0.20)         &                     &      (0.23)         &                     &      (0.24)         \\
\midrule
Hinge threshold \(k\)&                     &        0.23         &                     &        0.23         &                     &        0.23         \\
Observations        &     222,536         &     222,536         &     222,536         &     222,536         &     222,536         &     222,536         \\
\(R^{2}\)           &        0.04         &        0.08         &        0.04         &        0.07         &        0.04         &        0.07         \\
Quarter \& industry FEs&$\checkmark$         &$\checkmark$         &$\checkmark$         &$\checkmark$         &$\checkmark$         &$\checkmark$         \\
All controls        &$\checkmark$         &$\checkmark$         &$\checkmark$         &$\checkmark$         &$\checkmark$         &$\checkmark$         \\
\bottomrule
\end{tabular}
}
\caption*{\footnotesize \textit{Notes:} As Table~\ref{tab:field_vu_main}, using the cumulative market-adjusted return over the announcement window indicated in each column pair. Day 0 is the first date on which the earnings release can be traded.}
\end{table}

\FloatBarrier

 \paragraph{Alternative surprise measures.}
Appendix Table~\ref{tab:field_sue_scalings} compares the baseline price-scaled mean forecast error with four alternatives: the median forecast error divided by forecast standard deviation, the mean forecast error divided by absolute consensus, the dollar forecast error, and sign-preserving percentile ranks. To construct the latter, negative and positive baseline surprises are ranked separately in 50 equal-count groups, with zero retained as its own group. Continuous alternatives are winsorized at 1/99 and rescaled to the standard deviation of baseline SUE without recentering, so that the baseline hinge threshold has the same meaning across regressions.

\begin{figure}[!htbp]
\centering
\caption{Alternative Earnings-Surprise Measures by Valuation Uncertainty}
\label{fig:field_sue_scalings_by_vu}
\includegraphics[width=\linewidth]{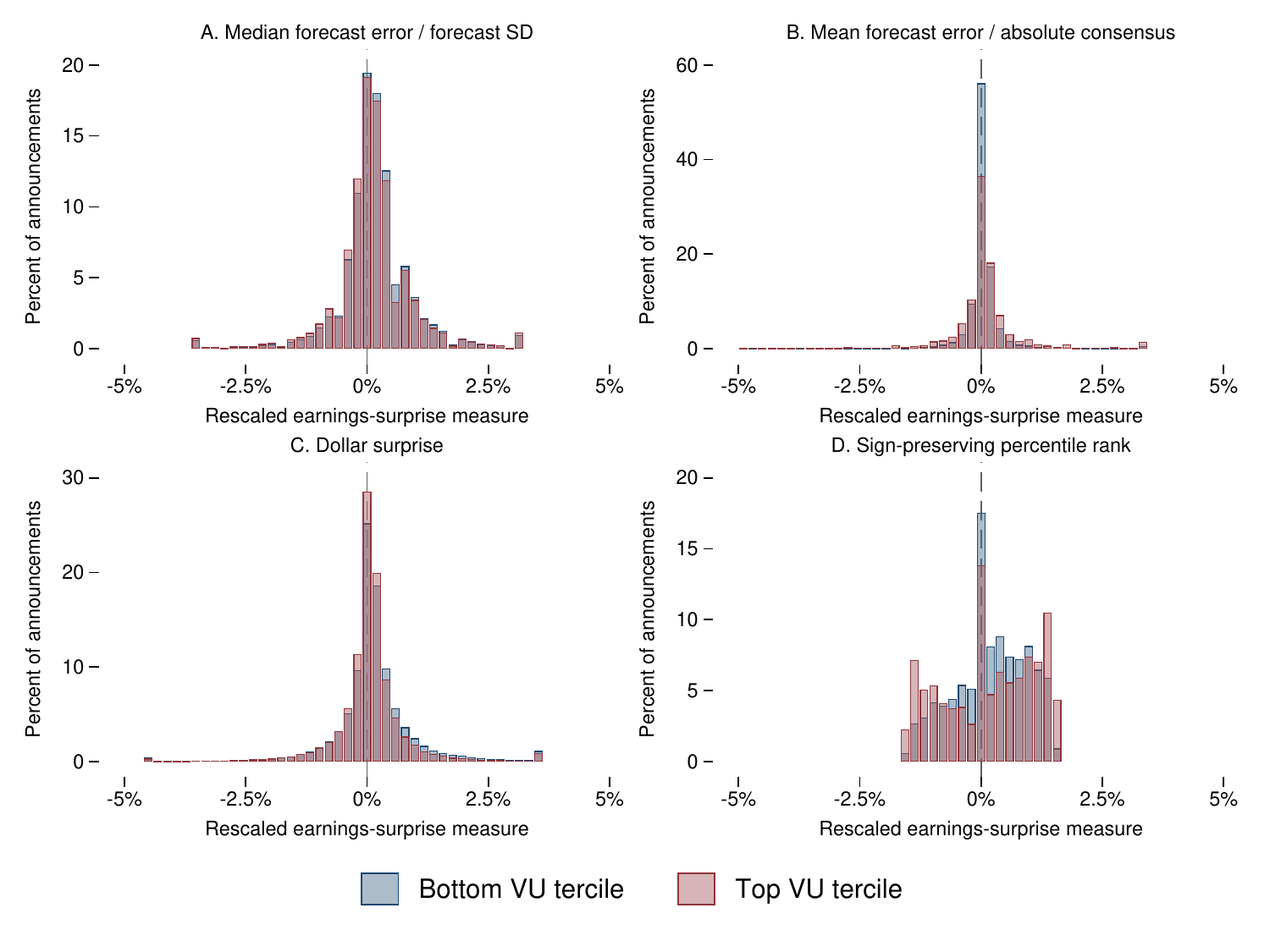}
\caption*{\footnotesize \textit{Notes:} Histograms for the four alternative surprise measures in Appendix Table~\ref{tab:field_sue_scalings}, comparing the bottom and top VU terciles within year-quarter and omitting the middle tercile. Each measure is rescaled as in that table and shown over $-5\%$ to $+5\%$. Every panel uses 51 equal-width intervals, so zero lies in the middle interval, which also contains nonzero surprises. Group-normalized frequencies are overlaid semi-transparently.}
\end{figure}

The definitions of some of these measures mean that they will comove somewhat mechanically with VU. For example, in that common sample, the mean analyst forecast standard deviation is $0.051$ dollars per share in the bottom VU tercile and $0.043$ in the top tercile (corresponding medians are $0.03$ and $0.02$). Appendix Figure~\ref{fig:field_sue_scalings_by_vu} plots the distribution of all these alternative measures by VU tercile.

\begin{table}[!htbp]
\centering
\caption{Market Earnings Response under Alternative Surprise Scalings}\label{tab:field_sue_scalings}
\fittable{\begin{tabular}{l*{10}{c}}
\toprule
& \multicolumn{10}{c}{\textit{Dependent variable: five-day market-adjusted return}} \\
& \multicolumn{2}{c}{\shortstack[c]{Scaled by\\price}} & \multicolumn{2}{c}{\shortstack[c]{Median forecast\\error / forecast\\SD}} & \multicolumn{2}{c}{\shortstack[c]{Mean forecast\\error / absolute\\consensus}} & \multicolumn{2}{c}{\shortstack[c]{Dollar\\surprise}} & \multicolumn{2}{c}{\shortstack[c]{Sign-preserving\\percentile ranks}} \\
\cmidrule(lr){2-3}\cmidrule(lr){4-5}\cmidrule(lr){6-7}\cmidrule(lr){8-9}\cmidrule(lr){10-11}
& (1) & (2) & (3) & (4) & (5) & (6) & (7) & (8) & (9) & (10) \\
\midrule
SUE                 &        2.32\sym{***}&        0.52\sym{***}&        2.74\sym{***}&        1.55\sym{***}&        1.83\sym{***}&        0.19\sym{***}&        2.20\sym{***}&        0.48\sym{***}&        2.89\sym{***}&        2.50\sym{***}\\
                    &      (0.11)         &      (0.08)         &      (0.09)         &      (0.06)         &      (0.07)         &      (0.04)         &      (0.10)         &      (0.05)         &      (0.10)         &      (0.12)         \\
$\Delta$ Small SUE  &                     &       12.74\sym{***}&                     &        7.36\sym{***}&                     &       15.19\sym{***}&                     &       11.86\sym{***}&                     &        1.83\sym{***}\\
                    &                     &      (0.47)         &                     &      (0.28)         &                     &      (0.51)         &                     &      (0.42)         &                     &      (0.33)         \\
VU                  &       -0.06         &       -0.06         &       -0.07         &       -0.07         &       -0.05         &       -0.06         &       -0.07         &       -0.08\sym{*}  &       -0.05         &       -0.06         \\
                    &      (0.05)         &      (0.05)         &      (0.05)         &      (0.05)         &      (0.05)         &      (0.05)         &      (0.05)         &      (0.05)         &      (0.05)         &      (0.05)         \\
VU $\times$ SUE     &       -0.29\sym{***}&       -0.13\sym{*}  &        0.34\sym{***}&        0.19\sym{***}&        0.02         &        0.10\sym{***}&        0.32\sym{***}&        0.05         &        0.12\sym{**} &       -0.02         \\
                    &      (0.08)         &      (0.07)         &      (0.06)         &      (0.06)         &      (0.04)         &      (0.03)         &      (0.06)         &      (0.04)         &      (0.06)         &      (0.11)         \\
VU $\times$ $\Delta$ Small SUE&                     &        1.25\sym{***}&                     &        1.14\sym{***}&                     &        0.51\sym{*}  &                     &        2.07\sym{***}&                     &        0.78\sym{**} \\
                    &                     &      (0.29)         &                     &      (0.20)         &                     &      (0.30)         &                     &      (0.26)         &                     &      (0.35)         \\
\midrule
Small SUE           &                     &13.26\sym{***}         &                     &8.90\sym{***}         &                     &15.38\sym{***}         &                     &12.34\sym{***}         &                     &4.33\sym{***}         \\
                    &                     &      (0.46)         &                     &      (0.30)         &                     &      (0.51)         &                     &      (0.41)         &                     &      (0.28)         \\
VU $\times$ Small SUE&                     &1.13\sym{***}         &                     &1.33\sym{***}         &                     &0.61\sym{**}         &                     &2.13\sym{***}         &                     &0.76\sym{***}         \\
                    &                     &      (0.27)         &                     &      (0.19)         &                     &      (0.29)         &                     &      (0.24)         &                     &      (0.26)         \\
\midrule
Corr. with baseline SUE&        1.00         &        1.00         &        0.58         &        0.58         &        0.60         &        0.60         &        0.67         &        0.67         &        0.75         &        0.75         \\
Hinge threshold \(k\)&                     &        0.23         &                     &        0.23         &                     &        0.23         &                     &        0.23         &                     &        0.23         \\
Observations        &     174,196         &     174,196         &     174,196         &     174,196         &     174,196         &     174,196         &     174,196         &     174,196         &     174,196         &     174,196         \\
\(R^{2}\)           &        0.05         &        0.07         &        0.07         &        0.08         &        0.04         &        0.07         &        0.05         &        0.07         &        0.07         &        0.07         \\
Quarter \& industry FEs&$\checkmark$         &$\checkmark$         &$\checkmark$         &$\checkmark$         &$\checkmark$         &$\checkmark$         &$\checkmark$         &$\checkmark$         &$\checkmark$         &$\checkmark$         \\
All controls        &$\checkmark$         &$\checkmark$         &$\checkmark$         &$\checkmark$         &$\checkmark$         &$\checkmark$         &$\checkmark$         &$\checkmark$         &$\checkmark$         &$\checkmark$         \\
\bottomrule
\end{tabular}
}
\caption*{\footnotesize \textit{Notes:} As Table~\ref{tab:field_vu_main}, using the earnings-surprise definition in each column pair. ``Median forecast error / forecast SD'' is defined as the actual EPS minus the median forecast of EPS divided by the standard deviation of forecasts.  ``Mean forecast error / absolute consensus'' divides actual EPS minus the mean forecast of EPS by the absolute value of the consensus forecast. Both are winsorized at the first and ninety-ninth percentiles. The SUE rank measure separately assigns negative and positive surprises to 50 equal-count groups, while putting all zero surprises in their own group. All alternatives are rescaled to the standard deviation of the paper's SUE. The common sample requires a nonzero forecast standard deviation and at least two analyst estimates. All columns have the same hinge scale of $k=0.23\%$.}
\end{table}

\subsection{Heterogeneity and Sources of Variation}\label{app:field_heterogeneity}

\paragraph{Firm size, profitability, and period.}
Appendix Table~\ref{tab:field_size_and_period} estimates the baseline specification separately by firm size, profitability, and period. Size terciles are formed within year-quarter. We define profit-making announcements as having nonnegative reported EPS, while loss-making announcements have negative EPS. 

Since we aim to study the effects of Reg FD (promulgated in August 2000 and entered into force in October 2000), which tightened disclosure practices in the US, we split the sample in 2000Q4, which is also roughly the middle of our sample. The table retains the full-sample hinge threshold and VU standardization, so that coefficients can be compared across periods.

\begin{table}[!htb]
\centering
\caption{Valuation Uncertainty by Firm Size, Profitability, and Period}\label{tab:field_size_and_period}
\fittable{\begin{tabular}{l r@{}l@{\hspace{0.3em}}l r@{}l@{\hspace{0.3em}}l r@{}l@{\hspace{0.3em}}l r}
\toprule
 & \multicolumn{3}{c}{Global} & \multicolumn{3}{c}{Local} & \multicolumn{3}{c}{Relative} & \\
 & \multicolumn{3}{c}{VU \(\times\) SUE} & \multicolumn{3}{c}{VU \(\times\) Small SUE} & \multicolumn{3}{c}{VU \(\times\) \(\Delta\) Small SUE} & Obs. \\
\midrule
\quad Full sample & -0.285 & \sym{***} & (0.065) & 0.880 & \sym{***} & (0.245) & 1.011 & \sym{***} & (0.257) & 224,350 \\
\multicolumn{11}{l}{\textit{Panel A: by within-quarter market-capitalization}} \\
\quad Smallest tercile & -0.330 & \sym{***} & (0.063) & -0.943 & \sym{**} & (0.429) & -0.811 & \sym{*} & (0.469) & 74,734 \\
\quad Middle tercile & 0.033 &  & (0.087) & 1.094 & \sym{***} & (0.320) & 1.033 & \sym{***} & (0.348) & 74,786 \\
\quad Largest tercile & 0.117 &  & (0.112) & 2.555 & \sym{***} & (0.255) & 2.760 & \sym{***} & (0.311) & 74,830 \\
\addlinespace
\multicolumn{11}{l}{\textit{Panel B: by the sign of reported earnings}} \\
\quad Profit-making & -0.037 &  & (0.083) & 2.357 & \sym{***} & (0.239) & 2.283 & \sym{***} & (0.253) & 183,898 \\
\quad Loss-making & -0.142 & \sym{***} & (0.051) & -0.300 &  & (0.375) & -0.185 &  & (0.415) & 40,452 \\
\addlinespace
\multicolumn{11}{l}{\textit{Panel C: before and after Regulation FD}} \\
\quad Before Regulation FD & 0.038 &  & (0.049) & 1.460 & \sym{***} & (0.222) & 1.480 & \sym{***} & (0.240) & 89,050 \\
\quad After Regulation FD & -0.786 & \sym{***} & (0.062) & -0.633 & \sym{**} & (0.267) & -0.277 &  & (0.313) & 135,300 \\
\addlinespace
\multicolumn{11}{l}{\textit{Panel D: by the sign of reported earnings, before and after Regulation FD}} \\
\quad Profit-making, before Regulation FD & 0.263 & \sym{***} & (0.059) & 2.363 & \sym{***} & (0.228) & 2.293 & \sym{***} & (0.257) & 76,431 \\
\quad Profit-making, after Regulation FD & -0.796 & \sym{***} & (0.080) & 1.339 & \sym{***} & (0.312) & 1.642 & \sym{***} & (0.356) & 107,467 \\
\quad Loss-making, before Regulation FD & -0.075 &  & (0.062) & -0.647 &  & (0.490) & -0.615 &  & (0.532) & 12,619 \\
\quad Loss-making, after Regulation FD & -0.281 & \sym{***} & (0.067) & -0.411 &  & (0.469) & -0.193 &  & (0.529) & 27,832 \\
\bottomrule
\end{tabular}
}
\caption*{\footnotesize \textit{Notes:} Each row re-estimates the linear and hinge specifications with all fixed effects and controls in Table~\ref{tab:field_vu_main} on the indicated subsample and reports global, local, and relative changes in sensitivity. The global attenuation measure ``VU $\times$ SUE'' comes from the linear specification, while the local measure ``VU $\times$ Small SUE'' and the relative measure ``VU $\times\Delta$ Small SUE'' comes from the hinge specification. Each row uses identical observations and controls, while also retaining the full-sample hinge threshold and VU standardization. Panel A forms market-capitalization terciles within year-quarter. Panel B splits by reported EPS sign, assigning zero to the profit-making group. Panel C splits at 2000Q4. Finally, Panel D jointly splits by profitability and period.}
\end{table}

\begin{figure}[!htbp]
\centering
\caption{Earnings Response by Profitability}\label{fig:field_sshape_by_profit_all}
\includegraphics[width=\linewidth]{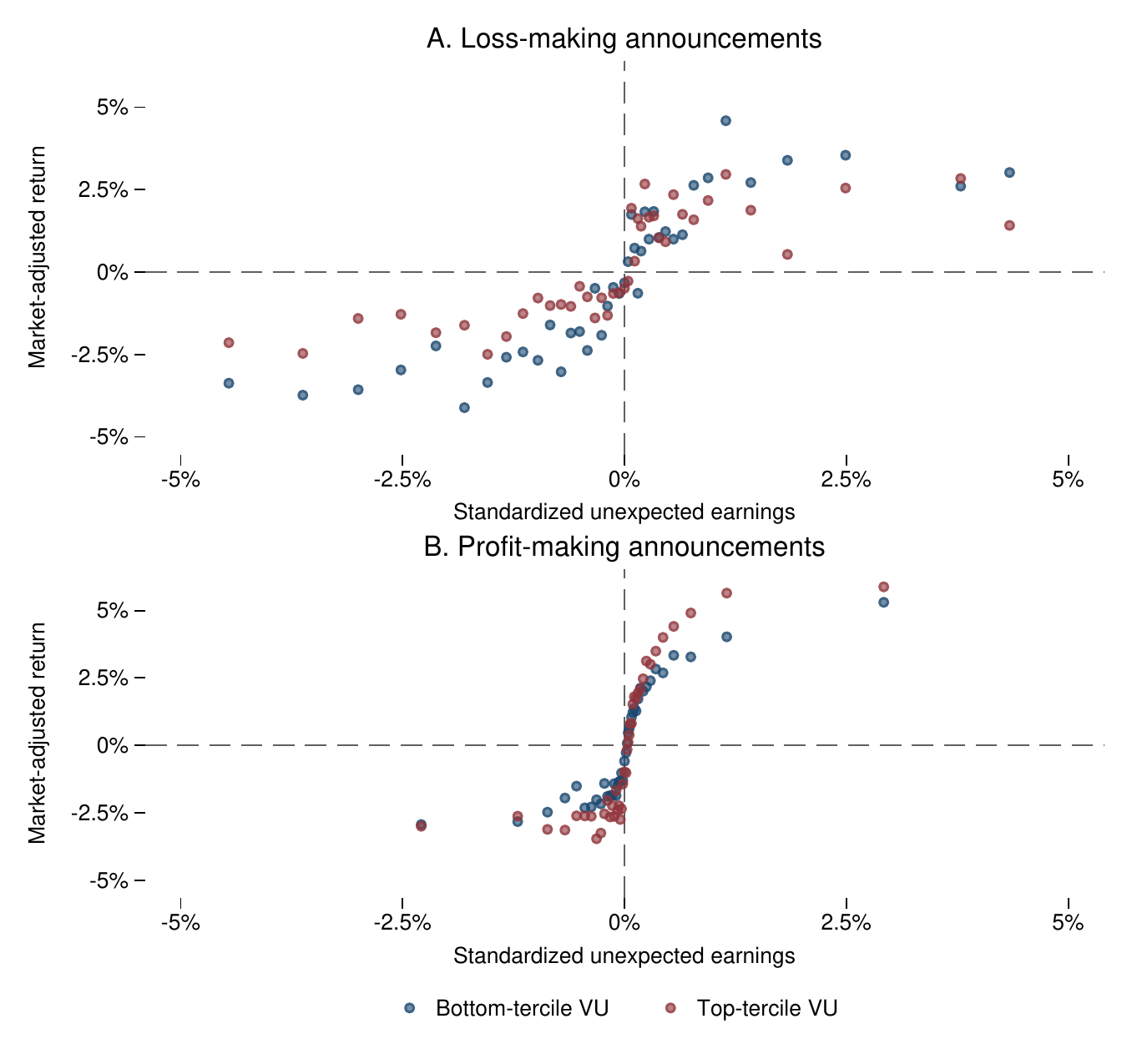}
\caption*{\footnotesize \textit{Notes:} Binned five-day market-adjusted returns for loss-making (Panel A) and profit-making (Panel B) announcements. Within each panel, announcements are split into bottom and top VU terciles within year-quarter and sorted into common equal-count SUE bins, with exact zeros in a separate bin. Fitted group levels are subtracted from returns. Note that zero reported EPS are included in the profit-making group.}
\end{figure}

Appendix Figure~\ref{fig:field_sshape_by_profit_all} plots the reaction function of returns separately for loss- and profit-making firms. It shows quite starkly that the $S$-shape is strong among profit-making firms, but impossible to make out among loss-making ones. In turn, this also makes the split by VU drive a clearly perceptible steepening among profit-making firms, but have no clear effect among loss-making ones.

This pattern has been observed as far back as \citet[][Table 6]{hayn1995information}, and identified more forcefully by \citet{gipper2020public}. In particular, the latter study the effects of PCAOB oversight on return reaction functions, and therefore split their analysis by profit- and loss-making firms. We find a similar interaction with the effect of VU on the reaction function, and hence also separate our analysis.

\begin{figure}[!htbp]
\centering
\caption{Share of Loss-Making Announcements}\label{fig:field_loss_share}
\includegraphics[width=\linewidth]{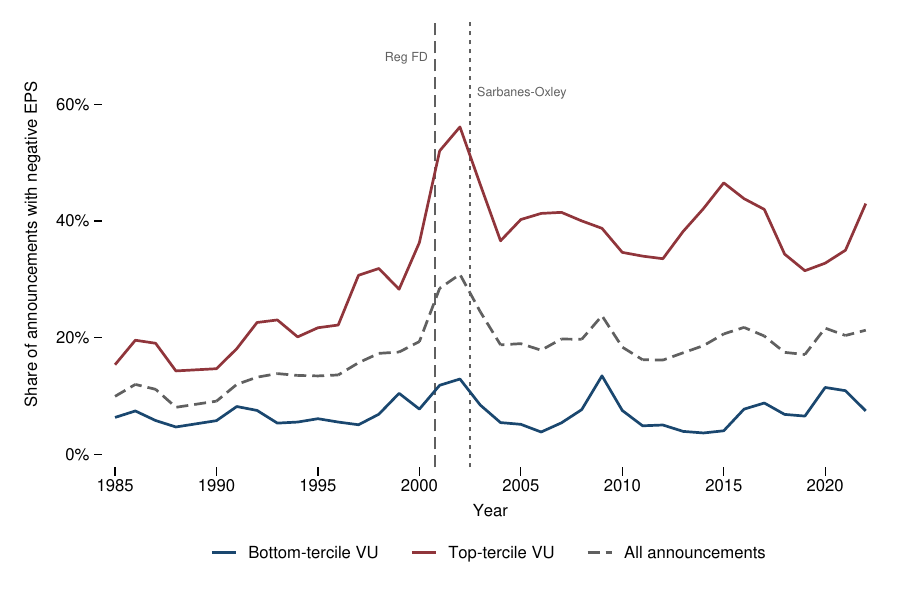}
\caption*{\footnotesize \textit{Notes:} Annual shares of announcements with negative reported EPS in the estimation sample, overall and in the bottom and top VU terciles formed within year-quarter. Years with fewer than four observed quarters are omitted. Vertical lines mark 2000Q4 (Regulation FD) and 2002Q3 (Sarbanes-Oxley).}
\end{figure}

\begin{figure}[!htbp]
\centering
\caption{Earnings Response by Profitability before and after Regulation FD}\label{fig:field_sshape_by_profit_period}
\includegraphics[width=\linewidth]{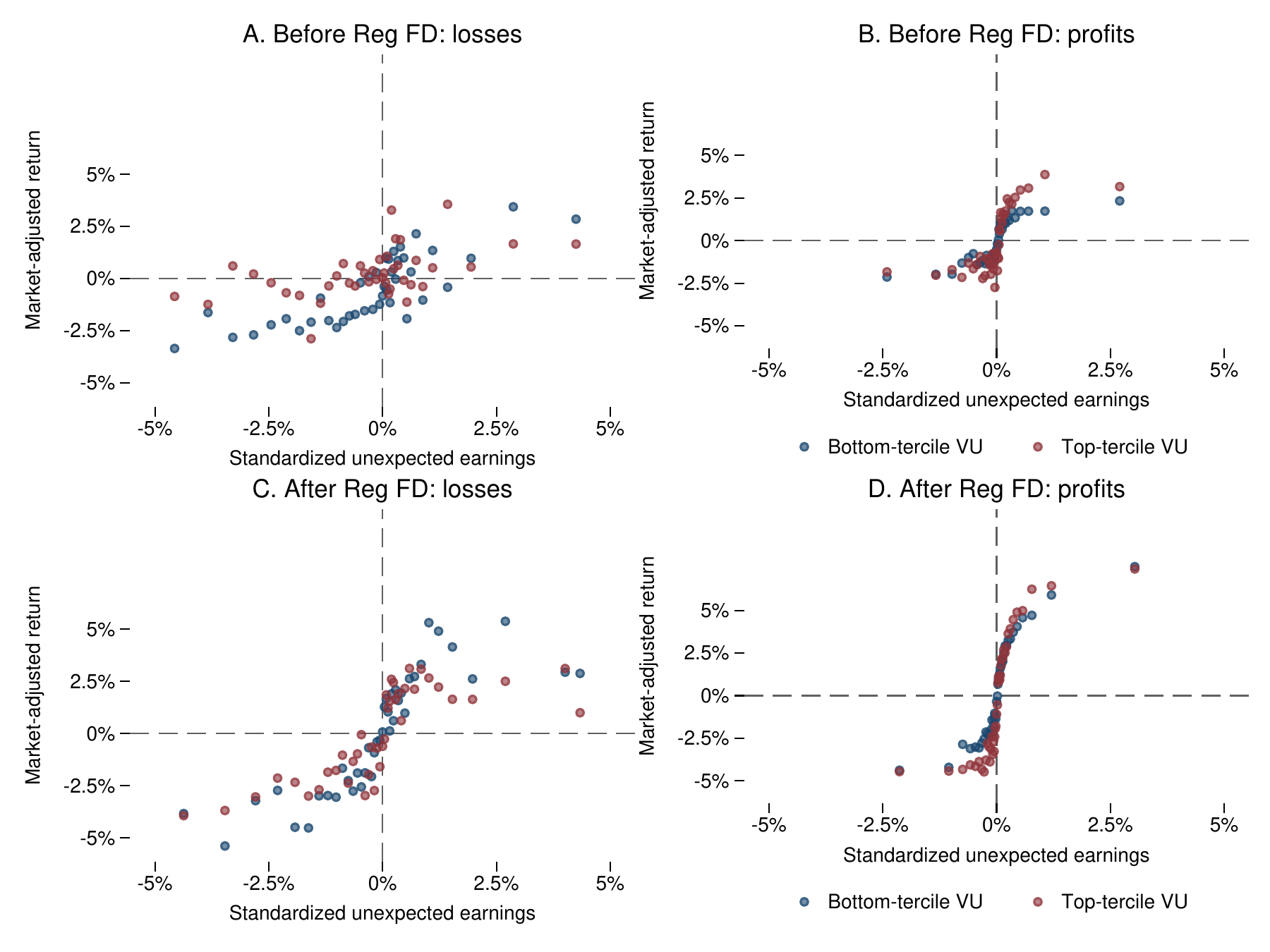}
\caption*{\footnotesize \textit{Notes:} As Appendix Figure~\ref{fig:field_sshape_by_profit_all}, with each profitability group split at 2000Q4. Panels A and B contain only observations from before Regulation FD, whereas Panels C and D contain only observations from after Regulation FD was passed. The left column contains announcements with $EPS<0$ while the right column contains announcements with $EPS \geq 0$. Bins and within-quarter terciles of VU are formed within each panel's sample.}
\end{figure}

\begin{figure}[!htbp]
\centering
\caption{Relative Central Sensitivity over Time, by Profitability}
\label{fig:field_coefficients_by_profit}
\includegraphics[width=\linewidth]{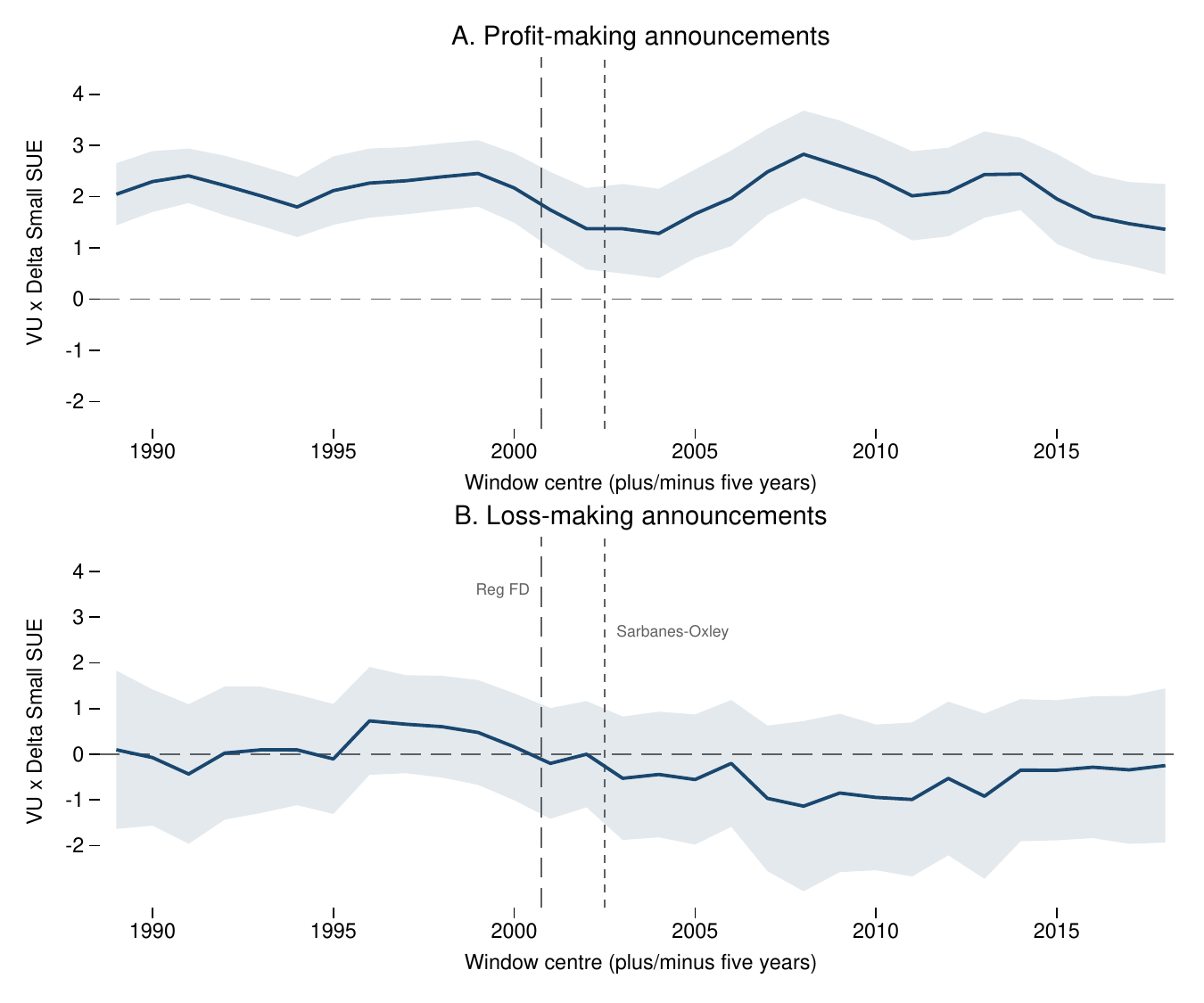}
\caption*{\footnotesize \textit{Notes:} The figure reports VU $\times\Delta$ Small SUE from regressions by profitability group in eleven-calendar-year windows $c-5$ through $c+5$, inclusive. Both groups retain the full-sample hinge threshold $k=0.23\%$, but VU is standardized separately within each group's full sample and held fixed across its rolling windows. Regressions include the baseline controls and industry and year-quarter fixed effects. Shaded regions are pointwise 95\% intervals using year-quarter-clustered standard errors. Vertical lines mark 2000Q4 (Regulation FD) and 2002Q3 (Sarbanes-Oxley). Note that zero EPS are counted towards the profit-making group.}
\end{figure}

Appendix Figure~\ref{fig:field_sshape_by_profit_period} hence investigates the apparent instability of the amplification of small surprises before and after Reg FD by plotting the $S$-shape separately by period and by profitability. It documents a stable effect within each category, in line with model predictions in the case of profit-making firms. Appendix Figure~\ref{fig:field_loss_share} then shows that the apparent instability reflects a compositional shift in the IBES sample: the share of loss-making firms increases over time, especially around the passage of Reg FD, and it does so more strongly for high-VU firms. Appendix Figure~\ref{fig:field_coefficients_by_profit} hence estimates our main specification in rolling windows of $\pm 5$ years, separately for profit- and loss-making firms. It shows the excess sensitivity to small surprises due to VU is remarkably stable over time across profit-making firms.

\FloatBarrier

\begin{table}[!htb]
\centering
\caption{Between- and Within-Quarter Valuation Uncertainty}\label{tab:field_variation_quarter}
\fittable{\begin{tabular}{l*{3}{c}}
\toprule
\multicolumn{4}{c}{\textit{Dependent variable: five-day market-adjusted return}} \\
                    &\multicolumn{1}{c}{No controls}&\multicolumn{1}{c}{Quarter \& industry FEs}&\multicolumn{1}{c}{All controls}\\
\midrule
SUE                 &        0.42\sym{***}&        0.43\sym{***}&        0.44\sym{***}\\
                    &      (0.08)         &      (0.08)         &      (0.08)         \\
$\Delta$ Small SUE  &       11.90\sym{***}&       12.17\sym{***}&       12.25\sym{***}\\
                    &      (0.83)         &      (0.91)         &      (0.92)         \\
Between-quarter VU $\times$ SUE&        0.16         &        0.18         &        0.22         \\
                    &      (0.19)         &      (0.20)         &      (0.19)         \\
Between-quarter VU $\times$ $\Delta$ Small SUE&        2.29\sym{*}  &        2.06         &        1.96         \\
                    &      (1.19)         &      (1.23)         &      (1.23)         \\
Within-quarter VU $\times$ SUE&       -0.24\sym{***}&       -0.25\sym{***}&       -0.25\sym{***}\\
                    &      (0.08)         &      (0.08)         &      (0.08)         \\
Within-quarter VU $\times$ $\Delta$ Small SUE&        0.92         &        0.93         &        0.94         \\
                    &      (1.07)         &      (1.12)         &      (1.13)         \\
\midrule
Small SUE           &12.32\sym{***}         &12.60\sym{***}         &12.69\sym{***}         \\
                    &      (0.83)         &      (0.90)         &      (0.91)         \\
Between-quarter VU $\times$ Small SUE&2.44\sym{**}         & 2.24\sym{*}         & 2.18\sym{*}         \\
                    &      (1.18)         &      (1.21)         &      (1.22)         \\
Within-quarter VU $\times$ Small SUE&        0.69         &        0.68         &        0.69         \\
                    &      (1.03)         &      (1.07)         &      (1.09)         \\
\midrule
Hinge threshold \(k\)&        0.23         &        0.23         &        0.23         \\
Observations        &     224,350         &     224,350         &     224,350         \\
\(R^{2}\)           &        0.06         &        0.07         &        0.07         \\
Quarter \& industry FEs&                     &$\checkmark$         &$\checkmark$         \\
All controls        &                     &                     &$\checkmark$         \\
\bottomrule
\end{tabular}
}
\caption*{\footnotesize \textit{Notes:} VU is decomposed into the mean within each year-quarter and the announcement's deviation from that mean. Both components are divided by the standard deviation of total VU. All columns use the baseline sample and hinge threshold. Standard errors are clustered by industry and year-quarter. Column (1) has no controls or fixed effects; column (2) adds industry and year-quarter fixed effects; column (3) adds all baseline controls.}
\end{table}

\begin{table}[!htbp]
\centering
\caption{Between- and Within-Industry Valuation Uncertainty}\label{tab:field_variation_industry}
\fittable{\begin{tabular}{l*{3}{c}}
\toprule
\multicolumn{4}{c}{\textit{Dependent variable: five-day market-adjusted return}} \\
                    &\multicolumn{1}{c}{No controls}&\multicolumn{1}{c}{Quarter \& industry FEs}&\multicolumn{1}{c}{All controls}\\
\midrule
SUE                 &        0.36\sym{***}&        0.37\sym{***}&        0.37\sym{***}\\
                    &      (0.08)         &      (0.08)         &      (0.08)         \\
$\Delta$ Small SUE  &       12.10\sym{***}&       12.40\sym{***}&       12.49\sym{***}\\
                    &      (0.62)         &      (0.66)         &      (0.68)         \\
Between-industry VU $\times$ SUE&       -0.54\sym{***}&       -0.58\sym{***}&       -0.59\sym{***}\\
                    &      (0.17)         &      (0.19)         &      (0.19)         \\
Between-industry VU $\times$ $\Delta$ Small SUE&        4.89\sym{***}&        5.11\sym{***}&        5.11\sym{***}\\
                    &      (0.69)         &      (0.76)         &      (0.77)         \\
Within-industry VU $\times$ SUE&        0.02         &        0.03         &        0.03         \\
                    &      (0.07)         &      (0.07)         &      (0.07)         \\
Within-industry VU $\times$ $\Delta$ Small SUE&       -0.70         &       -0.86\sym{*}  &       -0.87\sym{*}  \\
                    &      (0.44)         &      (0.45)         &      (0.45)         \\
\midrule
Small SUE           &12.46\sym{***}         &12.77\sym{***}         &12.86\sym{***}         \\
                    &      (0.64)         &      (0.68)         &      (0.69)         \\
Between-industry VU $\times$ Small SUE&4.35\sym{***}         &4.53\sym{***}         &4.52\sym{***}         \\
                    &      (0.76)         &      (0.81)         &      (0.84)         \\
Within-industry VU $\times$ Small SUE&       -0.67         &-0.84\sym{*}         &-0.84\sym{*}         \\
                    &      (0.43)         &      (0.44)         &      (0.44)         \\
\midrule
Hinge threshold \(k\)&        0.23         &        0.23         &        0.23         \\
Observations        &     224,350         &     224,350         &     224,350         \\
\(R^{2}\)           &        0.06         &        0.07         &        0.07         \\
Quarter \& industry FEs&                     &$\checkmark$         &$\checkmark$         \\
All controls        &                     &                     &$\checkmark$         \\
\bottomrule
\end{tabular}
}
\caption*{\footnotesize \textit{Notes:} VU is decomposed into the mean within each FF49 industry and the announcement's deviation from that mean. Both components are divided by the standard deviation of total VU. All columns use the baseline sample and hinge threshold. Standard errors are clustered by industry and year-quarter. Column (1) has no controls or fixed effects; column (2) adds industry and year-quarter fixed effects; column (3) adds all baseline controls.}
\end{table}

\FloatBarrier
\clearpage

\subsection{Relation to Alternative Explanations}\label{app:altexplanations}

\paragraph{Earnings persistence and future fundamentals.}

Panel A of Appendix Table~\ref{tab:field_future_fundamentals} adds future earnings measures to the baseline return regressions, since the literature going back to \cite{freeman1992nonlinear} has documented that these could explain the shape of the reaction function. We control for the future SUE at the next and fourth subsequent announcements, both rescaled by the price \textit{preceding each announcement}. We then additionally control for future EPS growth, i.e., the change from current EPS to EPS one and four quarters ahead, both scaled by the \textit{current} pre-announcement price. Each block interacts with VU.

\begin{table}[!htbp]
\centering
\caption{Future Fundamentals and Contemporaneous Guidance}
\label{tab:field_future_fundamentals}
\fittable{\begin{tabular}{l r@{}l@{\hspace{0.3em}}l r@{}l@{\hspace{0.3em}}l r@{}l@{\hspace{0.3em}}l r}
\toprule
 & \multicolumn{3}{c}{Global} & \multicolumn{3}{c}{Local} & \multicolumn{3}{c}{Relative} & \\
 & \multicolumn{3}{c}{VU \(\times\) SUE} & \multicolumn{3}{c}{VU \(\times\) Small SUE} & \multicolumn{3}{c}{VU \(\times\) \(\Delta\) Small SUE} & Obs. \\
\midrule
\quad Baseline (Table \ref{tab:field_vu_main}, columns 5--6) & -0.285 & \sym{***} & (0.065) & 0.880 & \sym{***} & (0.245) & 1.011 & \sym{***} & (0.257) & 224,350 \\
\multicolumn{11}{l}{\textit{Panel A: future fundamentals}} \\
\quad Baseline on the common sample & -0.253 & \sym{***} & (0.078) & 1.005 & \sym{***} & (0.253) & 1.086 & \sym{***} & (0.264) & 187,154 \\
\quad Controlling for future SUE and its VU interactions & -0.256 & \sym{***} & (0.077) & 0.977 & \sym{***} & (0.251) & 1.059 & \sym{***} & (0.263) & 187,154 \\
\quad Controlling for future EPS growth and its VU interactions & -0.322 & \sym{***} & (0.082) & 0.983 & \sym{***} & (0.252) & 1.135 & \sym{***} & (0.263) & 187,154 \\
\quad Controlling for both & -0.370 & \sym{***} & (0.082) & 0.979 & \sym{***} & (0.253) & 1.178 & \sym{***} & (0.265) & 187,154 \\
\addlinespace
\multicolumn{11}{l}{\textit{Panel B: contemporaneous guidance}} \\
\quad Excluding announcements with EPS guidance & -0.237 & \sym{***} & (0.062) & 0.923 & \sym{***} & (0.248) & 1.050 & \sym{***} & (0.262) & 197,240 \\
\bottomrule
\end{tabular}
}
\caption*{\footnotesize \textit{Notes:} The dependent variable is the five-day market-adjusted return. The global attenuation measure ``VU $\times$ SUE'' comes from the linear specification, while the local measure ``VU $\times$ Small SUE'' and the relative measure ``VU $\times\Delta$ Small SUE'' comes from the hinge specification. Each row uses identical observations and controls, while also retaining the baseline hinge threshold and VU standardization. Future SUE controls are the reported EPS minus the consensus forecast at the next and fourth subsequent announcements, divided by the price before those announcements. EPS-growth controls are the changes from current EPS to EPS one and four quarters ahead, divided by the current pre-announcement price. Each block also enters its interaction with VU. The future variables are winsorized at the first and ninety-ninth percentiles. Panel B excludes contemporaneous management EPS-guidance announcements matched in IBES. Standard errors are clustered by year-quarter.}
\end{table}

\paragraph{Information incorporation.}
Appendix Table~\ref{tab:field_preannouncement} enters VU jointly with two information-incorporation proxies: the absolute first-tradable-day announcement return divided by the standard deviation of returns over the preceding 22 trading days \citep{sammon2025passive}, and the price-jump ratio of \citet{weller2018algorithmic}. Both measures enter as main effects and through interactions with the response terms, in the format of Table~\ref{tab:field_joint_attenuation} \citep{atiase1985predisclosure}.

\begin{table}[!htbp]
    \centering
\caption{Valuation Uncertainty and Information-Incorporation Proxies}
    \label{tab:field_preannouncement}
    \fittable{\begin{tabular}{l*{4}{Y}}
\toprule
& \multicolumn{4}{c}{\textit{Dependent variable: five-day market-adjusted return}} \\
& \multicolumn{2}{c}{\shortstack[c]{Absolute return,\\normalized}} & \multicolumn{2}{c}{\shortstack[c]{Price\\jump}} \\
\cmidrule(lr){2-3}\cmidrule(lr){4-5}
& (1) & (2) & (3) & (4) \\
\midrule
SUE                 &        2.35\sym{***}&        0.65\sym{***}&        3.82\sym{***}&        0.94\sym{***}\\
                    &      (0.05)         &      (0.06)         &      (0.10)         &      (0.09)         \\
$\Delta$ Small SUE  &                     &       12.39\sym{***}&                     &       20.88\sym{***}\\
                    &                     &      (0.33)         &                     &      (0.51)         \\
VU $\times$ SUE     &       -0.39\sym{***}&       -0.14\sym{***}&       -0.73\sym{***}&       -0.37\sym{***}\\
                    &      (0.04)         &      (0.05)         &      (0.08)         &      (0.09)         \\
VU $\times$ $\Delta$ Small SUE&                     &        0.24         &                     &        1.40\sym{***}\\
                    &                     &      (0.23)         &                     &      (0.36)         \\
Candidate $\times$ SUE&        2.41\sym{***}&        0.62\sym{***}&        2.74\sym{***}&        0.59\sym{***}\\
                    &      (0.06)         &      (0.05)         &      (0.08)         &      (0.09)         \\
Candidate $\times$ $\Delta$ Small SUE&                     &       14.14\sym{***}&                     &       16.98\sym{***}\\
                    &                     &      (0.32)         &                     &      (0.46)         \\
\midrule
Small SUE           &                     &13.04\sym{***}         &                     &21.82\sym{***}         \\
                    &                     &      (0.31)         &                     &      (0.47)         \\
VU $\times$ Small SUE&                     &        0.11         &                     &1.03\sym{***}         \\
                    &                     &      (0.20)         &                     &      (0.31)         \\
Candidate $\times$ Small SUE&                     &14.75\sym{***}         &                     &17.57\sym{***}         \\
                    &                     &      (0.30)         &                     &      (0.42)         \\
\midrule
Hinge threshold \(k\)&                     &        0.23         &                     &        0.23         \\
Observations        &     210,095         &     210,095         &      78,716         &      78,716         \\
\(R^{2}\)           &        0.10         &        0.16         &        0.14         &        0.22         \\
Quarter \& industry FEs&$\checkmark$         &$\checkmark$         &$\checkmark$         &$\checkmark$         \\
All controls        &$\checkmark$         &$\checkmark$         &$\checkmark$         &$\checkmark$         \\
\bottomrule
\end{tabular}
}
\caption*{\footnotesize \textit{Notes:} As Table~\ref{tab:field_joint_attenuation}, entering VU jointly with the measure named in each column pair. The volatility-normalized return is the absolute market-adjusted return on the first tradable announcement day divided by the standard deviation of market-adjusted daily returns over the preceding 22 trading days \citep{sammon2025passive}. The price jump is $\mathrm{CAR}_{[-1,+2]}/\mathrm{CAR}_{[-22,+2]}$, using market-adjusted returns. Note that its ``non-event filter'' omits observations with small total-window CAR relative to prior volatility \citep{weller2018algorithmic,sammon2025passive}. Both measures are winsorized at the first and ninety-ninth percentiles, and then standardized.}
\end{table}

\FloatBarrier

\subsection{Post-Announcement Returns}\label{app:field_postannouncement}

\begin{figure}[!htbp]
\centering
\caption{Post-Announcement Returns by Valuation Uncertainty}
    \label{fig:field_response_pead}
    \includegraphics[width=\linewidth]{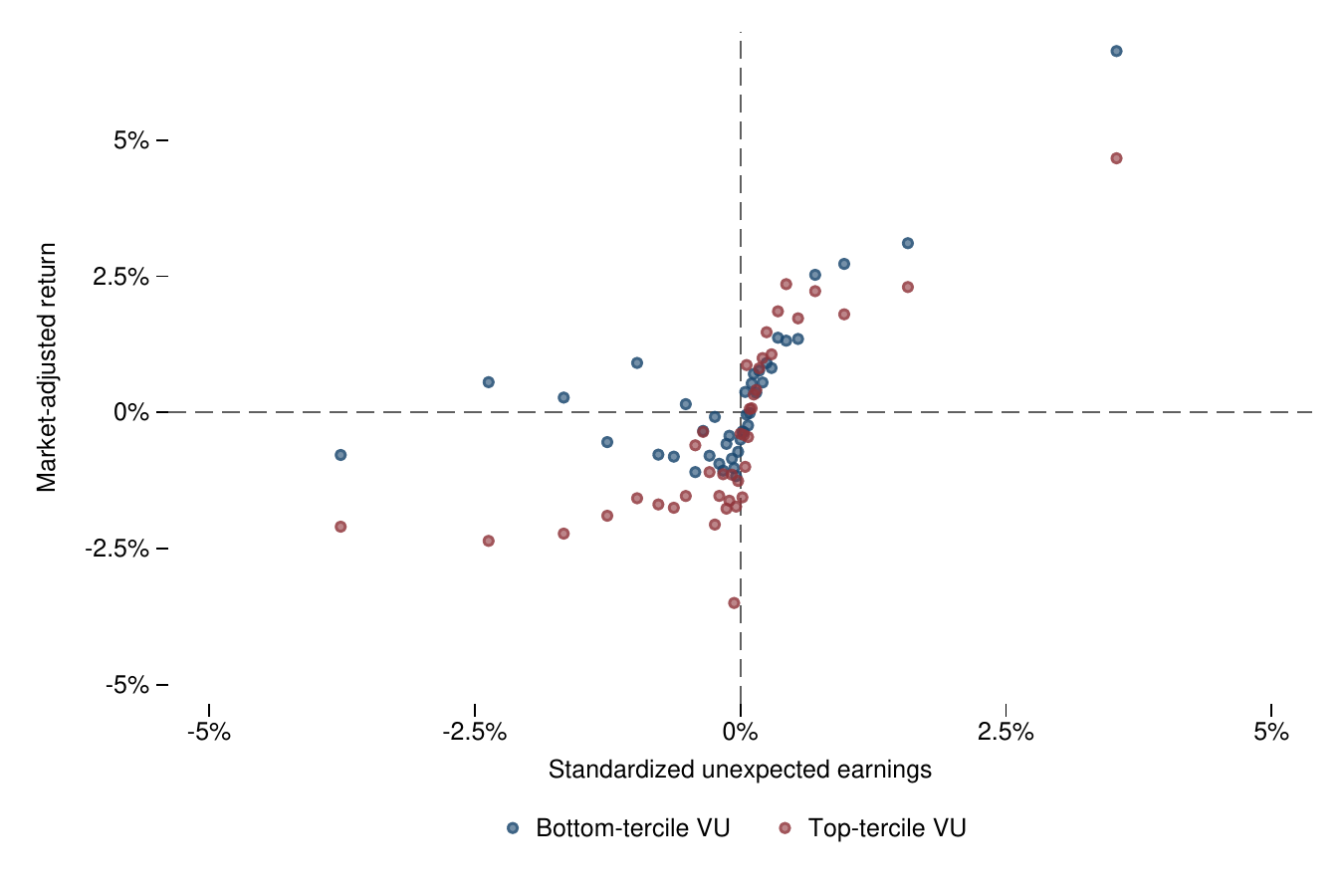}
\caption*{\footnotesize \textit{Notes:} As Appendix Figure~\ref{fig:field_sshape_by_vu_unfitted}, replacing the announcement-window return with the cumulative market-adjusted return over days $[5,59]$ after the announcement. Announcements are split into bottom and top valuation-uncertainty terciles within year-quarter using common SUE bins spanning $-5\%$ to $+5\%$, and group-specific fitted levels are removed from both series for comparable vertical positions. No response shape is fitted.}
\end{figure}

Appendix Figure~\ref{fig:field_response_pead} plots market-adjusted returns over trading days 5--59 without fitting any functional form, which show strong unconditional post-earnings announcement drift but no clear pattern by VU. To check the robustness of the main result, Appendix Table~\ref{tab:field_pead_start_days} repeats the hinge regressions of Table~\ref{tab:field_pead_horizons} with windows beginning on day 1 or day 2 and ending on day 21, 43, or 59. The table uses the sample and hinge threshold of Table~\ref{tab:field_pead_horizons}. As usual, returns are winsorized at the first and ninety-ninth percentiles.

\begin{table}[!htbp]
\centering
\caption{Valuation Uncertainty and Post-Announcement Returns by Start Day}
\label{tab:field_pead_start_days}
\fittable{\begin{tabular}{l*{6}{Y}}
\toprule
& \multicolumn{6}{c}{\textit{Dependent variable: post-announcement market-adjusted return}} \\
& \multicolumn{3}{c}{Window begins on day 1} & \multicolumn{3}{c}{Window begins on day 2} \\
\cmidrule(lr){2-4}\cmidrule(lr){5-7}
& \shortstack[c]{Days\\1--21} & \shortstack[c]{Days\\1--43} & \shortstack[c]{Days\\1--59} & \shortstack[c]{Days\\2--21} & \shortstack[c]{Days\\2--43} & \shortstack[c]{Days\\2--59} \\
& (1) & (2) & (3) & (4) & (5) & (6) \\
\midrule
SUE                 &        0.43\sym{***}&        0.56\sym{***}&        0.64\sym{***}&        0.35\sym{***}&        0.48\sym{***}&        0.55\sym{***}\\
                    &      (0.07)         &      (0.10)         &      (0.12)         &      (0.07)         &      (0.10)         &      (0.12)         \\
$\Delta$ Small SUE  &        2.86\sym{***}&        4.49\sym{***}&        5.76\sym{***}&        1.44\sym{***}&        3.09\sym{***}&        4.36\sym{***}\\
                    &      (0.39)         &      (0.63)         &      (0.76)         &      (0.33)         &      (0.58)         &      (0.71)         \\
VU                  &       -0.05         &        0.06         &       -0.01         &       -0.01         &        0.09         &        0.03         \\
                    &      (0.10)         &      (0.17)         &      (0.22)         &      (0.09)         &      (0.17)         &      (0.21)         \\
VU $\times$ SUE     &       -0.06         &       -0.09         &       -0.06         &       -0.04         &       -0.06         &       -0.04         \\
                    &      (0.07)         &      (0.10)         &      (0.13)         &      (0.07)         &      (0.10)         &      (0.12)         \\
VU $\times$ $\Delta$ Small SUE&       -0.03         &       -0.11         &       -0.08         &       -0.19         &       -0.28         &       -0.21         \\
                    &      (0.32)         &      (0.46)         &      (0.61)         &      (0.29)         &      (0.44)         &      (0.59)         \\
\midrule
Small SUE           &3.29\sym{***}         &5.06\sym{***}         &6.39\sym{***}         &1.79\sym{***}         &3.57\sym{***}         &4.92\sym{***}         \\
                    &      (0.34)         &      (0.57)         &      (0.68)         &      (0.29)         &      (0.51)         &      (0.63)         \\
VU $\times$ Small SUE&       -0.10         &       -0.20         &       -0.14         &       -0.23         &       -0.34         &       -0.26         \\
                    &      (0.27)         &      (0.40)         &      (0.54)         &      (0.25)         &      (0.39)         &      (0.53)         \\
\midrule
Hinge threshold \(k\)&        0.23         &        0.23         &        0.23         &        0.23         &        0.23         &        0.23         \\
Observations        &     222,536         &     222,536         &     222,536         &     222,536         &     222,536         &     222,536         \\
\(R^{2}\)           &        0.05         &        0.06         &        0.07         &        0.05         &        0.06         &        0.07         \\
Quarter \& industry FEs&$\checkmark$         &$\checkmark$         &$\checkmark$         &$\checkmark$         &$\checkmark$         &$\checkmark$         \\
All controls        &$\checkmark$         &$\checkmark$         &$\checkmark$         &$\checkmark$         &$\checkmark$         &$\checkmark$         \\
\bottomrule
\end{tabular}
}
\caption*{\footnotesize \textit{Notes:} As Table~\ref{tab:field_pead_horizons}, using cumulative market-adjusted returns over post-announcement windows beginning on day 1 or day 2 and ending on day 21, 43, or 59, winsorized at the first and ninety-ninth percentiles. The sample and hinge threshold are those of Table~\ref{tab:field_pead_horizons}. Standard errors are clustered by year-quarter.}
\end{table}

\begin{table}[!htbp]
    \centering
    \caption{Valuation Uncertainty and Cumulative Returns}
    \label{tab:field_cumulative_returns}

    \fittable{\begin{tabular}{l*{6}{Y}}
\toprule
& \multicolumn{6}{c}{\textit{Dependent variable: cumulative market-adjusted return from announcement day}} \\
                    &\multicolumn{1}{c}{Days 0--1}&\multicolumn{1}{c}{Days 0--2}&\multicolumn{1}{c}{Days 0--4}&\multicolumn{1}{c}{Days 0--21}&\multicolumn{1}{c}{Days 0--29}&\multicolumn{1}{c}{Days 0--59}\\
\midrule
SUE                 &       0.319\sym{***}&       0.336\sym{***}&       0.386\sym{***}&       0.704\sym{***}&       0.794\sym{***}&       0.940\sym{***}\\
                    &     (0.057)         &     (0.059)         &     (0.067)         &     (0.106)         &     (0.119)         &     (0.158)         \\
$\Delta$ Small SUE  &      11.819\sym{***}&      12.143\sym{***}&      12.386\sym{***}&      13.479\sym{***}&      13.702\sym{***}&      16.276\sym{***}\\
                    &     (0.420)         &     (0.436)         &     (0.456)         &     (0.489)         &     (0.551)         &     (0.720)         \\
VU                  &      -0.055\sym{*}  &      -0.079\sym{**} &      -0.118\sym{***}&      -0.050         &      -0.051         &      -0.006         \\
                    &     (0.032)         &     (0.036)         &     (0.044)         &     (0.106)         &     (0.134)         &     (0.220)         \\
VU $\times$ SUE     &      -0.112\sym{***}&      -0.101\sym{**} &      -0.121\sym{**} &      -0.171\sym{*}  &      -0.189\sym{*}  &      -0.171         \\
                    &     (0.043)         &     (0.049)         &     (0.055)         &     (0.093)         &     (0.100)         &     (0.141)         \\
VU $\times$ $\Delta$ Small SUE&       1.146\sym{***}&       1.000\sym{***}&       1.019\sym{***}&       1.064\sym{***}&       0.986\sym{**} &       1.000         \\
                    &     (0.217)         &     (0.235)         &     (0.254)         &     (0.380)         &     (0.384)         &     (0.642)         \\
\midrule
Small SUE           &12.14\sym{***}         &12.48\sym{***}         &12.77\sym{***}         &14.18\sym{***}         &14.50\sym{***}         &17.22\sym{***}         \\
                    &      (0.42)         &      (0.43)         &      (0.45)         &      (0.46)         &      (0.50)         &      (0.63)         \\
VU $\times$ Small SUE&1.03\sym{***}         &0.90\sym{***}         &0.90\sym{***}         &0.89\sym{**}         &0.80\sym{**}         &        0.83         \\
                    &      (0.22)         &      (0.23)         &      (0.24)         &      (0.35)         &      (0.35)         &      (0.58)         \\
\midrule
Hinge threshold \(k\)&       0.230         &       0.230         &       0.230         &       0.230         &       0.230         &       0.230         \\
Observations        &     222,536         &     222,536         &     222,536         &     222,536         &     222,536         &     222,536         \\
\(R^{2}\)           &       0.079         &       0.074         &       0.070         &       0.072         &       0.075         &       0.082         \\
Quarter \& industry FEs&$\checkmark$         &$\checkmark$         &$\checkmark$         &$\checkmark$         &$\checkmark$         &$\checkmark$         \\
All controls        &$\checkmark$         &$\checkmark$         &$\checkmark$         &$\checkmark$         &$\checkmark$         &$\checkmark$         \\
\bottomrule
\end{tabular}
}

    \caption*{\footnotesize \textit{Notes:}
    As Table~\ref{tab:field_pead_horizons}, using cumulative
    market-adjusted returns from announcement day~0 through day 1, 2, 4, 21, 29 or 59. Returns are winsorized
    at the first and ninety-ninth percentiles. The sample and hinge scale are exactly as in Table~\ref{tab:field_pead_horizons}. Standard errors are clustered by year-quarter.}
\end{table}

\begin{figure}[p]
    \centering
    \caption{Valuation Uncertainty and the Persistence of the Earnings Response}
    \label{fig:field_cumulative_returns}

    \includegraphics[height=.85\textheight]{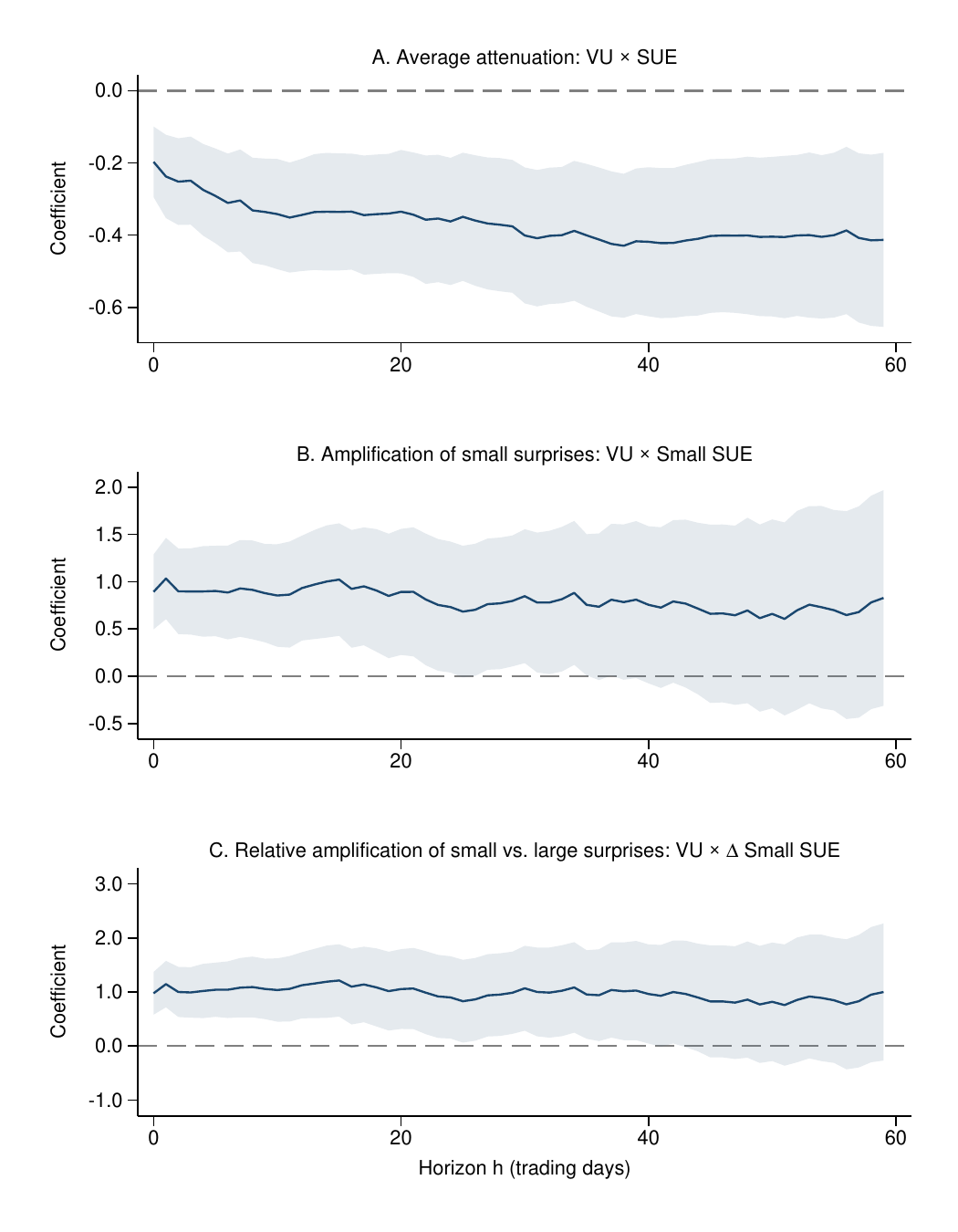}

    \caption*{\footnotesize \textit{Notes:}
    Cumulative return reaction to VU and SUE, beginning on announcement day~0 and ending on the trading day indicated on the horizontal axis. Panel~A regresses the return on SUE and VU $\times$ SUE, as well as controls plus quarter and industry FEs, so that its 5-day return coefficient matches column~(5) of Table~\ref{tab:field_event_windows}. Panels~B and~C  regress the return on SUE, $\Delta$ Small SUE, VU, VU $\times$ SUE and VU $\times$ $\Delta$ Small SUE, plus controls and FEs, so that their 2-, 3-, 5-, 22-, 30- and 60-day coefficients match those of Table~\ref{tab:field_cumulative_returns}. Shaded bands indicate 95\% confidence intervals, with standard errors clustered by year-quarter.}
\end{figure}

\end{document}